\documentclass[rmp,amsmath,nofootinbib, amssymb,aps, onecolumn,notitlepage]{revtex4-1}

\usepackage[colorlinks=true, citecolor = brown, linktocpage=true]{hyperref}

\usepackage{wrapfig}
\usepackage{geometry}  
\usepackage{enumitem}
\usepackage{natbib}
\usepackage{ulem}

\usepackage{color}

\usepackage{microtype}

\usepackage{amsthm}
\usepackage{dsfont}

\usepackage{graphicx}
\usepackage{epstopdf}
\usepackage{dcolumn}
\usepackage{bm}

\usepackage{bbm}

\def\EQ#1{\begin{eqnarray}#1\end{eqnarray}}

\def\ket#1{\left| #1 \right\rangle}
\def\bra#1{\left\langle #1 \right|}

\def\dm#1{\left|#1 \right\rangle \left\langle #1 \right|}

\usepackage[dvipsnames]{xcolor}
\definecolor{ao}{rgb}{0.0, 0.5, 0.0}

\usepackage{tcolorbox}

\makeatletter
\newcommand{\boxTLDR}[1]{%
  \setbox0=\hbox{#1}%
  \setlength{\@tempdima}{\dimexpr\wd0+13pt}%
  \begin{tcolorbox}[colframe=blue,boxrule=0.5pt,arc=4pt,
      left=6pt,right=6pt,top=6pt,bottom=6pt,boxsep=0pt,width=\textwidth]
 \underline{\textbf{Executive summary:}}  #1
  \end{tcolorbox}
}
\makeatother

\makeatletter
\newcommand{\boxPretty}[1]{%
  \setbox0=\hbox{#1}%
  \setlength{\@tempdima}{\dimexpr\wd0+13pt}%
  \begin{tcolorbox}[colframe=brown,boxrule=0.3pt,arc=2pt,
      left=6pt,right=6pt,top=6pt,bottom=6pt,boxsep=0pt,width=\textwidth]
    #1
  \end{tcolorbox}
}
\makeatother

\makeatletter
\newcommand{\boxPrettyTwo}[1]{%
  \setbox0=\hbox{#1}%
  \setlength{\@tempdima}{\dimexpr\wd0+13pt}%
  \begin{tcolorbox}[colframe=brown,boxrule=0.3pt,arc=2pt,
      left=6pt,right=6pt,top=6pt,bottom=6pt,boxsep=0pt,width=0.53\textwidth]
    #1
  \end{tcolorbox}
}
\makeatother

\makeatletter
\newcommand{\boxPrettyThree}[1]{%
  \setbox0=\hbox{#1}%
  \setlength{\@tempdima}{\dimexpr\wd0+13pt}%
  \begin{tcolorbox}[colframe=brown,boxrule=0.3pt,arc=2pt,
      left=6pt,right=6pt,top=6pt,bottom=6pt,boxsep=0pt,width=0.82\textwidth]
    #1
  \end{tcolorbox}
}
\makeatother

\makeatletter
\newtheorem*{rep@theorem}{\rep@title}
\newcommand{\newreptheorem}[2]{%
\newenvironment{rep#1}[1]{%
 \def\rep@title{#2 \ref{##1}}%
 \begin{rep@theorem}}%
 {\end{rep@theorem}}}
\makeatother

\begin{abstract}
\textbf{Abstract.} 
Quantum information technologies, on the one side, and intelligent learning
systems, on the other, are both emergent technologies that will likely have a
transforming impact on our society in the future.
The respective underlying fields of basic research -- quantum information (QI) versus
machine learning and artificial intelligence (AI)
-- have their own specific questions and challenges, which have hitherto been
investigated largely independently. However, in a growing body of recent work,
researchers have been probing the question to what extent these fields can indeed
learn and benefit from each other. QML explores the interaction between quantum
computing and machine learning, investigating how results and techniques from one
field can be used to solve the problems of the other. In recent time, we have
witnessed significant breakthroughs in both directions of influence. For instance,
quantum computing is finding a vital application in providing speed-ups for machine
learning problems, critical in our ``big data'' world. Conversely, machine learning
already permeates many cutting-edge technologies, and may become instrumental in
advanced quantum technologies. Aside from quantum speed-up in data analysis, or
classical machine learning optimization used in quantum experiments, quantum
enhancements have also been (theoretically) demonstrated for interactive learning
tasks, highlighting the potential of quantum-enhanced learning agents. Finally,
works exploring the use of artificial intelligence for the very design of quantum
experiments, and for performing parts of genuine research autonomously, have
reported their first successes. Beyond the topics of mutual enhancement -- exploring what
ML/AI can do for quantum physics, and vice versa -- researchers have also broached
the fundamental issue of quantum generalizations of learning and AI concepts. This
deals with questions of the very meaning of learning and intelligence in a world
that is fully described by quantum mechanics. In this review, we describe the main
ideas, recent developments, and progress in a broad spectrum of research
investigating machine learning and artificial intelligence in the quantum domain.
\end{abstract}

\begin{document}

\title{Machine learning \& artificial intelligence in the quantum domain}

\author{Vedran Dunjko}
\address{\mbox{Institute for Theoretical Physics, University of Innsbruck,
Innsbruck 6020, Austria}}
\address{
\mbox{Max Planck Institute of Quantum Optics, Garching 85748, Germany}\\
Email: vedran.dunjko@mpq.mpg.de}
\author{Hans J. Briegel}

\address{\mbox{Institute for Theoretical Physics, University of Innsbruck
Innsbruck 6020, Austria}}
\address{
\mbox{Department of Philosophy, University of Konstanz, 
Konstanz 78457, Germany}\\
Email: hans.briegel@uibk.ac.at}

%
%




%

%

%

%

%

\hspace{3cm}

\maketitle
\tableofcontents
\setlength{\intextsep}{5pt}%
\setlength{\columnsep}{10pt}%

\newpage\pagebreak

\section{Introduction}

\label{sec:intro}
Quantum theory has influenced most branches of physical sciences. This influence ranges from minor corrections, to profound overhauls, particularly in fields dealing with sufficiently small scales. In the second half of the last century, it became apparent that genuine quantum effects can also be exploited in engineering-type tasks, where such effects enable features which are superior to those achievable using purely classical systems. The first wave of such engineering gave us, for example, the laser, transistors, and  nuclear magnetic resonance devices. The second wave, which gained momentum in the '80s, constitutes a broad-scale, albeit not fully systematic, investigation of the potential of utilizing quantum effects for various types of tasks which, at the base of it, deal with the processing of information. This includes the research areas of cryptography, computing, sensing and metrology, all of which now share the common language of quantum information science. Often, the research into such interdisciplinary programs was exceptionally fruitful. For instance, quantum computation,  communication, cryptography and metrology are now mature, well-established and impactful research fields which have, arguably, revolutionized the way we think about information and its processing. 
 In recent years, it has become apparent that the exchange of ideas between quantum information processing and the fields of artificial intelligence and machine learning has its own genuine questions and promises. Although such lines of research are only now receiving a broader recognition, the very first ideas were present already at the early days of QC, 
 and we have made an effort to fairly acknowledge such visionary works.
 
 In this review we aim to capture research at the interplay between machine learning, artificial intelligence and quantum mechanics in its broad scope, with a reader with a physics background in mind. To this end, we dedicate comparatively large amount of space to classical machine learning and artificial intelligence topics, which are often sacrificed in physics-oriented literature, while keeping the quantum information aspects concise.

The structure of the paper is as follows. In the remainder of this introductory section \ref{sec:intro},
we give quick overviews of the relevant basic concepts of the fields quantum information processing, and of machine learning and artificial intelligence. We finish off the introduction with a glossary of useful terms, list of abbreviations, and comments on notation.
Subsequently, in section \ref{sec:clas} we delve deeper into chosen methods, technical details, and the theoretical background of the classical theories. The selection of topics here is not necessarily balanced, from a classical perspective. We place emphasis on elements which either appear in subsequent quantum proposals, which can sometimes be somewhat exotic, or on aspects which can help put the relevance of the quantum results into proper context.
Section \ref{sec:map} briefly summarizes the topics covered in the quantum part of the review. Sections \ref{sec:MLtoQIP} -  \ref{sec:QAI} cover the four main topics we survey, and constitute the central body of the paper. We finish with a an outlook section \ref{sec:discussion}.

 \textbf{Remark:} The overall objective of this survey is to give a broad, ``birds-eye'' account of the topics which contribute to the development of various aspects of the interplay between quantum information sciences, and machine learning and artificial intelligence. Consequently, this survey does not necessarily present all the developments in a fully balanced fashion. Certain topics, which are in their very early stages of investigation, yet important for the nascent research area, were given perhaps a disproportional level of attention, compared to more developed themes. This is, for instance, particularly evident in section \ref{sec:QAI}, which aims to address the topics of quantum artificial intelligence, beyond mainstream data analysis applications of machine learning. This topic is relevant for a broad perspective on the emerging field, however it has only been broached by very few authors, works, including the authors of this review and collaborators. The more extensively explored topics of, e.g., quantum algorithms for machine learning and data mining, quantum computational learning theory, or quantum neural networks, have been addressed in more focused recent reviews \cite{2014_Wittek,2014a_Schuld,2016_Biamonte,2017_deWolf,2017_Ciliberto}.

\subsection{Quantum mechanics, computation and information processing}
\label{QIP}
\boxTLDR{\textbf{Quantum theory} leads to many counterintuitive and fascinating phenomena, including the results of the field of \textbf{quantum information processing}, and in particular, quantum computation.
This field studies the intricacies of quantum information, its communication, processing and use.
Quantum information admits a plethora of phenomena which do not occur in classical physics. For instance, quantum information cannot be cloned -- this \textbf{restricts the types of processing that is possible} for general quantum information. Other aspects \textbf{lead to advantages}, as has been shown for various \textbf{communication and computation tasks:} for solving algebraic problems, reduction of sample complexity in black-box settings, sampling problems and optimization. Even \textbf{restricted models} of quantum computing, amenable for near-term implementations, can solve interesting tasks.
Machine learning and artificial intelligence tasks can, as components, rely on the solving of such problems, \textbf{leading to an advantage.} 
}

Quantum mechanics, as commonly presented in quantum information, is based on few simple postulates: 1) the pure state of a quantum system is given by a unit vector $\ket{\psi}$ in a complex Hilbert space, 2) closed system pure state evolution is generated by a Hamiltonian $H$, specified by the linear Schr\"{o}dinger equation $H \ket{\psi} = i \hbar \frac{\partial}{\partial t} \ket{\psi}$, 3) the structure of composite systems is given by the tensor product, and 4) projective measurements (observables) are specified by, ideally, non-degenerate Hermitian operators, and the measurement process changes the description of the observed system from state $\ket{\psi}$ to an eigenstate $\ket{\phi}$, with probability given by the Born rule $p(\phi) = |\langle{\psi} \ket{\phi} |^2$ \cite{NC}. While the full theory still requires the handling of subsystems and classical ignorance\footnote{This requires more general and richer formalism of density operators, and leads to generalized measurements, completely positive evolutions, etc.}, already the few mathematical axioms of pure state closed system theory give rise to many quintessentially quantum phenomena, like superpositions, no-cloning, entanglement, and others, most of which stem from just the linearity of the theory. Many of these properties re-define how researchers in quantum information perceive what information is, but also have a critical functional role in say quantum enhanced cryptography, communication, sensing and other applications. One of the most fascinating consequences of quantum theory are, arguably, captured by the field of quantum information processing (QIP), and in particular quantum computing (QC), which is most relevant for our purposes.

QC has revolutionized the theories and implementations of computation. This field originated from the observations by Manin \cite{1980_Manin} and Feynman \cite{1982_Feynman} that the calculation of certain properties of quantum systems, as they evolve in time, may be intractable,  while the quantum systems themselves, in a manner of speaking, do perform that hard computation by merely evolving.  Since these early ideas, QC has proliferated, and indeed the existence of quantum advantages which are offered by scalable universal quantum computers have been demonstrated in many settings.  Perhaps most famously, quantum computers have been shown to have the capacity to efficiently solve algebraic computational problems, which are believed to be intractable for classical computers. This includes the famous problems of factoring large integers computing the discrete logarithms  \cite{1997_Shor}, but also many others such as Pell equation solving, some non-Abelian hidden subgroup problems, and others, see e.g. \cite{2010_Childs, 2016_Montanaro} for a review. 
Related to this, nowadays we also have access to a growing collection of quantum algorithms\footnote{In this review it makes sense to point out that the term ``quantum algorithm'' is a bit of a misnomer, as what we really mean is ``an algorithm for a quantum computer''. An algorithm -- an abstraction -- cannot per se be ``quantum'', and the term quantum algorithm could also have meant e.g.``algorithm for describing or simulating quantum processes''. Nonetheless, this term, in the sense of ``algorithm for a quantum computer'' is commonplace in QIP, and we use it in this sense as well.
The concept of ``quantum machine learning'' is, however, still ambiguous in this sense, and depending on the authors, can easily mean ``quantum algorithm for ML``, or ``ML applied to QIP''. }  for various linear algebra tasks, as given in e.g. \cite{2009_Harrow,2015_Childs,2016_Rebentrost}, which may offer speed-ups.

 \begin{wrapfigure}{l}{0.3\textwidth}
 \includegraphics[width=0.3\textwidth,clip=true,trim =300 150 300 150 ]{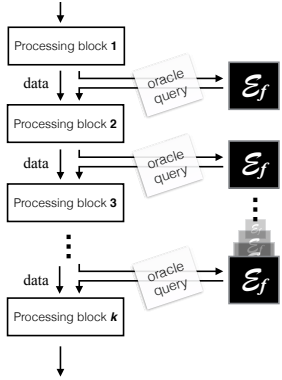}
\caption{\label{fig:oracle}\footnotesize{Oracular computation and query complexity: a (quantum) algorithm solves a problem by intermittently calling a black-box subroutine, defined only via its input-output relations. Query complexity of an algorithm is the number of calls to the oracle, the algorithm will perform.}}
 \end{wrapfigure}
 
Quantum computers can also offer improvements in many optimization and simulation tasks, for instance, computing certain properties of partition functions \cite{2009_Poulin}, simulated annealing \cite{2016_Crosson}, solving semidefinite programs \cite{2016_Brandao}, performing approximate optimization \cite{2014_Farhi}, and, naturally, in the tasks of simulating quantum systems \cite{2014_Nori}.

Advantages can also be achieved in terms of the efficient use of sub-routines and databases.
This is studied using oracular models of computation, where the quantity of interest is the number of calls to an \textit{oracle}, a black-box object with a well-defined set of input-output relations which, abstractly, stands in for a database, sub-routine, or any other information processing resource. The canonical example of a quantum advantage in this setting is the Grover's search \cite{1996_Grover} algorithm which achieves the, provably optimal, quadratic improvement in unordered search (where the oracle is the database). Similar results have been achieved in a plethora of other scenarios, such as spatial search \cite{2004_Childs},  search over structures (including various quantum walk-based algorithms \cite{2003_Kempe,2003_Childs, 2012_Reitzner}), NAND \cite{2009_Childs} and more general boolean tree evaluation problems \cite{2012_Zhan}, as well as more recent ``cheat sheet'' technique results \cite{2016_Aaronson} leading to better-than-quadratic improvements. Taken a bit more broadly, oracular models of computation can also be used to model communication tasks, where the goal is to reduce communication complexity (i.e. the number of communication rounds) for some information exchange protocols \cite{2002_deWolf}.
Quantum computers can also be used for solving sampling problems.
In sampling problems the task is to produce a sample according to an (implicitly) defined distribution, and they are important for both optimization and (certain instances of) algebraic tasks\footnote{Optimization and computation tasks can be trivially regarded as special cases of sampling tasks, where the target distribution is (sufficiently) localized at the solution.}.

For instance, Markov Chain Monte Carlo methods, arguably the most prolific set of computational methods in natural sciences, are designed to solve sampling tasks, which in turn, can be often used to solve other types of problems. For instance, in statistical physics, the capacity to sample from Gibbs distributions is often the key tool to compute properties of the partition function. A broad class of quantum approaches to sampling problems focuses on quantum enhancements of such Markov Chain methods \cite{2011_Temme,2012_Yung}. Sampling tasks have been receiving an ever increasing amount of attention in the QIP community, as we will comment on shortly.
Quantum computers are typically formalized in one of a few standard \textit{models of computation}, many of which are, computationally speaking, equally powerful\footnote{Various notions of ``equally powerful'' are usually expressed in terms of algorithmic reductions. In QIP, typically, the computational model B is said to be at least as powerful as the computational model A, if any algorithm of complexity $O(f(n))$ (where $f(n)$ is some scaling function, e.g. ``polynomial'' or ``exponential''), defined for model A, can be efficiently (usually this means in polynomial time) translated to an algorithm for B, which solves the same problem, and whose computational complexity is $O(poly(f(n)))$. Two models are then equivalent if A is as powerful as B and B is as powerful as A. Which specific reduction complexity we care about (polynomial, linear, etc.) depends on the setting: e.g. for factoring polynomial reductions suffice, since there seems to be an exponential separation between classical and quantum computation. In contrast, for search, the reductions need to be sub-quadratic to maintain a quantum speed-up, since only a quadratic improvement is achievable. }. 
Even if the models are computationally equivalent, 
they are conceptually different. Consequently, some are better suited, or more natural, for a given class of applications. Historically, the first formal model, the quantum Turing machine \cite{1985_Deutsch}, was preferred for theoretical and computability-related considerations. The quantum circuit model \cite{NC} is standard for algebraic problems. The measurement-based quantum computing (MBQC) model \cite{2001_Raussendorf, 2009_Briegel} is, arguably, best-suited for graph-related problems \cite{2016_Zhao}, multi-party tasks and distributed computation \cite{2016_Kashefi} and blind quantum computation \cite{2009_Broadbent}. 
Topological quantum computation \cite{2002_Freedman} was an inspiration for certain knot-theoretic algorithms \cite{2006_Aharonov}, and is closely related to algorithms for topological error-correction and fault tolerance.
The adiabatic quantum computation model \cite{2000_Farhi} is constructed with the task of ground-state preparation in mind, and is thus well-suited for optimization problems \cite{2017_Heim}.

\begin{wrapfigure}{r}{0.53\textwidth}
\boxPrettyTwo{
\begin{footnotesize}
\begin{tabular}{ll}
List of models & applications \\(BQP-complete)& (not exlusive)\\
\vspace{-0.2cm}\\
\hline
QTM&theory\\
QCircuits & algorithms\\
MBQC & distributed computing \\
Topological & knot-theoretic problems\\
Adiabatic & optimization problems\\
\vspace{-0.2cm}\\
\hline\hline
List of models & applications \\(restricted) & \\
\vspace{-0.2cm}\\
\hline
DQC1 & computing trace of unitary \\
Linear Optics & sampling\\
Shallow Random Q. Circuits & sampling\\
Commuting Q. Circuits & sampling\\
RestrictedAdiabatic & optimization tasks
 \end{tabular}
 \end{footnotesize}
  }
\caption{\label{fig:Models}Computational models}
 \end{wrapfigure}

Research into QIP also produced examples of interesting restricted models of computation: models which are in all likelihood not universal for efficient QC, however can still solve tasks which seem hard for classical machines. Recently, there has been an increasing interest in such models, specifically the linear optics model, the so-called low-depth random circuits model and the commuting quantum circuits model\footnote{Other restricted models exist, such as  the one clean qubit model (DQC1) where the input comprises only one qubit in a pure state, and others are maximally mixed. This model can be used to compute a function -- the normalized trace of a unitary specified by a quantum circuit -- which seems to be hard for classical devices. }.
 In \cite{2011_Aaronson} it was shown that the linear optics model can efficiently produce samples from a distribution specified by the permanents of certain matrices, and it was proven (barring certain plausible mathematical conjectures) that classical computers cannot reproduce the samples from the same distribution in polynomial time. Similar claims have been made for low-depth random circuits \cite{2016_Boixo,2017_Bravyi} and commuting quantum circuits, which comprise only commuting gates \cite{2009_Shepherd,2017_Bremner}.  
 Critically, these restricted models can be realized to sufficient size, as to allow for a demonstration of computations which the most powerful classical computers that are currently available cannot achieve, with near-term technologies. This milestone, referred to as \textit{quantum supremacy} \cite{2012_Preskill, 2017_Lund}, and has been getting a significant amount of attention in recent times. 
Another highly active field in QIP concentrates on (analogue) quantum simulations, with applications in quantum optics, condensed matter systems, and quantum many-body physics \cite{2014_Nori}. Many, if not most of the above mentioned aspects of quantum computation are finding a role in quantum machine learning applications. 

Next, we briefly review basic concepts from the classical theories of artificial intelligence and machine learning.

\subsection{Artificial intelligence and machine learning}
\boxTLDR{The field of artificial intelligence incorporates various methods, which are predominantly focused on solving problems which are \textbf{hard for computers, yet seemingly easy for humans}. 
Perhaps the most important class of such tasks pertain to \textbf{learning problems}. Various algorithmic aspects of learning problems are tackled by the field of machine learning, which evolved from the study of pattern recognition in the context of AI. Modern machine learning addresses a variety of learning scenarios, dealing with learning from data, e.g. supervised (data classification), and unsupervised (data clustering)  learning, or from interaction, e.g. reinforcement learning. Modern AI states, as its ultimate goal, the design of an \textbf{intelligent agent} which learns and thrives in unknown environments. Artificial agents that are intelligent in a general, \textit{human} sense must have the capacity to tackle all the individual problems addressed by machine learning and other more specialized branches of AI. They will consequently require a complex combination of techniques. }

\label{AIMLC}
In its broadest scope, the modern field of artificial intelligence (AI) encompasses a wide variety of sub-fields. Most of these sub-fields deal with the understanding and abstracting of aspects of various human capacities which we would describe as intelligent, and attempt to realize the same capacities in machines.
The term ``AI'' was coined at Dartmouth College conferences in the 1956 \cite{2009_Russel}, which were organized to develop ideas about machines that can think, and the conferences are often cited as the birthplace of the field. The conferences were aimed to ``find how to make machines use language, form abstractions and concepts, solve kinds of problems now reserved for humans, and improve themselves'' \footnote{Paraphrased from \cite{1955_Proposal}.}. The history of the field has been turbulent, with strong opinions on how AI should be achieved. For instance, over the course of its first 30 years, the field has crystalized into two main competing and opposite viewpoints \cite{2006_Eliasmith} on how AI may be realized: \textit{computationalism} -- holding that  that the mind functions by performing purely formal operations on symbols, in the manner of a Turing machine, see e.g. \cite{1976_Newell}), and \textit{connectionism}  -- which models mental and behavioral phenomena as the emergent processes of interconnected networks of simple units, mimicking the biological brain, see e.g. \cite{1998_Medler}). Aspects of these two viewpoints still influence approaches to AI. 
Irrespective of the underlying philosophy, for the larger part of the history of AI, the realization of ``genuine AI'' was, purportedly perpetually ``a few years away'' -- a feature often attributed also to quantum computers by critics of the field. In the case of AI, such runaway optimism had a severe calamitous effect on the field, in multiple instances, especially in the context of funding (leading to periods now dubbed ``winters of AI''). 
By the late 90s, the reputation of the field was low, and, even in hindsight, there was no consensus on the reasons why AI failed to produce human-level intelligence. Such factors played a vital role in the fragmentation of the field into various sub-fields which focused on specialized tasks, often appearing under different names.

A particularly influential perspective of AI, often called \textit{nouvelle} or \textit{embodied} AI, was advocated by Brooks, which posited that intelligence emerges from (simple) embodied systems which learn through interaction with their environments \cite{1990_Brooks}. In contrast to standard approaches of the time, Nouvelle AI insists on learning, rather than having properties pre-programmed, and on the embodiment of AI entities, as opposed to abstract entities like chess playing programs. To a physicist, this perspective that intelligence is embodied is reminiscent to the viewpoint that \textit{information is physical}, which had been ``the rallying cry of quantum information theory''\cite{1998_Steane}. Such embodied approaches are particularly relevant in robotics where the key issues involve \textbf{perception} (the capacity of the machine to interpret the external world using its sensors, which includes computer vision, machine hearing and touch), \textbf{motion} and \textbf{navigation} (critical in e.g. automated cars). Related to human-computer interfaces, AI also incorporates the field of \textbf{natural language processing} which includes language understanding -- the capacity of the machine to derive meaning from natural language, and language generation -- the ability of the machine to convey information in a natural language.

 \begin{wrapfigure}{l}{0.35\textwidth}
 \includegraphics[width=0.35\textwidth,clip=true,trim =220 20 220 20 ]{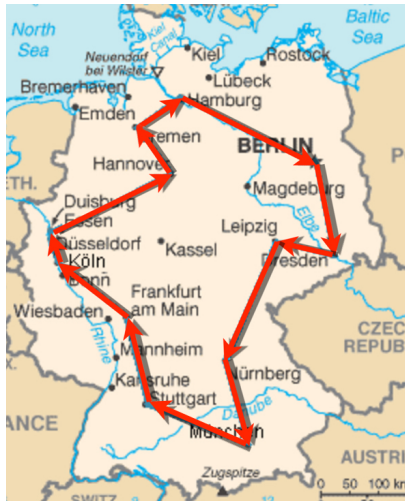}
\caption{\label{fig:TSP} TSP example: finding the shortest route visiting the largest cities in Germany.} 
 \end{wrapfigure}

Other general aspects of AI pertain to a few well-studied capacities of intelligent entities \cite{2009_Russel}. For instance, \textbf{automated planning} is related to decision theory\footnote{Not to be confused with decision problems, studied in algorithmic complexity.} and, broadly speaking, addresses the task of identifying strategies (i.e. sequences of actions) which need to be performed in order to achieve a goal, while minimizing (a specified) cost.

Already the simple class of so-called \textit{off-line} planning tasks, where the task, cost function, and the set of possible actions are known beforehand, contains genuinely hard problems, e.g. it  include, as a special case, the NP-complete\footnote{Roughly speaking, NP is the class of decision (yes, no) problems whose solutions can be efficiently verified by a classical computer in polynomial time. NP-complete problems are the hardest problems in NP in the sense that any other NP problem can be reduced to an NP complete problem via polynomial-time reductions. 
Note that the exact solutions to NP-compete problems are believed to be intractable even for quantum 
computers.}  
travelling salesman problem (TSP); for illustration see Fig. \ref{fig:TSP}
\footnote{ Figure \ref{fig:TSP} has been modified from \url{https://commons.wikimedia.org/wiki/File:TSP_Deutschland_3.png}.} .

In modern times, TSP itself would no longer be considered a genuine AI problem, but it is serves to illustrate how already very specialized, simple sub-sub-tasks of AI may be hard. 
More general planning problems also include \textit{on-line} variants, where not everything is known beforehand (e.g. TSP but where the ``map'' may fail to include all the available roads, and one simply has to actually travel to find good strategies). On-line planning overlaps with reinforcement learning, discussed later in this section. Closely related to planning is the capacity of intelligent entities for \textbf{problem solving}. In technical literature, problems solving is distinguished from planning by a lack of additional structure in the problem, usually assumed in planning -- in other words, problem solving is more general and typically more broadly defined than planning. The lack of structure in general problem solving establishes a clear connection to (also unstructured) searching and optimization: in the setting of no additional information or structure, problem solving is the search for the solution to a precisely specified problem. While general problem solving can be, theoretically, achieved by a general search algorithm (which can still be subdivided into classes such as depth-first, breath-first, depth-limited search etc.),
 more often there is structure to the problem, in which case an \textit{informed} search strategies -- often called \textit{heuristic search strategies} --  will be more efficient \cite{2009_Russel}. 
Human intelligence, to no small extent, relies on our knowledge. We can accumulate knowledge, reason over it, and use it to come to the best decisions, for instance in the context of problem solving and planning. An aspect of AI tries to formalize such \textbf{logical reasoning}, knowledge accumulation and \textbf{knowledge representation}, often relying on formal logic, most often first order logic. 

A particularly important class of problems central to AI, and related to knowledge acquisition, involve the capacity of the machine to learn through experience. This feature was emphasized already in the early days of AI, and the derived field of \textit{machine learning} (ML) now stands as arguably the most successful aspect (or spin-off) of AI, which we will address in more detail.

\subsubsection{Learning from data: machine learning}


 \begin{wrapfigure}{r}{0.6\textwidth}
 \includegraphics[width=0.6\textwidth,clip=true,trim =50 290 300 150 ]{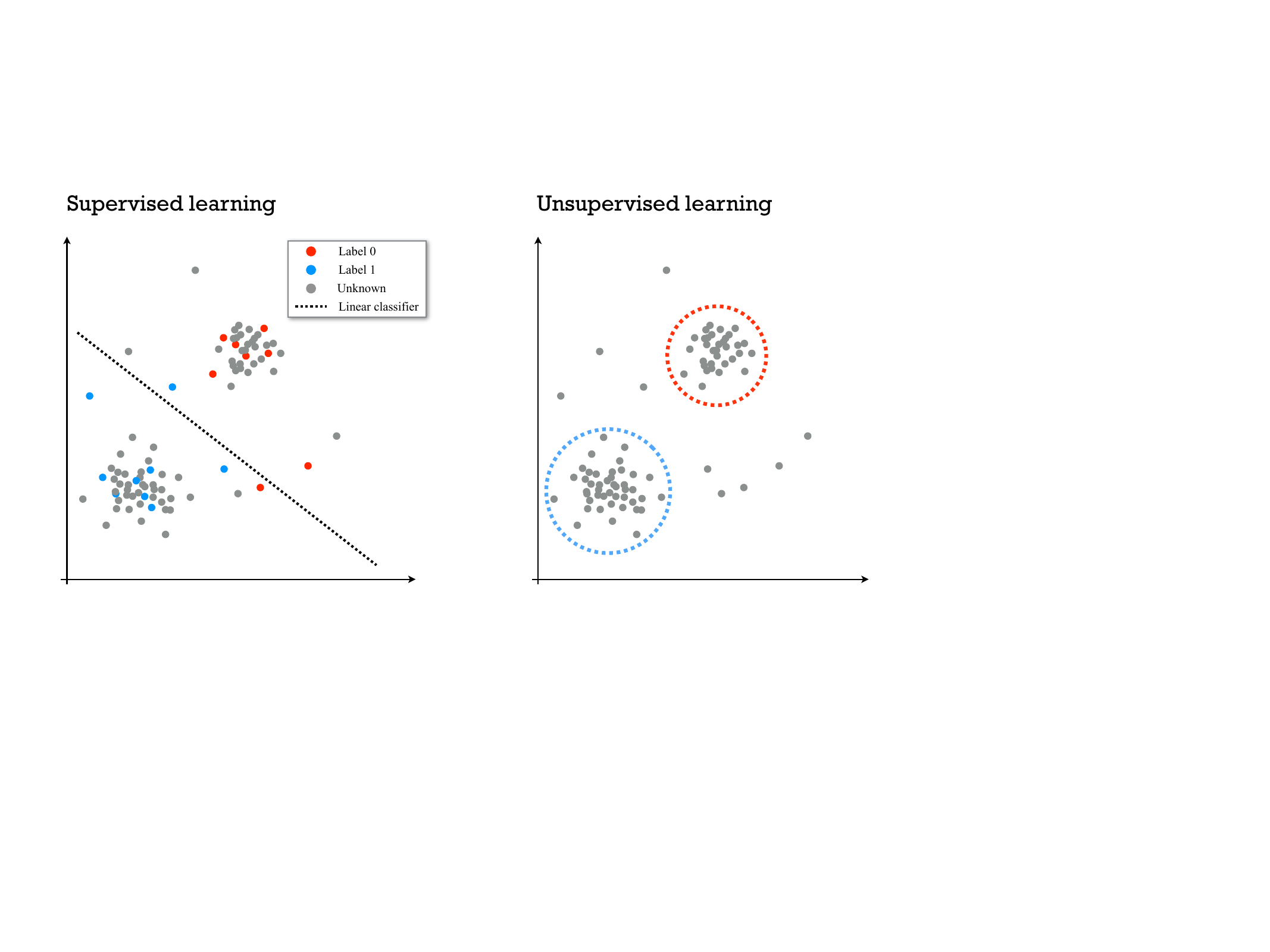}
\caption{\label{fig:unsup} Supervised (in this case, best linear classifier) and unsupervised learning (here clustering into two most likely groups and outliers) illustrated.  }
 \end{wrapfigure}
Stemming from the traditions of pattern recognition, such as recognizing handwritten text, and statistical learning theory (which places ML ideas in a rigorous mathematical framework), ML, broadly speaking, explores the construction of algorithms that can learn from, and make predictions about data.
Traditionally, ML deals with two main learning settings: supervised and unsupervised learning, which are closely related to data analysis and data mining-type tasks \cite{2014_Shalev}. A broader perspective \cite{2010_Alpaydin} on the field also includes reinforcement learning \cite{1998_Sutton}, which is closely related to learning as is realized by biological intelligent entities. We shall discuss reinforcement learning separately. 

In broad terms, \textbf{supervised learning} deals with learning-by-example: given a certain number of labeled points (so-called \textit{training set}) $\{ (x_i, y_i) \}_i$ where $x_i$ denote data points, e.g. $N-$dimensional vectors, and $y_i$ denote labels (e.g. binary variables, or real values), the task is to infer a ``labeling rule'' $x_i \mapsto y_i$ which allows us to guess the labels of previously \textit{unseen data}, that is, beyond the training set.
Formally speaking, we deal with the task of inferring the conditional probability distribution $P(Y=y | X=x)$ (more specifically, generating a labeling function which, perhaps probabilistically, assigns labels to points) based on a certain number of samples from the joint distribution $P(X,Y).$ For example, we could be inferring whether a particular DNA sequence belongs to an individual who is likely to develop diabetes. Such an inference can be based on the datasets of patients whose DNA sequences had been recorded, along with the information on whether they actually developed diabetes. In this example, the variable $Y$ (diabetes status) is binary, and the assignment of labels is not deterministic, as diabetes also depends on environmental factors. Another example could include two real variables, where $x$ is the height from which an object is dropped, and $y$ the duration of the fall. In this example, both variables are real-valued, and (in vacuum) the labeling relation will be essentially deterministic. 
In \textbf{unsupervised learning}, the algorithm is provided just with the data points without labels. Broadly speaking, the goal here is to identify the underlying distribution, or structure, and other informative features in the dataset.
In other words, the task is to infer properties of the distribution $P(X=x),$ based on a certain number of samples, relative to a user-specified guideline or rule. 
Standard examples of unsupervised learning are clustering tasks, where data-points are supposed be grouped in a manner which minimizes within-group mean-distance, while maximizing the distance between the groups. Note that the group membership can be thought of as a label, thus this also corresponds to a labeling task, but lacks ``supervision'': examples of correct labelings. 
In basic examples of such tasks, the number of expected clusters is given by the user, but this too can be automatically optimized. 

Other types of unsupervised problems include feature extraction and dimensionality reduction, critical in combatting the so-called \textbf{curse of dimensionality}. 
The curse of dimensionality refers to problems which stem from the fact that the raw representations of real-life data often occupy very high dimensional spaces. For instance, a standard resolution one-second video-clip at standard refresh frequency, capturing events which are extended in time maps to a vector in $\sim 10^8$ dimensional space\footnote{Each frame is cca. $10^6$ dimensional, as each pixel constitutes one dimension, multiplied with 30 frames required for the one-second clip.}, even though the relevant information it carries (say a licence-plate number of a speeding car filmed) may be significantly smaller. More generally, intuitively it is clear that, since geometric volume scales exponentially with the dimension of the space it is in, the number of points needed to capture (or \textit{learn}) general features of an $n-$dimensional object will also scale exponentially. In other words, learning in high dimensional spaces is exponentially difficult. Hence, a means of dimensionality reduction, from raw representation space (e.g. moving car clips), to the relevant feature space (e.g. licence-plate numbers) is a necessity in any real-life scenario.

These approaches the data-points to a space of significantly reduced dimension, while attempting to maintain the main features -- the relevant information -- of the structure of the data. A typical example of a dimensionality example technique is e.g. principal component analysis. In practice, such algorithms also constitute an important step in data pre-processing for other types of learning and analysis. Furthermore, this setting also includes generative models (related to density estimation), where new samples from an unknown distribution are generated, based on few exact samples. As humanity is \textbf{amassing data at an exponential rate} \cite{ExpData}  it becomes ever more relevant to extract genuinely useful information in an automated fashion. 
In modern world ubiquitous big data analysis and data mining are the central applications of supervised and unsupervised learning.

\subsubsection{Learning from interaction: reinforcement learning}
\label{subsubs:RL}
Reinforcement learning (RL) \cite{1998_Sutton, 2009_Russel} is, traditionally, the third canonical category of ML.
Partially caused by the relatively recent prevalence of (un)supervised methods in the contexts of the pervasive data mining and big data analysis topics, many modern textbooks on ML focus on these methods. RL strategies have mostly remained reserved for robotics and AI communities. Lately, however, the surge of interest in adaptive and autonomous devices, robotics, and AI have increased the prominence of RL methods. 

One recent celebrated result which relies on the extensive use of standard ML and RL techniques in conjunction is that of AlphaGo \cite{2016_Silver}, a learning system which mastered the game of Go, and achieved, arguably, superhuman performance, easily defeating the best human players. This result is notable for multiple reasons, including the fact that it illustrates the potential of learning machines over special-purpose solvers in the context of AI problems: while specialized devices which relied on programming over learning (such as Deep Blue) could surpass human performance in chess, they failed to do the same for the more complicated game of Go, which has a notably larger space of strategies. The learning system AlphaGo achieved this many years ahead of typical predictions.
The distinction between RL and other data-learning ML methods is particularly relevant from a quantum information perspective, which will be addressed in more detail in section \ref{QAE}. RL constitutes a broad learning setting, formulated within the general \textbf{agent-environment paradigm} (AE paradigm) of AI \cite{2009_Russel}. Here, we do not deal with a static database, but rather an interactive \textit{task environment}. 
The learning agent (or, a learning algorithm) learns through the interaction with the task environment. 
 \begin{wrapfigure}{r}{0.425\textwidth}
 \includegraphics[width=0.425\textwidth,clip=true,trim =130 180 200 160 ]{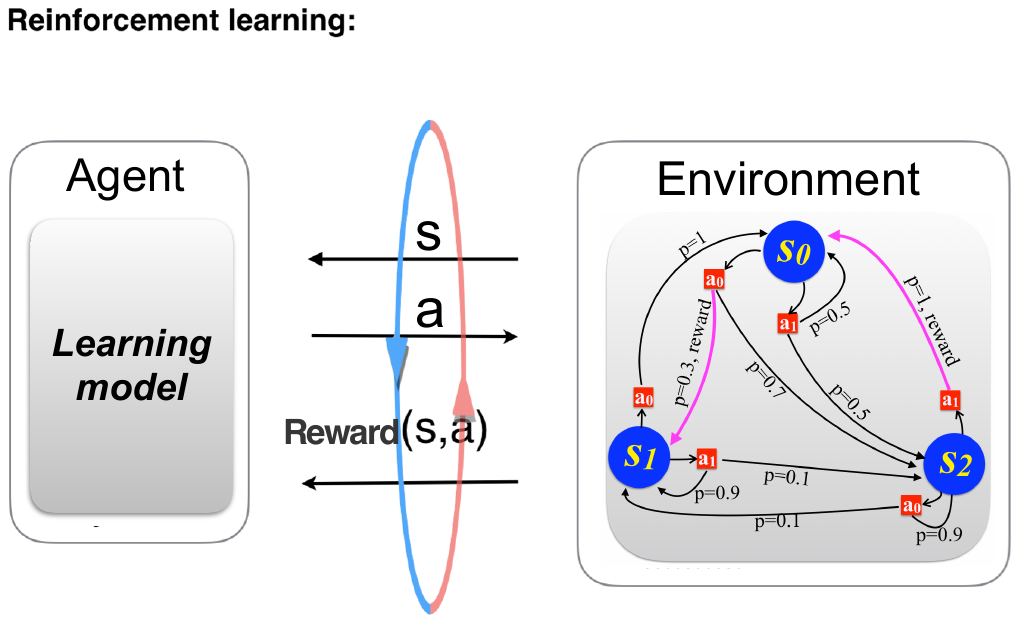}
 \vspace{-10pt}
\caption{\label{fig:RL}{An agent interacts with an environment by exchanging percepts and actions. In RL rewards can be issued. Basic environments are formalized by Markov Decision Processes (inset in Environment). Environments are reminiscent to oracles, see \ref{fig:oracle}, in that the agent only has access to the input-output relations. Further, figures of merit for learning often count the number of interaction steps, which is analogous to the concept of query complexity.}}
 \end{wrapfigure}

As an illustration, one can imagine a robot, acting on its environment, and perceiving it via its sensors -- the percepts being, say, snapshots made by its visual system, and actions being, say, movements of the robot -- as depicted in Fig. \ref{fig:RL} . 
The AE formalism is, however, more general and abstract. It is also unrestrictive as it can also express supervised and unsupervised settings. 
In RL, it is typically assumed that the goal of the process is manifest in a reward function, which, roughly speaking, rewards the agent, whenever the agents behavior was \textit{correct} (in which case we are dealing with \textit{positive reinforcement}, but other variants of operant conditioning are also used\footnote{More generally, we can distinguish four modes of such operant conditioning: positive reinforcement (reward when correct), negative reinforcement (removal of negative reward when correct), positive punishment (negative reward when incorrect) and negative punishment (removal of reward when incorrect).}). 
This model of learning seems to cover pretty well how most biological agents (i.e. animals) learn:  one can illustrate this through the process of training a dog to do a trick by giving it treats whenever it performs well. 
As mentioned earlier, RL is all about learning how to perform the ``correct'' sequence of actions, given the received percepts, which is an aspect of planning, 
in a setting which is fully on-line: the only way to learn about the environment is by interacting with it.

\subsubsection{Intermediary learning settings}
\label{intermediary:models}
While supervised, unsupervised and reinforcement learning constitute the three broad categories of learning, there are many variations and intermediary settings.
For instance, \textbf{semi-supervised learning} interpolates between unsupervised and supervised settings, where the number of labeled instances is very small compared to the total available training set. Nonetheless, even a small number of labeled examples have been shown to improve the bare unsupervised performance \cite{2010_Chapelle}, or, from an opposite perspective, unlabeled data can help with classification when facing a small quantity of labeled examples. In active supervised learning, the learning algorithm can further query the human user, or supervisor, for the labels of particular points which would improve the algorithm's performance. This setting can only be realized when it is operatively possible for the user to correctly label all the points, and may yield advantages when this exact labeling process is expensive.
Further, in supervised settings, one can consider so-called \textbf{inductive learning} algorithms which output a classifier function, based on the training data, which can be used to label all possible points. A \textbf{classifier} is simply a function which assigns labels to the points in the domain of the data. In contrast, in \textbf{transductive learning} \cite{2010_Chapelle} settings, the points that need to be labeled later are known beforehand -- in other  words, the classifier function is only required to be defined on a-priori known points. Next, a supervised algorithm can perform \textbf{lazy learning}, meaning that the whole labeled dataset is kept in memory in order to label unknown points (which can then be added), or \textbf{eager learning}, in which case, the (total) classifier function is output (and the training set is no longer explicitly required) \cite{2010_Alpaydin}. 
Typical examples of eager learning are linear classifiers, such as basic support vector machines, described in the next section, whereas lazy learning is exemplified by e.g. nearest-neighbour methods\footnote{For example, in $k-$nearest neighbour classification, the training set is split into disjoint subsets specified by the shared labels. Given a new point which is to be classified, the algorithm identifies $k$ nearest neighbour points from the data set to the new point. The label of the new point is decided by the majority label of these neighbours. The labeling process thus needs to refer to the entire training set.} . Our last example, \textbf{online learning} \cite{2010_Alpaydin}, can be understood as either an extension of eager supervised learning, or a special case of RL. Online learning generalizes standard supervised learning, in the sense that the training data is provided sequentially to the learner, and used to, incrementally, update the classifying function. In some variants, the algorithm is asked to classify each point, and is given the correct response afterward, and the performance is based on the guesses. The match/mismatch of the guess and the actual label can also be understood as a reward, in which case online learning becomes a restricted case of RL. 

\subsubsection{Putting it all together: the {{agent-environment paradigm}}}
\label{all-together}
The aforementioned specialized learning scenarios can be phrased in a unifying language, which also enables us to discuss how specialized tasks fit in the objective of realizing true AI. 

 In modern take on AI \cite{2009_Russel},  the central concept of the theory is that of an \textit{agent}. An agent is an entity which is defined relative to its environment, and which has the capacity to act, that is, \textit{do something}. 

  In computer science terminology the requirements for something to be an agent (or for something to \textit{act}) are minimal, and essentially everything can be considered an agent -- for instance, all non-trivial computer programs are also agents.


 AI concerns itself with agents which do more -- for instance they also perceive their environment, interact with it, and learn from experience. AI is nowadays defined\footnote{Over the course of its history, AI had many definitions, many of which invoke the notion of an agent, while some older, definitions talk about machines, or programs which ``think'', ``have minds'' and so on  \cite{2009_Russel}.

 As clarified, the field of AI has fragmented, and many of the sub-fields deal with specific computational problems, and the development of computational methodologies useful in AI related problems, for instance ML (i.e. its supervised and unsupervised variants). In such sub-fields with a more pragmatic computational perspective, the notion of agents is not used as often.} as the field which is aimed at designing \textit{intelligent agents}  \cite{2009_Russel}, which are autonomous, perceive their world using sensors, act on it using actuators, and choose their activities as to achieve certain goals -- a property which is also called \textit{rationality} in literature. 
 
      \begin{wrapfigure}{l}{0.3\textwidth}
 \includegraphics[width=0.3\textwidth,clip=true,trim =260 220 270 210 ]{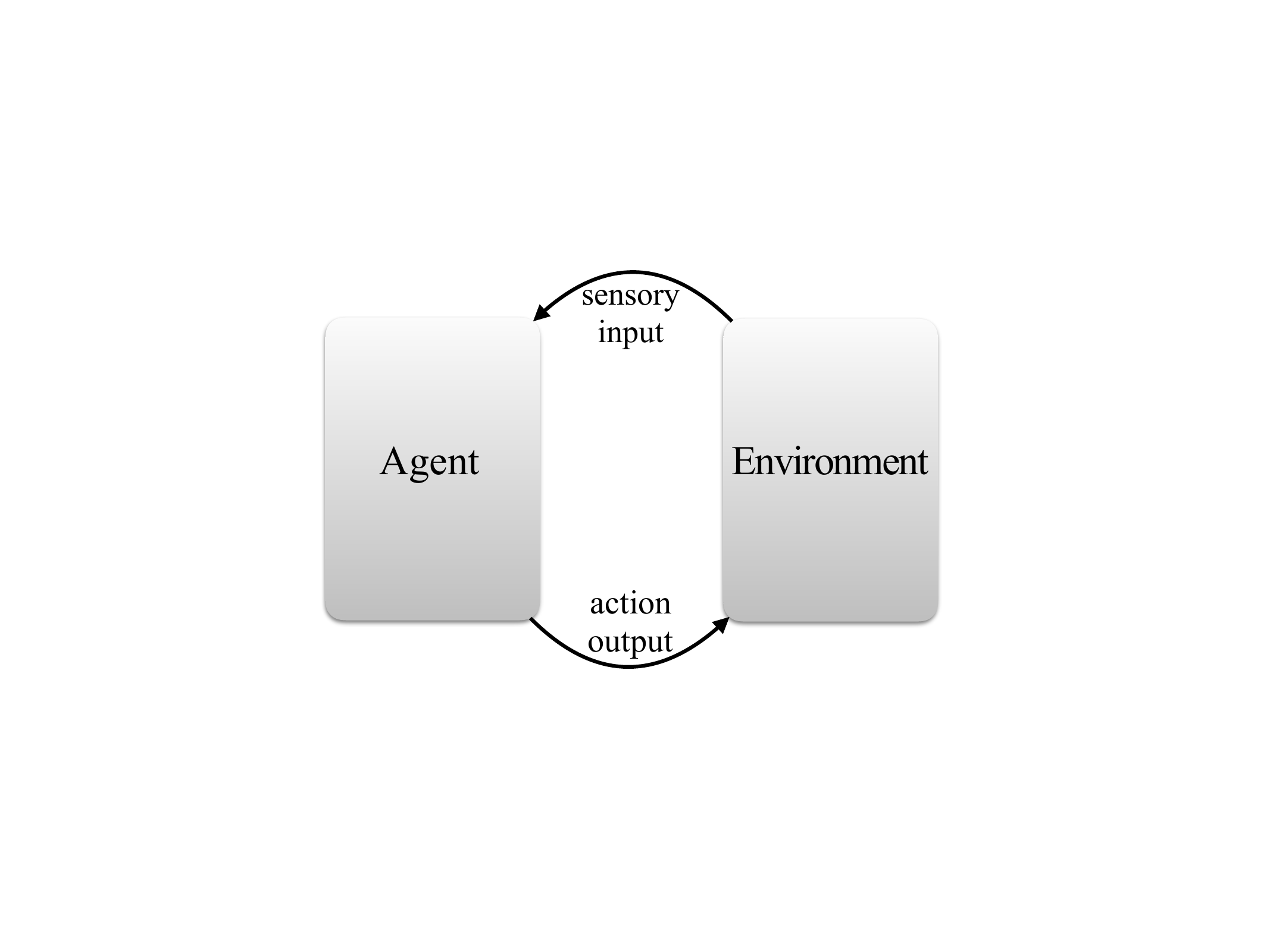}
 
 \caption{\label{fig:AE} Basic agent-environment paradigm.}
 \end{wrapfigure} 
 Agents only exist relative to an environment (more specifically a task environment), with which they interact, constituting the overall AE paradigm, illustrated in Fig. \ref{fig:AE}.
While it is convenient to picture robots when thinking about agents, they can also be more abstract and virtual, as is the case with computer programs ``living'' in the internet\footnote{The subtle topics of such virtual, yet embodied agents is touched again later in section \ref{QLI}.}. 
In this sense, any learning algorithm for any of the more specialized learning settings can also be viewed as a restricted learning agent, operating in a special type of an environment, e.g. a supervised learning environment may be defined by a training phase, where the environment produces examples for the learning agent, followed by a testing phase, where the environment evaluates the agent, and finally the application phase, where the trained and verified model is actually used. The same also obviously holds for more interactive learning scenarios such as the reinforcement-driven mode of learning -- RL --  we briefly illustrated in section \ref{subsubs:RL}, is natively phrased in the AE paradigm.
In other words, all machine learning models and settings can be phrased within the broad AE paradigm.
 
 Although the field of AI is fragmented into research branches with focus on isolated, specific goals, the ultimate motivation of the field remains the same: the design of true, general AI, sometimes referred to as \textit{artificial general intelligence (AGI)}\footnote{The field of AGI, under this label, emerged in mid 2000s, and the term is used to distinguish the objective of realizing intelligent agents from the research focusing more specialized tasks, which are nowadays all labeled AI. AGI is also referred to as \textbf{strong AI}, or, sometimes \textbf{full AI}.}, that is, the design of a ``truly intelligent'' agent  \cite{2009_Russel}.

The topic of what ingredients are needed to build AGI is difficult, and without a consensus.

One perspective focuses on the behavioral aspects of agents.
In literature, many features of intelligent behavior are captured by characterizing more specific types of agents: simple reflex agents,
model-based reflex agents, goal-based agents, utility-based agents, etc. Each type captures an aspect of intelligent behavior, much like the fragments of the field of ML, understood as a subfield of AI, capture specific types of problems intelligent agents should handle. 
For our purposes, the most important, overarching aspect of intelligent agents is the capacity to learn\footnote{A similar viewpoint, that essentially all AI problems/features map to a learning scenario, is also advocated in \cite{2005_Hutter}.}, and we will emphasize learning agents in particular.

The AE paradigm is particularly well suited for such an operational perspective, as it abstracts from the internal structure of agents, and focuses on behavior and input-output relations.

More precisely, the perspective on AI presented in this review is relatively simple. a) AI pertains to agents which behave intelligently in their environments, and, b) the central aspect of intelligent behaviour is that of learning.

While we, unsurprisingly, do not more precisely specify what intelligent behaviour entails, already this simple perspective on AI has non-trivial consequences. The first is that intelligence can be ascertained from the interaction history between the agent and its environment alone. Such a viewpoint on AI is also closely related to behavior-based AI and the ideas behind the Turing test \cite{1950_Turing}; it is in line with an embodied viewpoint on AI (see \textit{embodied AI} in section \ref{AIMLC}) and it has influenced certain approaches towards quantum AI, touched in section \ref{QAI}.
The second is that the development of better ML and other types of relevant algorithms does constitute genuine progress towards AI, conditioned only on the fact that such algorithms can be coherently combined into a whole agent. It is however important to note that to actually achieve this integration may be far from trivial.

In contrast to such strictly behavioral and operational points of view, an alternative approach towards whole agents (or complete intelligent agents) focuses on agent architectures and cognitive architectures \cite{2009_Russel}. 
In this approach to AI the emphasis is equally placed not only on intelligent behaviour, but also on forming a theory about the structure of the (human) mind. One of the main goals of a cognitive architecture is to design a comprehensive computational model which encapsulates various results stemming from research in cognitive psychology. The aspects which are predominantly focused on understanding human cognition are, however, not central for our take on AI.

 We discuss this further in section \ref{TQO}. 
\newpage
\pagebreak

%
%
%
%
%
%

\subsection{Miscellanea}

\paragraph{Abbreviations and acronyms\vspace{0.3cm}\\}
\boxPrettyThree{
\begin{tabular}{l|l|l}
Acronym & Meaning  & First occurrence\\
\hline
\hline
AE paradigm &	agent-environment paradigm & \ref{subsubs:RL} \\
AGI &	artificial general intelligence&\ref{all-together} \\
AI &	artificial intelligence&\ref{AIMLC}\\
ANN &	artificial neural network & \ref{ANNs}\\
BED &	Bayesian experimental design& \ref{GenHam}\\
BM	&Boltzmann machine & \ref{ANNs}\\
BQP	& bounded-error quantum polynomial time& \ref{QLI}\\
CAM & content-addressable memory & \ref{ANNs}\\
CLT	& computational learning theory& \ref{maths}\\
DME	& density matrix exponentiation & \ref{qcircuit} \\
DQC1	& one clean qubit model& \ref{QIP}\\
HN	& Hopfield network& \ref{ANNs}\\
MBQC & measurement-based quantum computation & \ref{QIP}\\
MDP	&Markov decision process & \ref{RL-section}\\
ML&	machine learning & \ref{AIMLC}\\
NN	& neural network & \ref{ANNs}\\
NP	& non-deterministic polynomial time & \ref{AIMLC}\\
PAC learning& probably approximately correct learning & \ref{CPAC} \\
PCA	&principal component analysis& \ref{qcircuit}\\
POMDP	& partially observable Markov decision process &\ref{RL-section}\\
PS	& projective simulation & \ref{RL-section} \\
QC	& quantum computation & \ref{QIP}\\
QIP	& quantum information processing & \ref{QIP}\\
QUBO	& quadratic unconstrained binary optimization& \ref{Adiabatic}\\
RL&	reinforcement learning & \ref{subsubs:RL}\\
rPS& reflective PS & \ref{QLI} \\
SVM	& support vector machine &  \ref{SVMs}\\
\end{tabular}
}

\paragraph{Notation}

Throughout this review paper, we have strived to use the notation specified in the reviewed works. To avoid a notational chaos, we, however keep the notation consistent within subsections -- this means that, within one subsection, we adhere to the notation used in the majority of works if inconsistencies arise. 

\section{Classical background}

\label{sec:clas}

The main purpose of this section is to provide the background regarding classical ML and AI techniques and concepts which are either addressed in quantum proposals we discuss in the following sections or important for the proper positioning of the quantum proposal in the broader learning context.
The concepts and models of this section include common models found in classical literature, but also certain more exotic models, which have been addressed in modern quantum ML literature.
While this section contains most of the classical background needed to understand the basic ideas of the quantum ML literature, to tame the length of this section, certain very specialized classical ML ideas are presented on-the-fly during the upcoming reviews.

We first provide the basics concepts related to common ML models, emphasizing neural networks in \ref{ANNs} and support vector machines in \ref{SVMs}. 
Following this, in \ref{othermodels}, we also briefly describe a larger collection of algorithmic methods, and ideas arising in the context of ML, including regression models, $k-$means/medians, decision trees, but also more general optimization and linear algebra methods which are now commonplace in ML. Beyond the more pragmatic aspects of model design for learning problems, in subsection \ref{maths} we provide the main ideas of the mathematical foundations of computational learning theory, which discuss learnability -- i.e.  the conditions under which learning is possible at all -- computational learning theory and the theory of Vapnik and Chervonenkis -- which rigorously investigate the bounds on learning efficiency for various supervised settings.
Subsection \ref{RL-section} covers the basic concepts and methods of RL. 

\nopagebreak

\subsection{Methods of machine learning}
\boxTLDR{{Two particularly famous models in machine learning are \textbf{artificial neural networks} -- inspired by biological brains, and \textbf{support vector machines} -- arguably the best understood supervised learning model. 
Neural networks come in many flavours, all of which model parallel information processing of a network of simple computational units, neurons. Feed-forward networks (without loops) are typically used for supervised learning. Most of the popular \textbf{deep learning} approaches fit in this paradigm.
Recurrent networks have loops -- this allows e.g. feeding information from outputs of a (sub)-network back to its own input . Examples include \textbf{Hopfield networks}, which can be used as \textbf{content-addressable memories}, and \textbf{Boltzmann machines}, typically used for unsupervised learning. These networks are related Ising-type models, at zero, or finite temperatures, respectively -- this sets the grounds for some of the proposals for quantization.
Support vector machines classify data in an Euclidean space, by identifying best separating hyperplanes, which allows for a comparatively  simple theory. The linearity of this model is a feature making it amenable to quantum processing.
The power of hyperplane classification can be improved by using \textbf{kernels} which, intuitively, map the data to higher dimensional spaces, in a non-linear way. ML naturally goes beyond these two models, and includes regression (data fitting) methods and many other specialized algorithms.
 }}

Since the early days of the fields of AI and ML, there have been many proposals on how to achieve the flavours of learning we described above. In what follows we will describe two popular models for ML, specifically artificial neural networks, and support vector machines. We highlight that many other models exist, and indeed, in many fields other learning methods (e.g. regression methods), are more commonly used.
A selection of such other models is briefly mentioned thereafter, along with examples of techiques which overlap with ML topics in a broader sense, such as matrix decomposition techniques, and which can be used for e.g. unsupervised learning.

Our choice of emphasis is, in part, again motivated by later quantum approaches, and by features of the models which are particularly well-suited for cross-overs with quantum computing.


\subsubsection{Artificial neural networks and deep learning}
\label{ANNs}
Artificial neural networks (artificial NNs, or just NNs) are a biologically inspired approach to tackling learning problems. Originating in 1943 \cite{1943_McCulloch}, the basic component of NNs is the artificial neuron (AN), which is, abstractly speaking, a real-valued function $AN: \mathbbmss{R}^k \rightarrow \mathbbmss{R}$ parametrized by a vector of real, non-negative weights $ (w_i)_i = \mathbf{w} \in\mathbbmss{R}^k$, and the activation function $\phi: \mathbbmss{R} \rightarrow \mathbbmss{R},$ given with
\EQ{
AN(\mathbf{x}) = \phi \left( \sum_{i} x_i w_i \right), \textup{with\ } \mathbf{x} =  (x_i)_i \in \mathbbmss{R}^k.
}
For the particular choice when the activation function is the threshold function $\phi_{\theta} (x) = 1$ if $x>\theta \in \mathbbmss{R}^+$ and $\phi_{\theta} (x) = 0$ otherwise, the AN is called a \textbf{perceptron} \cite{1957_Rosenblatt}, and has been studied extensively.
Already such simple perceptrons performing classification into subspaces specified by the hyperplane with the normal vector $\mathbf{w}$, and off-set $\theta$ (c.f. \textit{support vector machines} later in this section).  

Note, in ML terminology, a distinction should be made between \textit{artificial neurons} (ANs) and \textit{perceptrons} -- perceptrons are special cases of ANs, with the fixed activation function -- the step function --, and a specified update or training rule. ANs in modern times use various activation functions (often the differentiable sigmoid functions), and can use different learning rules. For our purposes, this distinction will not matter.The training of such a classifier/AN for supervised learning purposes consists in optimizing the parameters $\mathbf{w}$ and $\theta$ as to correctly label the training set -- there are various figures of merit particular approaches care about, and various algorithms that perform such an optimization, which are not relevant at this point.
By combining ANs in a network we obtain NNs (if ANs are perceptrons, we usually talk about multi-layered perceptrons). While single perceptrons, or single-layered perceptrons can realize only linear classification, already a three-layered network suffices to approximate any continuous real-valued function (precision depending on the number of neurons in the inner, so-called hidden, layer). Cybenko \cite{1989_Cybenko} was the first to prove this for sigmoid activation functions, whereas Hornik generalized this to show that the same holds for all non-constant, monotonically increasing and bounded activation functions \cite{1991_Hornik} soon thereafter.
This shows that if sufficiently many neurons are available, a three-layered ANN can be trained to learn any dataset, in principle\footnote{More specifically, there exists a set of weights doing the job, even though standard training algorithms may fail to converge to that point.}. Although this result seems very positive, it comes with the price of a large model complexity, which we discuss in section \ref{VCtheory}\footnote{Roughly speaking, models with high model complexity are more likely to ``overfit'', and it is more difficult to provide guarantees they will generalize well, i.e., perform well beyond the training set.}. 
In recent times, it has become apparent that using multiple, sequential, hidden feed-forward layers (instead of one large), i.e. \textbf{deep neural networks} (deep NNs), may have additional benefits. First, they may reduce the number of parameters \cite{2017_Poggio}. Second, the sequential nature of processing of information from layer to layer can be understood as a feature abstraction mechanism (each layer processes the input a bit, highlighting relevant features which are processed further). This increases the \textbf{interpretability of the model} (intuitively, the capacity for high level explanations of the model's performance) \cite{2016_Lipton}, which is perhaps best illustrated in so-called convolutional (deep) NNs, whose structure is inspired by the visual cortex.
One of the main practical disadvantages of such deep networks is the computational cost and computational instabilities in training (c.f.. \textit{the vanishing gradient problem} \cite{2001_Hochreiter}), and also the size of the dataset which has to be large \cite{2009_Larochelle}. With modern technology and datasets, both obstacles are becoming less prohibitive, which has lead to a minor revolution in the field of ML. 

Not all ANNs are feed-forward: \textbf{recurrent neural networks} (recurrent NNs) allow for the backpropagation of signals. Particular examples of such networks are so called \textbf{Hopfield networks} (HNs), and \textbf{Boltzmann machines} (BMs), which are often used for different purposes than feed-forward networks.
In HNs, we deal with one layer, where the outputs of all the neurons serve as inputs to the same layer. The network is initialized by assigning binary values (traditionally, $-1$ and $1$ are used, for reasons of convenience) to the neurons (more precisely, some neurons are set to fire, and some not), which are then processed by the network, leading to a new configuration. This update can be synchronous (the output values are "frozen" and all the second-round values are computed simultaneously) or asynchronous (the update is done one neuron at a time in a random order).
The connections in the network are represented by a matrix of weights $(w_{ij})_{ij},$ specifying the connection strength between the $i^{th}$ and the $j^{th}$ neuron. The neurons are perceptrons, with a threshold activation function, given with the local threshold vector $(\theta_i)_i$.  Such a dynamical system, under a few mild assumptions \cite{1982_Hopfield}, converges to a configuration (i.e. bit-string) which (locally) minimizes the energy functional 
\EQ{
E(\mathbf{s}) = -\dfrac{1}{2} \sum_{ij} w_{ij} s_i s_j + \sum_{i} \theta_i s_i, \label{ising}}
with $\ \mathbf{s} = (s_i)_i, \ s_i \in \{-1,1 \}$, that is, the Ising model. In general, this model has many local minima, which depend on the weights $w_{ij}$, and the thresholds, which are often set to zero. 
Hopfield provided a simple algorithm (called Hebbian learning, after D. Hebb for historic reasons \cite{1982_Hopfield}), which enables one to ``program'' the minima -- in other words, given a set of bitstrings $S$ (more precisely, strings of signs $+1/-1$), one can find the matrix $w_{ij}$ such that exactly those strings $S$ are local minima of the resulting functional $E$. Such programmed minima are then called \textit{stored patterns}. Furthermore, Hopfield's algorithm achieved this in a manner which is local (the weights $w_{ij}$ depend only on the $i^{th}$ and $j^{th}$ bits of the targeted strings, allowing parallelizability), incremental (one can modify the matrix $w_{ij}$ to add a new string without having to keep the old strings in memory), and immediate. Immediateness means that the computation of the weight matrix is not a limiting, but finite process. Violating e.g. incrementality would lead to a lazy algorithm (see section \ref{intermediary:models}), which can be sub-optimal in terms of memory requirements, but often also computational complexity\footnote{The lazy algorithm may have to process all the patterns/data-points the number of which may be large and/or growing. }. 
 It was shown that the minima of such a trained network are also attractive fixed-points, with a finite basin of attraction.
This means that if a trained network is fed a new string, and let run, it will (eventually) converge to a pattern which is closest to it (the distance measure that is used depends on the learning rule, but typically it is the Hamming distance, i.e. number of entries where the strings disagree).
Such a system then forms an \textbf{associative memory}, also called a \textbf{content-addressable memory} (CAM).
CAMs can be used for supervised learning (the ``labels'' are the stored patterns), and conversely, supervised learning machinery can be used for CAM\footnote{For this, one simply needs to add a look-up table connecting labels to fixed patterns.}.
An important feature of HNs is their capacity: how many distinct patterns it can store\footnote{Reliable storage entails that previously stored patterns will be also recovered without change (i.e they are energetic local minima of Eq. (\ref{ising}), but also that there is a basin of attraction -- a  ball around the stored patterns with respect to a distance measure (most commonly  the Hamming distance) for which the dynamical process of the network converges to the stored pattern. An issue with capacities is the occurrence of spurious patterns: local minima with a non-trivial basin of attraction which were not stored.  }. For the Hebbian update rule this number scales as $O(n/\log(n))$, where $n$ is the number of neurons, which Storkey \cite{1997_Storkey} improved to  $O(n/\sqrt{\log(n)})$. In the meantime, more efficient learning algorithms have been invented \cite{2014_Hillar}. Aside from applications as CAMs, due to the representation in terms of the energy functional in Eq. (\ref{ising}), and the fact that the running of HNs minimize it, they have also been considered for the tasks of optimization early on \cite{1985_Hopfield}. The operative isomorphism between Hopfield networks and the Ising model, technically, holds only in the case of a zero-temperature system. \textbf{Boltzmann machines} generalize this.
Here, the value of the $i^{th}$ neuron is set to $-1$ or $1$ (called ``off'' and ``on'' in literature, respectively) with probability 
\EQ{
p(i=-1) = \left(1+\exp \left(-\beta \Delta E_i \right) \right)^{-1},\ \textup{with }\Delta E_{i}=\sum _{j}w_{ij}\,s_{j}+\theta _{i}, \label{boltz}
}

where $\Delta E_{i}$ is the energy difference of the configuration with $i^{th}$ neuron being on or off, assuming the connections $\mathbf{w}$ are symmetric, and $\beta$ is the inverse temperature of the system. In the limit of infinite running time, the network's configuration is given by the (input-state invariant) Boltzmann distribution over the configurations, which depends on the weights $\mathbf{w}$, local thresholds (weights) $\bf{\theta}$ and the temperature. 
BMs are typically used in a generative fashions, to model, and sample from, (conditional) probability distributions.
In the simplest variant, 
the training of the network attempts to ensure that the limiting distribution of the network matches the observed frequencies in the dataset. This is achieved by the tuning of the parameters $\mathbf{w}$ and $\bf{\theta}$.
The structure of the network dictates how complicated a distribution can be represented.
To capture more complicated distributions, over say $k$ dimensional data, the BMs have $N>k$ neurons. $k$ of them will be denoted as visible units, and the remainder are called hidden units, and they capture latent, not directly observable, variables of the system which generated the dataset, and which we are in fact modelling. Training such networks consists in a gradient ascent  of the log-likelihood of observing the training data, in the parameter space. While this seems conceptually simple, it is computationally intractable, in part as it requires accurate estimates of probabilities of equilibrium distributions, which are hard to obtain.
In practice, this is somewhat mitigated by using \textbf{restricted BMs}, where the hidden and visible units form the partition of a bi-partite graph (so only connections between hidden and visible units exist). (Restricted) BMs have a large spectrum of uses, including 
providing generative models -- producing new samples from the estimated distribution,  as classifiers -- via conditioned generation,  as feature extractors -- a form of unsupervised clustering, and as building blocks of deep architectures \cite{2009_Larochelle}. However, their utility is mostly limited by the cost of training -- for instance, the cost of obtaining equilibrium Gibbs distributions, or by the errors stemming from heuristic training methods such as contrastive divergence
 \cite{2009_Larochelle,2009_Bengio,2014_Wiebe}.

\subsubsection{Support Vector Machines}
\label{SVMs}

Support Vector Machines (SVMs) form a family of perhaps best understood approaches for solving classification problems. The basic idea behind SVMs is that a natural way to classify points based on a dataset $\{\mathbf{x}_i, y_i \}_i,$ for binary labels $y_i \in \{-1,1\},$ is to generate a hyperplane separating the negative instances from the positive ones.
Such observations are not new, and indeed, perceptrons, briefly discussed in the previous section, perform the same function.

 Such a hyperplane can then be used to classify all points. 
 Naturally, not all sets of points allow this (those that do are called linearly separable), but SVMs are further generalized to deal with sets which are not linearly separable using so-called kernels. Kernels, effectively, realize non-linear mappings of the original dataset to higher dimensions where they may become separable, depending on a few technical conditions \footnote{Indeed, this can be supported by hard theory, see Cover's Theorem \cite{1965_Cover}.}), and by allowing a certain degree of misclassification, which leads to so-called ``soft-margin'' SVMs.

  \begin{wrapfigure}{l}{0.325\textwidth}
 \includegraphics[width=0.325\textwidth,clip=true,trim =50 300 630 190 ]{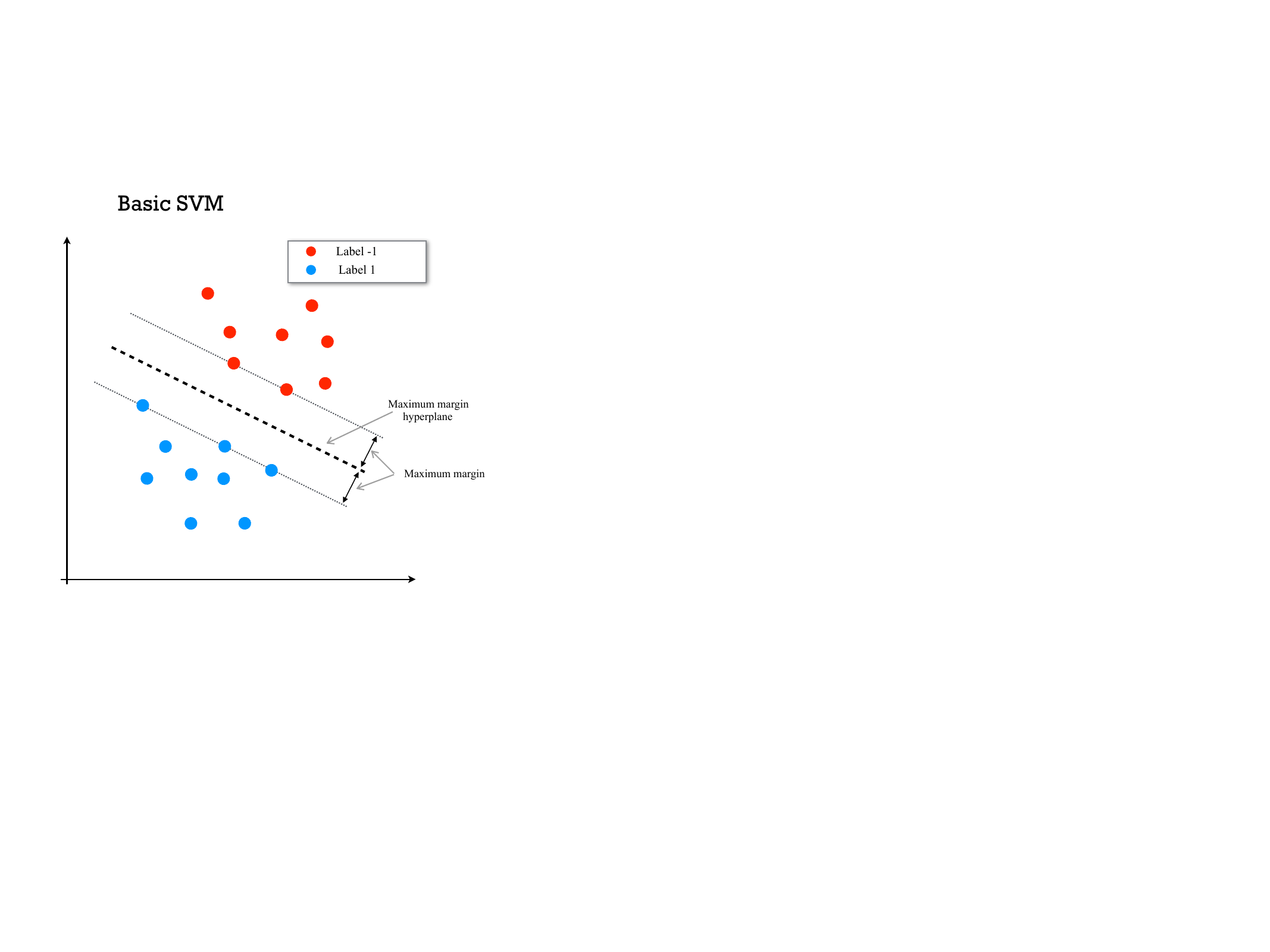}
\caption{\label{fig:SVMw}
Basic example of an SVM, trained on a linearly separable data-set.}
 \end{wrapfigure}

Even in the case the dataset is linearly separable, there will still be many hyperplanes doing the job. This leads to various variants of SVMs, but the basic variant identifies a hyperplane which: a) correctly splits the training points, and b) maximizes the so-called margin: the distance of the hyperplane to the nearest point (see Fig. \ref{fig:SVMw}).

The distance of choice is most often the geometric Euclidean distance, which leads to so-called \textit{maximum margin classifiers}. In high-dimensional spaces, in general the maximization of the margin ends in a situation where there are multiple $+1$ and $-1$ instances of training data points which are equally far from the hyperplane. These points are called \textbf{support vectors}.
The finding of a maximum margin classifier corresponds to finding a normal vector $\mathbf{w}$ and offset $b$ of the separating hyperplane, which corresponds to the optimization problem 

\EQ{
\mathbf{w}^\ast= \textup{argmin}_{\mathbf{w}, b} \frac{1}{2}\| \mathbf{w} \|^2\\
\textup{such that}\ y_i (\mathbf{w}.\mathbf{x}_i+b) \geq 1.
}

The formulation above is actually derived from the basic problem by noting that we may arbitrarily and simultaneously scale the pair $(\mathbf{w}, b)$ without changing the hyperplane. Therefore, we may always choose a scaling such that the realized margin is 1, in which case, the margin corresponds to  $\|\mathbf{w} \|^{-1}$, which simply maps a maximization problem to a minimization problem as above. The square ensures the problem is stated as a standard quadratic programming problem. This problem is often expressed in its Lagrange dual form, which reduces to
\EQ{
(\alpha^\ast_1 ,\ldots \alpha^\ast_N) = \textup{argmin}_{\alpha_1 \ldots \alpha_N} \left( \sum_{i} \alpha_i - \frac{1}{2} \sum_{i,j} \alpha_i  \alpha_j  y_i y_j\mathbf{x}_i.\mathbf{x}_j \right) \label{inner}\\
\textup{such that}\ \alpha_i\geq 0\ \textup{and}\ \sum_{i}\alpha_iy_i =0 \label{alphas},
}
where the solution of the original problem is given by 
\EQ{
\mathbf{w}^\ast = \sum_{i}y_i\alpha_i \mathbf{x}_i.
}
In other words, we have expressed $\mathbf{w}^\ast$ in the basis of the data-vectors, and the data-vectors $\mathbf{x}_i$ for which the corresponding coefficient $\alpha_i$ is non-zero are precisely the support vectors.
The offset $b^\ast$ is easily computed having access to one  support vector of, say, an instance $+1$, denoted $\mathbf{x}^{+}$, by solving  $\mathbf{w}^{\ast}.\mathbf{x}^{+}+b^{\ast} = 1$.

The class of a new point $\mathbf{z}$ can also be computed directly using the support vectors via the following expression
\EQ{
\mathbf{z} \mapsto \textup{sign} \left( \sum_{i}y_i\alpha_i \mathbf{x}_i.\mathbf{z}   + b^\ast \right).
}


The dual representation of the optimization problem is convenient when dealing with kernels. 
As mentioned, a way of dealing with data which is not linearly separable, is to first map all the points into a higher-dimensional space via a non-linear function $\phi: \mathbbmss{R}^{m} \rightarrow \mathbbmss{R}^{n} $, where $m<n$ is the dimensionality of the datapoints. 
As we can see, in the dual formulation, the data-points only appear in terms of inner products $\mathbf{x}_i.\mathbf{x}_j$. This leads to the notion of the \textit{kernel function} $k$ which, intuitively, measures the similarity of the points \textit{in the larger space}, typically defined with $k(\mathbf{x}_i,\mathbf{x}_j) = \phi(\mathbf{x}_i)^{\tau} \phi(\mathbf{x}_j)$. In other words, to train the SVM according to a non-trivial kernel $k$, induced by the non-linear mapping $\phi$,
the optimization line  Eq. (\ref{inner}) will be replaced with
$ \textup{argmin}_{\alpha_1 \ldots \alpha_N} \left( \sum_{i} \alpha_i - \frac{1}{2} \sum_{i,j} \alpha_i  \alpha_j  y_i y_j k(\mathbf{x}_i,\mathbf{x}_j) \right)\label{kernel}$. 
The offset is computed analogously, using just one application of $\phi$.
The evaluation of a new point is given in the same way with $\mathbf{z} \mapsto \textup{sign} \left( \sum_{i}y_i\alpha_i k(\mathbf{x}_i,\mathbf{z})   + b^\ast \right).$
 In other words, the data-points need not be explicitly mapped via $\phi$, as long as the map-inducing inner product $k(\cdot, \cdot)$ can be computed more effectively. The choice of the kernel is critical in the performance of the classifier, and the finding of good kernels is non-trivial and often solved by trial-and-error.
 
While increasing the dimension of the extended space (co-domain of $\phi$) may make data-points more linearly separable (i.e. fewer mismatches for the optimal classifier), in practice they will not be fully separable (and furthermore, increasing the kernel dimension comes with a cost which we elaborate on later). To resolve this, SVMs allow for misclassification, with various options for measuring the ``amount'' of misclassification, inducing a penalty function.
A typical approach to this is to introduce so-called ``slack variables'' $\xi_i \geq 0$ to the original optimization task, so:
\EQ{
\mathbf{w}^\ast= \textup{argmin}_{\mathbf{w}, b} \left( \frac{1}{2}\| \mathbf{w} \|^2 + C \sum_{i} \xi_i \right)\\
\textup{such that}\ y_i (\mathbf{w}.\mathbf{x}_i+b) \geq 1 - \xi_i.
}
If the value $\xi_i$ of the optimal solution is between 0 and 1, the point $i$ is correctly classified, but is within the margin, and $\xi_i>1$ denotes a misclassification. The (hyper)parameter $C$ controls the relative importance we place on minimizing the margin norm, versus the importance we place on misclassification.
Interestingly, the dual formulation of the above problem is near-identical to the hard-margin setting discussed thus far, with the small difference that the parameters $\alpha_i$ are now additionally constrained with $\alpha_i \leq C$ in Eq. (\ref{alphas}).
SVMs, as described above, have been extensively studied from the perspective of computational learning theory, and have been connected to other learning models. In particular, their  generalization performance, which, roughly speaking, characterizes how well a trained model\footnote{In ML, the term model is often overloaded. Most often it refers to a classification system which has been trained on a dataset, and in that sense it ``models'' the actual labeling function. Often, however, it will also refer to a class of learning algorithms (e.g. the SVM learning model).} will perform \textit{beyond the training set} can be analyzed. This is the most important feature of a classifying algorithm. We will briefly discuss generalization performance in section \ref{VCtheory}. We end this short review of SVMs by considering a non-standard variant, which is interesting for our purposes as it has been beneficially quantized. SVMs as described are trained by finding the maximal margin hyperplane. Another model, called least-squares SVM (LS-SVM) takes a regression (i.e. data-fitting) approach to the problem, and finds a hyperplane which, essentially, minimizes the least square distance of the vector of labels, and the vector of distances from the hyperplane, where the $i^{th}$ entry of the vector is given with $(\mathbf{w}.\mathbf{x}_i+b).$
This is effected by a small modification of the soft-margin formulation: 
\EQ{
\mathbf{w}^\ast_{LS}= \textup{argmin}_{\mathbf{w}, b} \frac{1}{2}\| \mathbf{w} \|^2 + C \sum_{i} \xi_i^2\\
\textup{such that}\ y_i (\mathbf{w}.\mathbf{x}_i+b) =1 - \xi_i,
} 
where the only two differences are that the constraints are now equalities, and the slack variables are squared in the optimization expression.
This seemingly innocuous change causes differences in performance, but also in the training. The dual formulation of the latter optimization problem reduces to a linear system of equations:
\EQ{
\left[{\begin{matrix}0&1_{}^{T}\\1_{N}&\Omega +\gamma ^{-1}I_{}\end{matrix}}\right]\left[{\begin{matrix}b\\\bm{\alpha} \end{matrix}}\right]=\left[{\begin{matrix}0\\Y\end{matrix}}\right], \label{LSSVM:sys}
}
where $1$ is an ``all ones'' vector, $Y$ is the vector of labels $y_i$, $b$ is the offset, $\gamma$ is a parameter depending on $C$. $\bm{\alpha}$ is the vector of the Lagrange multipliers yielding the solution. This vector again stems from the dual problem which we omitted due to space constraints, and which can be found in \cite{1999_Suykens}. Finally, $\Omega$ is the matrix collecting the (mapped) ``inner products'' of the training vectors so $\Omega_{i,j} = k(\mathbf{x}_i,\mathbf{x}_j),$ where $k$ is a kernel function, in the simplest case, just the inner product. The training of LS-SVMs is thus simpler (and particularly convenient from a quantum algorithms perspective), but the theoretical understanding of the model, and its relationship to the well-understood SVMs, is still a matter of study, with few known results (see e.g. \cite{2007_Ye}).

\subsubsection{Other models} 
\label{othermodels}

While NNs and SVMs constitute two popular approaches for ML tasks (in particular, supervised learning), many other models exist, suitable for a variety of ML problems. Here we very briefly list and describe some of such models which have also appeared in the context of quantum ML. While classification typically assigns discrete labels to points, in the case when the labeling function has a continuous domain (say the segment $[0,1]$) we are dealing with function approximation tasks, often dealt with by using \textbf{regression} techniques.
Typical examples here include \textbf{linear regression}, which approximate the relationship of points and labels with a linear function, most often minimizing the least-squares error. More broadly, such techniques are closely related to \textbf{data-fitting}, that is, fitting the parameters of a parametrized function such as to best fit observed (training) data. 
The \textit{k-nearest neighbour} algorithm is an intuitive classification algorithm which given a new point considers the $k$ nearest training points (with respect to a metric of choice), and assigns the label by the majority vote (if used for classification), or by averaging (in the case of regression, i.e. continuous label values). 
The mutually related \textbf{k-means} and \textbf{k-medians} algorithms are typically used for clustering: the $k$ specifies the number of clusters, and the algorithm defines them in a manner which minimizes the within-cluster distance to the mean (or median) point. 

Another method for classification and regression optimizes \textbf{decision trees}, where each dimension, or entry (or more generally a feature\footnote{Features, however, have a more generic meaning in the context of ML. A data vector is a vector of features, where what a feature is depends on the context. For instance, features can be simply values at particular positions, or more global properties: e.g. a feature of data vectors depicting an image may be ``contains a circle'', and all vectors corresponding to pictures with circles have it. Even more generically, features pertain to observable properties of the objects the data-points represent (``observable'' here simply means that the property can be manifested in the data vector).}) of the new data point influences a move on a \textit{decision tree}. The depth of the tree is the length of the vector (or number of features), and the degree of each node depends on the possible number of distinct features/levels per entry\footnote{For instance, we can classify humans, parrots, bats and turtles, by binary features $can\_fly$ and $is\_mamal$. E.g. choosing the root $can\_fly$ leads to the branch $can\_fly=no$ with two leaves decided by $is\_mamal = yes$ pinpointing the human, whereas $is\_mamal = no$ would specify the turtle. Parrots and bats would be distinguished by the same feature in the  $can\_fly=yes$ subtree.}.
 The vertices of the tree specify an arbitrary feature of interest, which can influence the classification result, but most often they consider the overlaps with geometrical regions of the data-point space. Decision trees are in principle maximally expressive (can represent any labeling function), but very difficult to train without constraints.

More generally, classification tasks can be treated as the problem of \textbf{finding a hypothesis} $h: Data \rightarrow Labels$ (in ML, the term hypothesis is essentially synonymous to the term {classifier}, also called a {learner}) which is from some family $H$, which minimizes error (or \textit{loss}) under some loss function. For instance, the hypotheses realized by SVMs are given by the hyperplanes (in the kernel space), and in neural nets they are parametrized by the parameters of the nets: geometry, thresholds, activation functions, etc. 
Additional to loss terms, the minimization of which is called \textbf{empirical risk minimization}, ML applications benefit from adding an additional component to the objective function: the \textbf{regularization term}, the purpose of which is to penalize complex functions, which could otherwise lead to poor generalization performance, see section \ref{VCtheory}.
The choices of loss functions, regularization terms, and classes of hypotheses lead to different particular models, and training corresponds to optimization problems given by the choice of the loss function and the hypothesis (function) family. 
Furthermore, it has been shown that essentially any learning algorithm which requires only convex optimization for training leads to poor performance under noise. Thus non-convex optimization is necessary for optimal learning (see e.g. \cite{2010_Long, 2011_Manwani}).

An important class of meta-algorithms for classification problems are \textbf{boosting algorithms}.
The basic idea behind boosting algorithms is the highly non-trivial observation, first proven via the seminal AdaBoost algorithm \cite{1997_Freund}, that multiple \textbf{weak classifiers}, which perform better than random on distinct parts of the input space, can be combined into an overall better classifier.
More precisely, given a set of (weak) hypotheses/classifiers $\{h_j\}, h_j:  \mathbbmss{R}^{n}\rightarrow \{-1,1\}$, under certain technical conditions, there exists a set of weights $\{w_i\}, w_i  \in \mathbbmss{R}$, such that the composite classifier of the form $hc_{\mathbf{w}}(\mathbf{x}) = \textup{sign}(\sum_i w_i h_i(\mathbf{x}))$ performs better.
 Interestingly, a single (weak) learning model can be used to generate the weak hypotheses needed for the construction of a better composite classifier -- one which, in principle, can achieve arbitrary high success probabilities, i.e. a \textit{strong learner}. The first step of this process is achieved by altering the frequencies at which the training labeled data-points appear, one can effectively alter the distributions over the data (in a black-box setting, these can be obtained by e.g. rejection sampling methods). The training of one and the same model on such differentially distributed datasets can generate distinct weak learners, which emphasize distinct parts of the input space. Once such distinct hypotheses are generated, optimization of the weight $w_i$ of the composite model is performed. In other words, \textbf{weak learning models can be boosted\footnote{It should be mentioned that the above description only serves to illustrate the intuition behind boosting ideas. In practice, various boosting methods have distinct steps, e.g. they may perform the required optimizations in differing orders, using training phases in parallel etc. which is beyond the needs of this review.}. }

Aside from the broad classes of approaches to solve various ML tasks, ML is also often conflated with specific computational tools which are used to solve them. A prominent example of this is the development of algorithms for optimization problems, especially those arising in the training of standard learning models. 
This includes e.g. particle swarm optimization, genetic and evolutionary algorithms, and even variants of stochastic gradient descent.
ML also relies on other methods including linear algebra tools, e.g. matrix decomposition methods, such as singular value decomposition, QR-, LU- and other decompositions, derived methods such as principal component analysis 
, and various techniques from the field of signal analysis (Fourier, Wavelet, Cosine, and other transforms). 
The latter set of techniques serves to reduce the effective dimension of the data set, and helps combat the curse of dimensionality.  
The optimization, linear algebra, and signal processing techniques, and their interplay with quantum information is an independent body of research with enough material to deserve a separate review, and we will only reflect on these methods when needed. 


\subsection{{Mathematical theories of supervised and inductive learning}}

\boxTLDR{Aside from proposing learning models, such as NNs or SVMs, learning theory also provides formal tools to identify the limits of learnability. \textbf{No free lunch} theorems provide sobering arguments that na\"ive notions of ``optimal'' learning models cannot be obtained, and that all learning must rely on some prior assumptions. \textbf{Computational learning theory} relies on ideas from computational complexity theory, to formalize many settings of supervised learning, such as the task of approximating or identifying an unknown (boolean) function -- a \textbf{concept} --which is just the binary labeling function. The main question of the theory is the quantification of the number of invocations of the black-box -- i.e. of the function (or of the oracle providing examples of the function's values on selected inputs) -- needed to reliably approximate the (partially) unknown concept to desired accuracy. In other words, computational learning theory considers the \textbf{sample complexity bounds} for various learning settings, specifying the concept families and type of access. The theory of Vapnik and Chervonenkis, or simply \textbf{VC theory}, stems from the tradition of statistical learning. One of the key goals of the theory is to provide theoretical guarantees on \textbf{generalization performance}. This is what is asked for in the following question: given a learning machine trained on a dataset of size $N$, stemming from some process, with a measured \textit{empirical risk} (error on the training set) of some value $R$, what can be said about its future performance on other data-points which may stem from the same process? One of the key results of VC theory is that this can be answered, with the help of a third parameter -- the \textbf{model complexity} of the learning machine. Model complexity, intuitively, captures how complicated functions the learner can learn: the more complicated the model, the higher chance of ``overfitting'', and consequently, the weaker the guarantees on performance beyond the training set. \textbf{Good learning models can control their model complexity}, leading to a learning principle of \textbf{structural risk minimization}.
The art of ML is \textbf{a juggling act}, balancing \textbf{sample complexity}, \textbf{model complexity}, and the \textbf{computational complexity} of the learning algorithm\hyperref[wit]{\textsuperscript{27}}.}
%
%
%
\footnotetext{\label{wit} While the dichotomies between sample complexity and computational complexity are often considered in literature, the authors have first heard the trichotomic setting, including model complexity from \cite{Wittek}. Examples of such balancing, and its failures can be observed in sections \ref{quantum-concepts}, and \ref{qqpac}.}

\label{maths}
Although modern increase of interests in ML and AI are mostly due to applications, aspects of ML and AI do have strong theoretical backgrounds. 
Here we focus on such foundational results which clarify what learning is, and which investigate the questions of what learning limits are. We will very briefly sketch some of the basic ideas.

The first collection of results, called No Free Lunch theorems place seemingly pessimistic bounds on the conditions under which learning is at all possible \cite{1996_Wolpert}. No Free Lunch theorems are, essentially, a mathematical formalization of Hume's famous \textbf{problem of induction} \cite{1739_Hume,2016_Vickers}, which deals with the justification of inductive reasoning. One example of inductive reasoning occurs during generalization. Hume points out that, without a-priori assumptions, concluding any property concerning a class of objects based on any number of observations\footnote{An exception to this would be the uninteresting case when the class was finite and all instances had been observed.} is not justified. 

In a similar vein, learning based on experience cannot be justified without further assumptions: expecting that a sequence of events leads to the same outcome as it did in the past, is only justified if we assume a uniformity of nature. 
The problems of generalization and of uniformity can be formulated in the context of supervised learning and RL, with (not uncontroversial) consequences (c.f. \cite{NFL-web}).
For instance, one of the implications is that the expected performance of any two learning algorithms beyond the training set must be equal,  if one uniformly averages over all possible labeling functions, and analogous statements hold for RL settings -- in other words, without assumptions on environments/datasets, the expected performance of any two learning models will be essentially the same, and two learning models cannot be meaningfully compared in terms of performance without making statements about the task environments in question.
In practice, we, however, always have \textit{some} assumptions on the dataset and environment: for instance the principle of parsimony (i.e. Occam's razor), asserting that simpler explanations tend to be correct, prevalent in science, suffices to break the symmetries required for NFLs to hold in their strongest form \cite{2011_Lattimore,2010_Hutter,2011_BenDavid}. 

No review of theoretical foundations of learning theory should circumvent the works of Valiant, and the general computational learning theory (CLT), which stems from a computer science tradition, initiated by Valiant \cite{1984_Valiant}, and the related VC theory of Vapnik and Chervonenkis, developed from a statistical viewpoint \cite{1995_Vapnik}. We present the basic ideas of these theories in no particular order.
\setcounter{paragraph}{0}
\subsubsection{Computational learning theory}
\label{CPAC}
CLT can be understood as a  rigorous formalization of supervised learning, and which stems from a computational complexity theory tradition.
The most famous model in CLT is that of \textit{probably approximately correct} (PAC) learning. We will explain the basic notions of PAC learning on a simple example: optical character recognition. Consider the task of training an algorithm to decide whether a given image (given as a black and white bitmap) of a letter corresponds to the letter ``A'', by supplying a set of examples and counterexamples: a collection of images. Each image $\mathbf{x}$ can be encoded as a binary vector in $\{0,1\}^{n}$ (where $n=$height$\times$width of the image).

Assuming that there exists an univocally correct assignment of label $0$ (not ``A") and $1$ to each image implies there exists a characteristic function $f: \{0,1\}^{n} \rightarrow \{0,1\}$ which discerns letters $A$ from other images. Such an underlying characteristic function (or, equivalently, the subset of bitstrings for which it attains value "1") is, in computational learning theory, called a \textbf{concept}.
Any (supervised) learning algorithm will first be supplied with a collection of $N$ examples $(\mathbf{x}_i, f((\mathbf{x}_i))_i$. In some variants of PAC learning, it is assumed that the data-points ($\mathbf{x}$) are drawn from some distribution $D$ attaining values in  $\{0,1\}^{n}$. Intuitively, this distribution can model the fact that in practice, the examples that are given to the learner stem from its interaction with the world, which specifies what kinds of ``A''s we are more likely to see\footnote{For instance, in modern devices, the devices are (mostly) trained for the handwriting of the owner, which will most of the time be distinct from other persons handwritings, although the device should in principle handle any (reasonable) handwriting.}.
PAC learning typically assumes \textit{inductive} settings, meaning that the learning algorithm, given a sample set $S_N$ (comprising $N$ identically independently distributed  samples from D) outputs a hypothesis $h:\{0,1\}^{n} \rightarrow \{0,1\}$ which is, intuitively, the algorithms ``best guess'' for the actual concept $f$.
The quality of the guess is measured by the total \textbf{error} (also known as loss, or regret), 
\EQ{
err_D(h_{S_N}) = \sum_{\mathbf{x}} P(D = \mathbf{x}) |h_{S_N}(\mathbf{x}) -  f(\mathbf{x})|, \label{errorPAC}
}
averaged according to the same (training) distribution $D$, where $h_{S_N}$ is the hypothesis the (deterministic) learning algorithm outputs given the training set $S_N$.
Intuitively, the larger the training set is ($N$), the smaller the error will be, but this also depends on the actual examples (and thus $S_N$ and $D$).
PAC theory concerns itself with probably ($\delta$), approximately $(\epsilon)$ correct learning, i.e. with the following expression:
\EQ{
P_{S_{N} \sim D^{N}}\left[    err_D(h_{S_N})  \leq  \epsilon \right] \geq 1-\delta,\label{PAC}
}
where $S \sim D$ means $S$ was drawn according to the distribution $D$.
The above expression is a statement certifying that the learning algorithm, having been trained on the dataset sampled from $D,$ will, except with probability $\delta$, have a total error below $\epsilon$.
We say a concept $f$ is $(\epsilon, \delta)$-learnable, under distribution $D$, if there exists 
a learning algorithm, and an $N$, such that Eq. (\ref{PAC}) holds, and simply learnable, if it is $(\epsilon, \delta)$-learnable for all $(\epsilon, \delta)$ choices. The functional dependence of $N$ on $(\epsilon, \delta)$ (and on the concept and distribution $D$) is called the \textbf{sample complexity}. 
In PAC learning, we are predominantly concerned with identifying tractable problems, so a concept/distribution pair $f,D$ is PAC-learnable if there exists an algorithm for which the sample complexity is polynomial in $\epsilon^{-1}$ and $\delta^{-1}$.
These basic ideas are generalized in many ways. First, in the case the algorithm cannot output all possible hypotheses, but only a restricted set $H$ (e.g. the hypothesis space is smaller than the total concept space), we can look for the best case solution by substituting the actual concept $f$ with the optimal choice $h^{\ast} \in H$ which minimizes the error in (\ref{errorPAC}), in all the expressions above. Second, we are typically not interested in just distinguishing the letter ``A'' from all other letters, but rather recognizing all letters. In this sense, we typically deal with a \textbf{concept class} (e.g. ``letters''), which is a set of concepts, and it is (PAC) learnable if there exists an algorithm for which each of the concepts in the class are (PAC) learnable. If, furthermore, the same algorithm also learns for all distributions $D$, then the class is said to be (distribution-free) learnable. 

CLT contains other models, generalizing PAC. For instance, concepts may be noisy or stochastic.
 In the \textbf{agnostic learning} model,
the labeled examples $(\mathbf{x},y)$ are sampled from a distribution $D$ over $\{ 0,1\}^{n}\times\{0,1\},$ which also models probabilitstic concepts\footnote{Note that we recover the standard PAC setting once the conditional probability distribution of $P_D(y|\mathbf{x})$ where the values of the first $n$ bits (data-points) are fixed, is Kronecker-delta -- i.e. the label is deterministic.}. Further, in agnostic learning, we define a set of concepts $C \subseteq \{c | c: \{ 0,1\}^n \rightarrow \{ 0,1\} \}$, and given $D,$ we can identify the best deterministic approximation of $D$ in the set $C$, given with $opt_C = min_{c\in C} err_D(c)$. The goal of learning is to produce a hypothesis $h \in C$ which performs not much worse than the best approximation $opt_C,$ in the PAC sense -- the algorithm is a $(\epsilon, \delta)-$\textit{agnostic learner} for $D$ and $C,$ if given access to samples from $D$ it outputs a hypothesis $h\in C$ such that $err_D(c) \leq \epsilon + opt_C$, except with probability $\delta$.

Another common model in CLT is, the \textbf{exact learning from membership queries} model \cite{1988_Angluin}, which is, intuitively, related to active supervised learning (see section \ref{intermediary:models}). Here, we have access to an \textit{oracle}, a black-box, which outputs the concept value $f(\textbf{x})$ when queried with an example $\textbf{x}$. The basic setting is exact, meaning we are required to output a hypothesis which makes no errors whatsoever, however with a bounded probability (say 3/4). In other words, this is PAC learning where $\epsilon = 0,$ but we get to choose which examples we are given, adaptively, and $\delta$ is bounded away from 1/2. The figure of merit usually considered in this setting is \textbf{query complexity,} which denotes the number of calls to the oracle the learning algorithm uses, and is for most intents and purposes synonymous to sample complexity\footnote{When the oracle allows non-trivial inputs, one typically talks about query complexity. Sample complexity deals with the question of ``how many samples'' which suggest the setting where the oracle only produces outputs, without taking inputs. The distinction is not relevant for our purposes and is more often a matter of convention of the research line.}. This, in spirit, corresponds to an active supervised learning setting.

Much of PAC learning deals with identifying examples of interesting concept classes which are learnable (or proving that relevant classes are not), but other more general results exist connecting this learning framework. For instance, we can ask whether we can achieve a finite-sampling universal learning algorithm: that is, an algorithm that can learn any concept, under any distribution using some fixed number of samples $N$. The No Free Lunch theorems we mentioned previously imply that this is not possible: for each learning algorithm (and $\epsilon, \delta$), and any $N$ there is a setting (concept/distribution) which requires more than $N$ samples to achieve $(\epsilon, \delta)$-learning.

Typically, the criterion for a problem to be learnable assumes that there exists a classifier whose performance is essentially arbitrarily good -- that is, it assumes the classifier is strong. The \textbf{boosting} result in ML, already touched upon in section \ref{othermodels}, shows that settling on weak classifiers, which perform only slightly better than random classification, does not generate a different concept of learnability \cite{1990_Schapire}. 
 
Classical CLT theory has also been generalized to deal with concepts with continuous ranges. In particular, so called \textit{p-concepts} have range in $[0,1]$ \cite{1994_Kearns}. The generalization of the entire CLT to deal with such continuous-valued concepts is not without problems, but nonetheless, some of the central results, for instance quantities which are analogs of the VC-dimension, and analogous theorems relating this to generalization performance, can still be provided (see \cite{2007_Aaronson} for an overview given in the context of the learning of quantum states discussed in section \ref{QGML}).

Computational learning theory is closely related to the statistical learning theory of Vapnik and Chervonenkis   (VC theory) which we discuss next.
\subsubsection{VC theory}
\label{VCtheory}
The statistical learning formalism of Vapnik and Chervonenkis was developed over the course of more than 30 years, and in this review we are forced to present just a chosen aspect of the total theory, which deals with generalization performance guarantees.
In the previous paragraph on PAC learning, we have introduced the concept of total error, which we will refer to as (total) risk. It is defined as the average over all the data points, which is, for a hypothesis $h$, given with $R(h) = error(h) =  \sum_{\mathbf{x}} P(D = \mathbf{x}) |h(\mathbf{x}) -  f(\mathbf{x})| $ (we are switching notation to maintain consistency with literature of differing communities). However, this quantity cannot be evaluated in practice, as in practice we only have access to the training data. This leads us to the notion of the \textbf{empirical risk} given with 
\EQ{\hat{R}(h) =  \dfrac{1}{N}\sum_{\mathbf{\mathbf{x} \in S_N}} |h(\mathbf{x}) -  f(\mathbf{x})|, }
 where $S_N$ is the training set drawn independently from the underlying distribution $D$.
 
 The quantity $\hat{R}(h)$ is intuitive and directly measurable. However, the problem of finding learning models which optimize empirical risk alone is not in it self interesting as it is trivially resolved with a look-up table.
 From a learning perspective, the more interesting and relevant quantity is the performance beyond the training set, which is contained in the \textbf{un}measurable $R(h),$ and indeed the task of inductive supervised learning is identifying $h$ which minimizes $R(h)$, given only the finite training set $S_N$.
 Intuitively, the hypothesis $h$ which minimizes the empirical risk should also be our best bet for the hypothesis which minimizes $R(h),$ but this can only make sense if our hypothesis family is somehow constrained, at least to a family of total functions: again, a look-up table has zero empirical risk, yet says nothing about what to do beyond.
 One of the key contributions of VC theory is to establish a rigorous relationship between the observable quantity $\hat{R}(h)$ -- the empirical risk, the quantity we actually wish to bound $R(h)$ -- the total risk, and the family of hypotheses our learning algorithm can realize.
 Intuitively, if the function family is too flexible (as is the case with just look-up tables) a perfect fit on the examples says little. In contrast, having a very restrictive set of hypotheses, say just one (which is independent from the dataset/concept and the generating distribution), suggest that the empirical risk is a fair estimate of the total risk (however bad it may be), as nothing has been tailored for the training set. This brings us to the notion of the \textbf{model complexity} of the learning model, which has a few formalizations
 , and here we focus on the Vapnik-Chervonenkis dimension of the model (VC dimension)\footnote{Another popular measure of model complexity is e.g. Rademacher complexity \cite{2003_Bartlett}.}.

 The VC-dimension is an integer number assigned to a set of hypotheses $H \subseteq \{h | h: \mathbf{S} \rightarrow \{0,1 \} \},$ (e.g. the possible classification functions our learning algorithm can even in principle be trained to realize), where $\mathbf{S}$ can be, for instance,  the set of bitstrings $\{ 0,1\}^{n},$ or, more generally, say real vectors in $\mathbbmss{R}^{n}$.
In the context of basic SVMs, the set of hypotheses are ``all hyperplanes''\footnote{Naturally, a non-trivial kernel function enriches the set of hypotheses realized by SVMs.}.
Consider now a subset $C_k$ of $k$ points in $\mathbbmss{R}^{n}$ in a general position\footnote{General position implies that no sub-set of points is co-planar beyond what is necessary, i.e. points in $S \mathbbmss{R}^n$ are in general position if no hyperplane in $\mathbbmss{R}^n$ contains more than $n$ points in $S$.}. These points can attain binary labels in $2^k$ different ways. The hypothesis family $H$ is said to \textbf{shatter} the set $C,$ if for any labeling $\ell$ of the set $C_k,$ there exists a hypothesis $h \in H$ which correctly labels the set $C_k$ according to $\ell$. In other words, using functions from $H$ we can learn any labeling function on the set $C_k$ of $k$ points in a general position perfectly. 
The \textbf{VC dimension} of $H$ is then the largest $k_{max}$ such that there exists the set $C_{k_{max}}$ of points in general position which is shattered (perfectly ``labelable'' for any labeling) by $H$. For instance, for $n=2$, ``rays'' shatter three points but not 4 (imagine vertices of a square where diagonally opposite vertices share the same label), and in $n=N$, ``hyperplanes'' shatter $N+1$ points. While it is beguiling to think that the VC dimension corresponds to the number of free parameters specifying the hypothesis family, this is not the case\footnote{The canonical counterexample is the family specified by the partition of the real plane, halved by the graph of the two-parametric function $h_{\alpha,\beta}(x)=\alpha \sin( \beta x),$ which can be proven to shatter any finite number of points in $n=2$. The fact that the number of parameters of a function does not fully capture the complexity of the function should not be surprising as any (continuous) function over $k+n$ variables (parameters + dimension) can be encoded as a function over $1+n$ variables. }.
The VC theorem (in one of its variants) \cite{1996_Devroye} then states that the empirical risk matches total risk, up to a deviation which decays in the number of samples, but grows in the VC-dimension of the model, more formally:
\EQ{
P\left(  \hat{R}(h_{S_N}) - R(h_{S_N}) \leq \epsilon \right) = 1- \delta\\
\epsilon = \sqrt{\dfrac{d \left( \log(2N/d) +1 \right)}{N}  - \dfrac{\log(\delta/4)}{N}  }, \label{VCth}
}
where $d$ is the VC-dimension of the model, $N$ number of samples, and $ h_{S_N}$ is the hypothesis output by the model, given the training set $S_N,$ which is sampled from the underlying distribution $D$. The underlying distribution $D$ implicitly appears also in the total risk $R$.
Note that the chosen acceptable probability of incorrectly bounding the true error, that is, probability $\delta$,  contributes only logarithmically to the misestimation bound $\epsilon$, whereas the VC dimension and the number of samples contribute (mutually inversely) linearly to the square of $\epsilon$.

The VC theorem suggests that the ideal learning algorithm would have a low VC dimension (allowing a good estimate of the relationship of the empirical and total risk), while at the same time, performing well on the training set. This leads to a learning principle called \textbf{structural risk minimization}. Consider a parametrized learning model (say parametrized by an integer $l \in \mathbbmss{l}$) such that each $l$ induces a hypothesis family $H^{l},$ each more expressive then the previous, so $H^{l} \subseteq H^{l+1}$.
Structural risk minimization (contrasted to empirical risk minimization which just minimizes empirical risk) takes into account that in order to have (a guarantee on) good generalization performance we need to have both good observed performance (i.e. low empirical risk) and low model complexity. High model complexity induces the risk stemming from the structure of the problem, manifested in common issues such as data overfitting.
In practice, this is achieved by considering (meta-)parametrized models, like $\{ H^{l}\}, $where we minimize a combination of $l$ (influencing the VC-dimension) and the empirical risk associated to $H^{l}$.
In practice, this is realized by adding a regularization term to the training optimization, so generically the (unregularized) learning process resulting in $\textup{argmin}_{h \in H} \hat{R}(h)$ is updated to $\textup{argmin}_{h^{l} \in H^{l}} \left(\hat{R}(h) + reg(l)  \right)$, where  $reg(\cdot)$ penalizes the complexity of the hypothesis family, or just the given hypothesis.

VC dimension is also a vital concept in PAC learning, connecting the two frameworks. Note first that a concept class $C,$ which is a set of concepts is also a legitimate set of hypotheses, and thus has a well-defined VC dimension $d_{C}.$ Then, the sample complexity of $(\epsilon, \delta)-$(PAC)-learning of $C$ is given with $O\left( (d_C + \ln{1/\delta}) \epsilon^{-1} \right).$

Many of the above results can also be applied in the contexts of unsupervised learning, however the theory of unsupervised (or structure learning), is mostly concerned with the understanding of particular methodologies, the topic of which is beyond this review paper.

%
%

\subsection{{Basic methods and theory of reinforcement learning}}
\label{subsec:RL}
\boxTLDR{While RL, in all generality, studies learning in and from interactive task environments, perhaps the best understood models consider more restricted settings. Environments can often be characterized by \textbf{Markov Decision Processes}, i.e. they states, which can be observed by the agent. The agent can cause transitions from states to states, by its actions, but the \textit{rules of transitions} are not known beforehand. Some of the transitions are rewarded. The agent learns which actions to perform, given that the environment is in some state, such that it receives the highest value of rewards (expected return), either in a fixed time frame (\textbf{finite-horizon}) or over (asymptotically) long time periods, where future rewards are geometrically depreciated (\textbf{infinite-horizon}).
Such models can be solved by estimating \textbf{action-value functions}, which assign expected return to actions given states, for which the agent must explore the space of strategies, but other methods exist. In more general models, the state of the environment need not be fully observable, and such settings are significantly harder to solve.
RL settings can also be tackled by models from the so-called \textbf{Projective Simulation} framework for the design of learning agents, inspired by physical stochastic processes. While comparatively new, this model is of particular interest as it had been designed with the possibilities of beneficial quantization in mind.
Interactive learning methods include models beyond textbook RL, including partially observable settings, which require generalization and more. Such extensions, e.g. generalization, typically require techniques from non-interactive learning scenarios, but also lead to agents with an ever increasing level of autonomy. In this sense, RL forms a bridge between ML and general AI models.
}

\label{RL-section}
Broadly speaking, RL deals with the problem of learning how to optimally behave in unknown environments.  In the basic textbook formalism, we deal with a \textbf{task environment}, which is specified by a Markov decision process (MDP). MDPs are  labeled, directed graphs with additional structures, comprising a discrete and finite sets of \textbf{states} $\mathcal{S} = \{ s_i\}$ and \textbf{actions} $\mathcal{A} = \{ a_i\}$, which denote the possible states of the environment, and the actions the learning agent can perform on it, respectively.

   \begin{wrapfigure}{r}{0.4\textwidth}
 \includegraphics[width=0.4\textwidth,clip=true,trim =310 140 240 200]{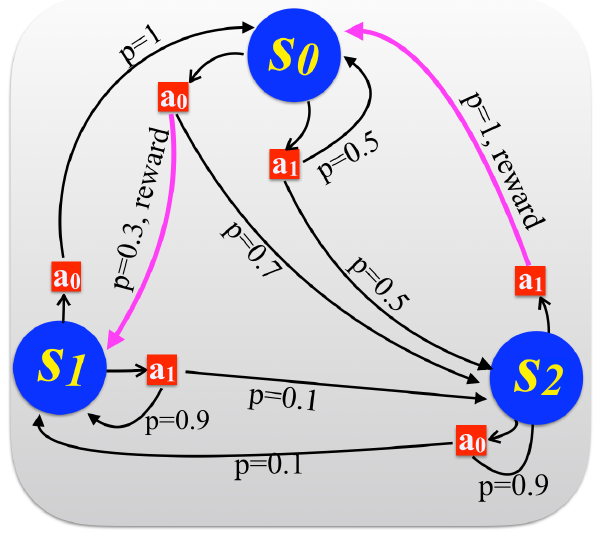}
 \vspace{-0.5cm}
\caption{\label{fig:MDP}A three state, two-action MDP. }
 \end{wrapfigure}
 The choice of the actions of the agent change the state of the environment, in a manner which is specific to the environment (MDP), and which may be probabilistic.
This is captured by a \textbf{transition rule} $\textup{P}(s | s',a),$ denoting the probability of the environment ending up in the state $s,$ if the action $a$ had been performed in the state $s'$. Technically, this can be viewed as a collection of action-specific Markov transition matrices $\{ P^{a} \}_{a \in \mathcal{A}}$ that the learner can apply on the environment by performing an action. 

 These describe the dynamics of the environment conditioned on the actions of the agent. 
 The final component specifying the environment is a \textbf{reward function} $R: \mathcal{S} \times \mathcal{A} \times \mathcal{S} \rightarrow \Lambda $, where $\Lambda$ is a set of rewards, often binary. In other words, the environment rewards certain transitions\footnote{Rewards can also be probabilistic. This can be modelled by explicitly allowing stochastic reward functions, or by extending the state space, to include rewarding and non-rewarding instances of states (note, the reward depends on current state, action \textit{and} the reached state) in which case the probability of the reward is encoded in the transition probabilities.}.
 At each time instance, the action of the learner is specified by a \textbf{policy:} a conditional probability distribution $\pi(a | s),$ specifying the probability of the agent outputting the action $a$ provided it is in the state $s$. Given an MDP, intuitively the goal is finding good policies, i.e. those which yield high rewards. This can be formalized in many non-equivalent ways.

 Given a policy $\pi$ and some initial state $s$ we can e.g. define \textbf{finite-horizon} expected total reward after $N$ interaction steps with $R_{N}^{s}(\pi) = \sum_{i=1}^{N} r_i,$ where
 $r_i$ is the expected reward under policy $\pi$ at time-step $i,$ in the given environment, and assuming we started from the state $s$.
 If the environment is finite and strongly connected\footnote{In this context this means that the underlying MDP has finite return times for all states, that is, there is a finite probability of going back to the initial state from any state for some sequence of actions.}, the finite-horizon rewards diverge as the horizon $N$ grows. However, by adding a geometrically depreciating factor (rate $\gamma$) 
 we obtain an always bounded expression
$
 R_{\gamma}(\pi) = \sum_{i=1}^{\infty} \gamma^i r_i,
 $
 called the \textbf{infinite horizon expected reward} (parametrized by $\gamma$), which is more commonly studied in literature.
 The expected rewards in finite or infinite horizons form the typical \textbf{figures of merit} in solving MDP problems, which come in two flavors.
First, in decision theory, or \textbf{planning} (in the context of AI), the typical goal is finding the policy $\pi^{opt}$ which optimizes the (in)finite horizon reward in \textit{a given MDP}, formally:
 given the (full or partial) specification of the MDP $M$, solve
$
 \pi^{opt} = \textup{argmax}_{\pi}  R_{N/\gamma}(\pi),
 $
where $R$ is the expected reward in finite (for $N$ steps) or infinite horizon (for a given depreciation $\gamma$) settings, respectively.  
Such problems can be solved by {dynamic} and {linear programming}.
In RL \cite{1998_Sutton}, the specification of the environment (the MDP), in contrast, is not given, but rather can be explored by interacting with it dynamically. The agent can perform an action, and receive the subsequent state (and perhaps a reward).
The ultimate goal here comes in two related (but conceptually different) flavours. One is to design an agent which will over time learn the optimal policy $ \pi^{opt}$, meaning the policy can be read out from the memory of the agent/program. Slightly differently, we wish an agent which will, over time gradually alter its behaviour (\textit{policy}) as to \textbf{act according to the optimal policy}. While in theory these two are closely related, e.g. in robotics these are quite different as the reward rate before convergence (perfect learning) also matters \footnote{These two flavours are closely related to the notions of on-policy and off-policy learning. These labels typically pertain to how the estimates of the optimal policy are internally updated, which may be in accordance to the actual current policy and actions of the agent, or independently from the executed action, respectively. For more details see e.g. \cite{1998_Sutton}. }. 
First of all, we point out that RL problems as given above can be solved reliably whenever the MDP is finite and strongly connected: a trivial solution is to stick to a random policy until a reliable tomography of the environment can be done, after which the problem is resolved via dynamic programming \footnote{If the environment is not strongly connected, this is not possible: for instance the first move of the learner may lead to ``good'' or ``bad'' regions from which there is no way out, in which case optimal behaviour cannot be obtained with certainty.}. 
Often, environments actually have additional structure, so-called initial and terminal states: if the agent reaches the terminal state, it is ``teleported'' to the fixed initial state. Such structure is called \textbf{episodic}, and can be used as a means of ensuring the strong connectivity of the MDP.

One way of obtaining solutions is by tracking so-called {value functions} $V_{\pi}(s):\mathcal{S} \rightarrow \mathbbmss{R}$ which assign expected reward under policy $\pi$ assuming we start from state $s$; this is done recursively: the value of the current state is the current reward plus the averaged value of the subsequent state (averaged under the stochastic transition rule of the environment $P(s| a, s')$). Optimal policies optimize these functions, and this too is achieved sequentially by modifying the policy as to maximize the value functions.
This, however, assumes the knowledge of the transition rule $P(s| a, s')$.
In further development of the theory, it was shown that tracking action-value functions $Q_{\pi}(s,a),$ given by 
 \EQ{ Q_{\pi}(s,a)=\sum _{s'}P(s'| a, s)(\Lambda(s,a,s')+\gamma V_{\pi}(s'))
}
assigning the value not only to the state, but the subsequent action as well can be modified into an online learning algorithm\footnote{This rule is inspired by the the Bellman optimality equation, 
$Q^\ast(s, a) := \mathbbmss{E}[ R(s, a) ] + \gamma \mathbbmss{E}[ max_{a'} Q\ast(s',a')]$, where the expected values are taken over the randomness MDP transition rule and the reward function, which has as the solution -- the fixed point -- the optimal $Q-$value function. This equation can be used when the specification of the  environment is fully known. Note that the optimal Q-values can be found without actually explicitly identifying an optimal policy. 
}. In particular, the Q-values can be continuously estimated by weighted averaging the current reward (at timestep $t$) for an action-value, and the estimate of the highest possible Q-value of the subsequent action-value:

\EQ{
 Q^{t+1}(s_{t},a_{t}) = \underbrace {Q^{t}(s_{t},a_{t})} _{\rm {old~value}}+\underbrace {\alpha _{t}} _{\rm {learning~rate}}\cdot \left(\overbrace {\underbrace {r_{t+1}} _{\rm {reward}}+\underbrace {\gamma } _{\rm {discount}} \cdot \underbrace{\max _{a}Q^{t}(s_{t+1},a)} _{\rm {{estimate~of~optimal \atop future~value}}}} ^{\rm {learned~value}}-\underbrace {Q^{t}(s_{t},a_{t})} _{\rm {old~value}}\right). \label{q-value}
}
Note that having access to the optimal Q-values suffices to find the optimal policy: given a state, simply pick an action with the highest Q-value, but the algorithm above says nothing about which policy the agent should employ while learning.
In \cite{1992_Watkins} it was shown that the algorithm, specified by the update rule of Eq. \ref{q-value}, called \textit{Q-learning} indeed converges to optimal $Q$ values as long as the agent employs any fixed policy which has non-zero probabilities for all actions given any state (the parameter $\alpha_t$, which is a function of time, has to satisfy certain conditions, and $\gamma$ should be the $\gamma$ of the targeted figure of merit $R_{\gamma}$)\footnote{Q-learning is an example of an off-policy algorithm as the estimate of the future value in Eq. \ref{q-value} is not evaluated relative to the actual policy of the agent (indeed, it is not necessarily even defined), but rather relative to the so-called ``greedy-policy'', which takes the action with the maximal value estimate (note the estimate appears with a maximization term). }.
 
In essence, this result suffices for solving the first flavour of RL, where the optimal policy is ``learned'' by the agent in the limit, but, in principle, never actually used.
The convergence of the Q-learning update to the optimal Q-values, and consequently to the optimal behaviour, has been proven for all learning agents using \textit{Greedy-in-the-limit, infinite exploration} (GLIE) policies. As the name suggests, such policies, in the asymptotic limit perform actions with the highest value estimated\footnote{To avoid any confusion, we have introduced the concept \textit{policy} to refer to the conditional probability distributions specifying what the agent will do given a state. However, the same term is often overloaded to also refer to the specification of the effective policy an agent will use given some state/time-step. For instance, ``$\epsilon-$greedy policies'' refer to behaviour in which, given a state, the the agent outputs the action with the highest corresponding $Q-$value -- i.e. acts 
\textit{greedily} -- with probability $1-\epsilon$, and produces a random action otherwise. Clearly, this rule specifies a policy at any given time step, given the current Q-value table of the agent. One can also think of \textit{time-dependent} policies, which mean that the policy also explicitly depends on the time-step. An example of a such a time-dependant and a (slowly converging) GLIE policy is an $\epsilon-$greedy policy, where $\epsilon = \epsilon(t) = 1/t$ is a function of the time-step, converging to zero.}. 

At the same time, infinite exploration means that, in the limit all state/action combinations will be tried out infinitely many times ensuring true optimal action values are found, and that the local minima are avoided. In general, the optimal trade off between these two competing properties, the \textit{exploration of the learning space}, and the \textit{exploitation of obtained knowledge} is quintessential for RL.
There are many other RL algorithms which are based on state value, or action-value optimizations, such as SARSA\footnote{SARSA is the acronym for state-action-reward-state-action.}, various value iteration methods, temporal difference methods etc. \cite{1998_Sutton}.
In more recent times, progress has been achieved by using parametrized approximations of state-action-value-functions -- a cross-breed between function approximation and reinforcement learning -- which reduces the search space of available Q-functions. Here, the results which combine deep learning for value function approximation with RL have been particularly successful \cite{2015_Mnih} and the same approach also underpins the AlphaGo \cite{2016_Silver} system. 
This brings us to a different class of methods which do not optimize state, or action-value functions, but rather learn complete policies, often by performing an estimate of gradient descent, or other means of \textbf{direct optimization in policy space}. This is feasible whenever the policies are specified indirectly, by a comparably small number of parameters, and can in some cases be faster \cite{2001_Peshkin}.

The methods we discussed thus far consider special cases of environments, where the environment is Markovian, or, related to this, fully observable. The most common generalization of this are so-called \textbf{partially observable} MDPs (POMDP), where the underlying MDP structure is extended to include a set of \textbf{observations} $\mathcal{O}$ and a stochastic function defined with the conditional probability distribution $P_{POMDP}(o \in \mathcal{O} | s \in \mathcal{S}, a \in \mathcal{A})$. The set of states of the environment are no longer directly accessible to the agent, but rather the agent perceives the observations from the set $\mathcal{O}$, which indirectly and, in general, stochastically depend on the actual unobservable environmental state, as given by the distribution $P_{POMDP}$, and the action the agent took last. POMDPs are expressive enough to capture many real world problems, and are thus a common \textit{world model} in AI, but are significantly more difficult to deal with compared to MDPs \footnote{For instance, the problem of finding optimal infinite-horizon policies, which was solvable via dynamical programming in the fully observable (MDP) case becomes, in general, uncomputable.}. 

   \begin{wrapfigure}{l}{0.4\textwidth}
 \includegraphics[width=0.4\textwidth,clip=true,trim =0 0 0 0]{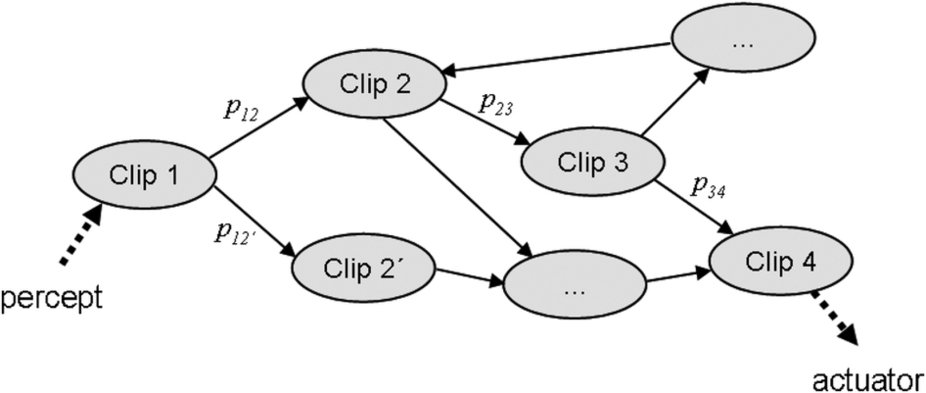}
 \vspace{-0.5cm}
\caption{\label{fig:PSbasic}{Illustration of the structure of the episodic and compositional memory in PS, comprising clips (episodes) and probabilistic transitions. The actuator of the agent performs the action. Adapted from \cite{2012_Briegel}.}  \vspace{-0cm}}
 \end{wrapfigure}
As mentioned, the setting of POMDPs moves us one step closer to arbitrary environment settings, which is the domain of artificial (general) intelligence\footnote{To comment a bit on how RL methods and tasks may be generalized towards general AI, one can consider learning scenarios where one has to combine standard data-learning ML to handle the realistic percept space (which is effectively infinite) with RL techniques. An example of this as was done e.g. in the famous AlphaGo system \cite{2016_Silver}. Further, one could also consider more general types of interaction, beyond the strict turn-based metronomic model. For instance in \textit{active reinforcement learning}, the interaction occurs relative to an external clock, which intertwines computational complexity and learning efficiency of the agent (see section \ref{QLI}).
 Further, the interaction may occur in fully continuous time.
 This setting is also not typically studied in the basic theory of AI, but occurs in the closely related problem of control theory \cite{2010_Wiseman}, which may be more familiar to physicists. Such generalizations are at the cutting edge of research, also in the classical realm, and also beyond the scope of this paper.}.

 The context of AGI is often closely related to modern view on robotics, where the structure of what can be observed, and what actions are possible stems not only from the nature of the environment, but also (bodily) constraints of the agent: e.g. a robot is equipped with sensors, specifying and limiting what the robot can observe or perceive, and actuators, constraining the possible actions.
In such an agent-centric viewpoint, we typically talk about the set of percepts -- signals that the agent can perceive -- which may correspond to full states, or partial observations, depending on the agent-environment setting -- and the set of actions\footnote{In this sense, a particular agent/robot, may perceive the full state of the environment in some environments (making the percepts identical to states), whereas in other environments, the sensors fail to observe everything, in which case the percepts correspond to observations. }.

This latter viewpoint, that the percept/action structure stems from the physical constitution of the agent and the environment, which we will refer to as an \textbf{embodied perspective}, was one of the starting points of the development of the projective simulation (PS) model for AI.
PS is a physics-inspired model for AI which can be used for solving RL tasks. The centerpiece of the model is the so-called Episodic and Compositional Memory (ECM),
 which is a stochastic network of clips, see Fig. \ref{fig:PSbasic}.

  Clips are representations of short autobiographical episodes, i.e. memories of the agent. Using the compositional aspects of the memory, which allows for a rudimentary notion of creativity, the agent can also combine actual memories to generate fictitious, conceivable clips which need not have actually occurred. 
More formally, clips can be defined recursively as either memorized percepts or actions, or otherwise structures (e.g. sequences) of clips.  Given a current percept, the PS agent calls its ECM network to perform a stochastic random walk over its clip space (the structure of which depends on the history of the agent) projecting itself into conceivable situations, before committing to an action.
Aspects of this model have been beneficially quantized, and also used both in quantum experiments and in robotics and we will focus more on this model in section \ref{QLI}.

\paragraph{Learning efficiency and learnability for RL}

As mentioned in the introduction to this section, No Free Lunch theorems also apply to RL, and any statement about learning requires us to restrict the space of possible environments. For instance, ``finite-space, time-independent MDPs'' is a restriction which allows perfect learning relative to some of the standard figures of merit, as was first proven by the Q-learning algorithm. Beyond learnability, in more recent times, notions of sample complexity for RL tasks have also been explored, addressing the problem from different perspectives.  
The theory of sample complexity for RL settings is significantly more involved than for supervised learning, although the very basic desiderata remain the same: how many interaction steps are needed before the agent learns. Learning can naturally mean many things, but most often what is meant is that the agent learns the optimal policy. 
Unlike supervised learning, RL has the additional temporal dimension in the definitions of optimality (e.g. finite or infinite horizons), leading to an even broader space of options one can explore. 
Further details on this important field of research are beyond the scope of this review, and we refer the interested reader to e.g. the thesis of Kakade \cite{2003_Kakade} which also does a good job of reviewing some of the early works, and finds sample complexity bounds for RL for many basic settings, or e.g. \cite{2013_Lattimore, 2015_Dann} for some of the newer results.

\section{Quantum mechanics, learning, and AI}
\label{sec:map}
Quantum mechanics has already had profound effect on the fields of computation and information processing. However, its impact on AI and learning has, up until very recently, been modest. Although the fields of ML and AI have a strong connection to theory of computation, these fields are still different, and not all progress in (quantum) computation implies qualitative progress in AI.
For instance, although it has been more than 20 years, still the arguably most celebrated result in QC is that of Shor's factoring algorithm \cite{1997_Shor}, which, on the face of it, has no impact on AI\footnote{In fact, this is not entirely true -- certain proofs of separation between PAC learnability in the quantum and classical model assume hardness of factoring of certain integers (see section \ref{LMQ}).}. Nonetheless, other, less famous results may have application to various aspects of AI and learning. 
The field of QIP has thus, from its early stages had a careful and tentative interplay with various aspects of AI, although it is only recently that this line of research has received a broader attention.
Roughly speaking, we can identify \textbf{four main directions} covering the interplay between ML/AI summarized in  in Fig. \ref{fig:Topics}.

\begin{figure}[!htb]
\boxPretty{
    \centering
    \begin{minipage}{.485\textwidth}
        \centering
     \begin{itemize}[leftmargin=*]

\item[] \hspace{-0.5cm} \textit{Applications of ML in quantum physics}
\begin{enumerate}
\item Estimation and metrology
\item Quantum control and gate design
\item Controlling quantum experiments, and machine-assisted research
\item Condensed matter and many body physics
\end{enumerate}

\item[] \hspace{-0.5cm} \textit{Quantum enhancements for ML}
\begin{enumerate}
\item Quantum perceptrons and neural networks 
\item Quantum computational learning theory
\item Quantum enhancement of learning capacity
\item Quantum computational algorithmic speed-ups for learning
\end{enumerate}

\end{itemize}

    \end{minipage}%
    \begin{minipage}{0.485\textwidth}
 \centering
  \begin{itemize}

\item[] \hspace{-0.5cm} \textit{Quantum generalizations of ML-type tasks} 
\begin{enumerate}
\item Quantum generalizations: machine learning of quantum data 
\item (Quantum) learning of quantum processes
\end{enumerate}

 \item[] \hspace{-0.5cm} \textit{Quantum learning agents and elements of quantum AI}
\begin{enumerate}
\item Quantum-enhanced learning through interaction
\item Quantum agent-environment paradigm
\item Towards quantum AI
\end{enumerate}
\end{itemize}

    \end{minipage}}
    \caption{\label{fig:Topics}Table of topics investigating the overlaps between quantum physics, machine learning, and AI.    }
\end{figure}

Historically speaking, the first contacts between aspects of QIP and learning theory occurred in terms of the direct application of statistics and statistical learning in light of the quantum theory, which forms the first line: \textbf{classical machine learning applied in quantum theory and experiment} reviewed in section \ref{sec:MLtoQIP}. In this first topic, ML techniques are applied to data stemming from quantum experiments. The second topic, in contrast, machine learning over genuinely quantum data: \textbf{quantum generalization of machine learning-type tasks}, discussed in section \ref{QgenML}.
This brings us to the topic which has been receiving substantial interest in recent times: \textbf{can quantum computers genuinely help in machine learning problems}, addressed in section \ref{QeML}. 
The final topic we will investigate considers aspects of QIP which extend beyond machine learning (taken in a narrow sense), such as generalizations of RL, and which can be understood as stepping-stones towards quantum AI. This is reflected upon in section \ref{QAI}

It is worthwhile to note that there are many possible natural classifications of the comprehensive field we discuss in this review. Our chosen classification is motivated by two subtly differing perspectives on the classification of quantum ML, discussed further in section \ref{Taxonomy}.

\section{Machine learning applied to (quantum) physics}

\label{sec:MLtoQIP}

In this section we review works and ideas where ML methods have been either directly utilized, or have otherwise been instrumental for QIP results. To do so, we are however, facing the ungrateful task of specifying the boundaries of what is considered a ML method.
In recent times, partially due to its successes, ML has become a desirable key word, and consequently an umbrella term for a broad spectrum of techniques. This includes algorithms for solving genuine learning problems, but also methods and techniques designed for indirectly related problems.
From such an all-encompassing viewpoint, ML also includes aspects of (parametric) statistical learning, the solving of black-box (or derivative-free) optimization problems, but also the solving of hard optimization problems in general\footnote{Certain optimization problems, such as online optimization problems where information is revealed incrementally, and decisions are made before all information is available, are more clearly related to ``quintessential'' ML problems such as supervised, unsupervised, or reinforcement learning. }. 

As we do not presume to establish hard boundaries, we adopt a more inclusive perspective.
The collection of all works which utilize such methods, which could conceivably fit in broad-scope ML, for QIP applications cannot be covered in one review. Consequently, we place emphasis on pioneering works, and works where the authors themselves advertise the ML flavour of used methodologies, thereby emphasizing the potential of such ML/QIP interdisciplinary endeavors.

The use of ML in the context of QIP, understood as above, has been considerable, with an effective explosion of related works in the last few years. ML has been shown to be effective in a great variety of QIP related problems: in quantum signal processing, quantum metrology, Hamiltonian estimation, and in problems of quantum control. In recent times, the scope of applications has been significantly extended, ML and involved techniques have also been applied to combatting noise in the process of performing quantum computations, problems in condensed-matter and many-body physics, and in the design of novel quantum optical experiments.
Such results suggest that advanced ML/AI techniques will play an integral role in quantum labs of the future, and in particular, in the construction of advanced quantum devices and, eventually, quantum computers.
In a complementary direction, QIP applications have also engaged many of the methods of ML, showing that QIP may also become a promising proving ground for cutting edge ML research.

Contacts between statistical learning theory (as a part of the theoretical foundations of ML) and quantum theory come naturally due to the statistical foundations of quantum theory.
Already the very early theories of quantum signal processing \cite{1969_Helstrom},  probabilistic aspects of quantum theory and quantum state estimation \cite{1982_Holevo}, and early works \cite{1994_Braunstein_Caves}  which would lead to modern quantum metrology \cite{2011_Giovanetti} included statistical analyses which establish tentative grounds for more advanced ML/QIP interplay. 
Related early works further emphasize the applicability of statistical methods, in particular maximum likelihood estimation, to quantum tomographic scenarios, such as the tasks of state estimation \cite{1997_Hradil}, the estimation of quantum processes \cite{2001_FiurasekHradil} and measurements \cite{2001_Fiurasek} and the reconstruction of quantum processes from incomplete tomographic data \cite{2005_Ziman_Stelmahivic}\footnote{Interestingly, such techniques allow for the identification of optimal approximations of \textit{unphysical processes} which can be used to shed light on the properties of quantum operations.}.
The works of this type generically focus on physical scenarios where clean analytic theory can be applied. However, in particular in experimental, or noisy (thus, realistic) settings, many of the assumptions, which are crucial for the pure analytic treatment, fail. This leads to the first category of ML applications to QIP we consider.

\newpage
\pagebreak

\subsection{Hamiltonian estimation and metrology}
\boxTLDR{Metrological scenarios can involve \textbf{complex measurement strategies}, where, e.g., the
 measurements which need to be performed may depend on previous outcomes. Further, the physical system under 
 analysis may be controlled with the help of additional parameters -- so-called controls -- which can be sequentially modified, leading to a more complicated space of possibilities.
\textbf{ML techniques} can help us \textbf{find optima} in such a complex space of strategies, under various constraints, which are often \textbf{pragmatically and experimentally motivated} constraints.
}

The identifying of properties of physical systems, be it dynamic properties of evolutions (e.g. process tomography), or properties of the states of given systems (e.g. state tomography), is a fundamental task. Such tasks are resolved by various (classical) metrological theories and methods, which can identify optimal strategies, characterize error bounds, and which have also been quite generally exported to the quantum realm. For instance, quantum metrology studies the estimation of the parameters of quantum systems, and, generally, identifies \textbf{optimal measurement strategies}, for their estimation. Further, quantum metrology places particular emphasis on scenarios where genuine quantum phenomena  -- a category of phenomena associated to and sometimes even defined by the need for complex, and difficult-to-implement quantum devices for their realization -- yield an advantage over simpler, classical strategies. 

The specification of optimal strategies, in general, constitute the problem of \textbf{planning}\footnote{More specifically, most metrology settings problems constitute  instances of off-line planning, and thus not RL, as the ``environment specification'' is fully specified -- in other words, there is no need to actually run an experiment, and the optimal strategies can be found off-line. See section \ref{AIMLC} for more detail.}, for which various ML techniques can be employed.
The first examples of ML applications for finding measurement strategies originate from the problem of phase estimation, a special case of Hamiltonian estimation. Interestingly, already this simple case, provides a fruitful playground for ML techniques: analytically optimal measurement strategies are relatively easy to find, but are experimentally unfeasible. In turn, if we limit ourselves to a set of ``simple measurements'', near-optimal results are possible, but they require difficult-to-optimize adaptive strategies -- the type of problem ML is good for. Hamiltonian estimation problems have also been tackled in more general settings, invoking more complex machinery. We first briefly describe basic Hamiltonian estimation settings and metrological concepts. Then we will delve deeper in these results combining ML with metrology problems.  

\subsubsection{Hamiltonian estimation}

The generic scenarios of Hamiltonian estimation, a common instance of metrology in the quantum domain, consider a quantum system governed by a (partially unknown) Hamiltonian within a specified family $H(\bm{\theta}),$ where $\bm{\theta} = (\theta_1, \ldots, \theta_n),$ is a set of parameters $\bm{\theta}.$ 
Roughly speaking, Hamiltonian estimation deals with the task of identifying the optimal methods (and the performance thereof) for estimating the Hamiltonian parameters. 

This amounts to optimizing the choice of initial states (\textit{probe states}), which will evolve under the Hamiltonian, and the choice of the subsequent measurements, which uncover the effect the Hamiltonian had, and thus, indirectly, the parameter values\footnote{Technically, the estimation also involves the use of a suitable estimator function, but these details will not matter.}. 
This prolific research area considers many restrictions, variations and generalizations of this task.
For instance, one may assume settings in which  we either have control over the Hamiltonian evolution time $t$, or it is fixed so that $t=t_0$, which are typically referred to as frequency, and phase estimation, respectively.
Further, the efficiency of the process can be measured in multiple ways. In a frequentist approach, one is predominantly interested in estimation strategies which, roughly speaking, allow for the best scaling of precision of the estimate, as a function of the number of measurements. The quantity of interest is the so-called quantum Fisher information, which bounds and quantifies the scaling.
Intuitively, in this setting, also called the local regime, many repetitions of measurements are typically assumed.
Alternatively, in the Bayesian, or single-shot, regime the prior information, which is given as a distribution over the parameter to be estimated, and its update to the posterior distribution given a measurement strategy and outcome, are central objects  \cite{2015_Jarzyna}. 
 The objective here is the identification of preparation/measurement strategies which optimally reduce the average variance of the posterior distribution, which is computed via Bayes' theorem.   
 
One of the key interests in this problem is that the utilization of, arguably, {genuine quantum features}, such as entanglement, squeezing etc. in the structure of the probe states and measurements may lead to provably more efficient estimation than is possible by so-called classical strategies for many natural estimation problems. 
Such quantum-enhancements are potentially of immense practical relevance \cite{2011_Giovanetti}. 
The identification of optimal scenarios has been achieved in certain ``clean'' theoretical scenarios, which are, however, often unrealistic or impractical. It is in this context that ML-flavoured optimization, and other ML approaches can help.
\subsubsection{Phase estimation settings}
Interesting estimation problems, from a ML perspective, can already be found in the simple examples of a phase shift in an optical interferometer, where one of the arms of an otherwise balanced interferometer contains a phase shift of $\theta$. Early on, it was shown that given an optimal probe state, with mean photon number $N$, and an optimal (so-called \textit{canonical}) measurement, the asymptotic phase uncertainty can decay as $N^{-1}$\cite{1995_SandersMilburn}\footnote{This is often also expressed in terms of the variance $(\Delta \theta)^2$, so as $N^{-2}$, rather than the standard deviation.} , known as the Heisenberg limit.  In contrast, the restriction to ``simple measurement strategies'' (as characterized by the authors) , involving only photon number measurements in the two output arms, achieve a quadratically weaker scaling of $\sqrt{N^{-1}},$ referred to as the standard quantum limit.
This was proven in more general terms: the optimal measurements cannot be achieved by the classical post-processing of photon number measurements of the output arms, but constitute an involved, experimentally unfeasible POVM \cite{2000_BerryWiseman}. However in \cite{2000_BerryWiseman} it was shown how this can be circumvented by using ``simple measurements'', provided they can be altered in run-time. Each measurement consists of a photon number measurement of the output arms, and is parametrized by an additional, controllable phase shift of $\phi$ in the free arm -- equivalently, the unknown phase can be tweaked by a chosen $\phi$. The optimal measurement process is an \textit{adaptive strategy}: an entangled N-photon state is prepared (see e.g. \cite{2001_BerryWisemanBreslin}), the photons are sequentially injected into the interferometer, and photon numbers are measured. At each step, the measurement performed is modified by choosing a differing phase shift  $\phi$, which depends on previous measurement outcomes. {In \cite{2000_BerryWiseman,  2001_BerryWisemanBreslin}, an explicit strategy was given, which achieves the Heisenberg scaling of the optimal order $O(1/N)$. However, for  $N>4$ it was shown this strategy is not \textit{strictly} optimal.}

 This type of planning is hard as it reduces to the solving of non-convex optimization problems\footnote{The non-convexity stems from the fact that the effective input state at each stage depends on previous measurements performed. As the entire interferometer set-up can be viewed as a one-subsystem measurement, the conditional states also depend on unknown parameters, and these are used in the subsequent stages of the protocol \cite{2010_HentschelSanders}. }. The field of ML deals with such planning problems as well, and thus many optimization techniques have been developed for this purpose. The applications of such ML techniques, specifically particle swarm optimization were first suggested in pioneering works \cite{2010_HentschelSanders, 2011_HentschelSanders}, and later in \cite{2012_Sergeevich}.
 In subsequent work, perhaps more well-known methods of differential evolution have been demonstrated to be superior and more computationally efficient \cite{2013_Lovett_Sanders}. 

\subsubsection{Generalized Hamiltonian estimation settings}
\label{GenHam}
ML techniques can also be employed in significantly more general settings of quantum process estimation.
More general Hamiltonian estimation settings consider a partially controlled evolution given by $H_C(\bm{\theta}),$ where $C$ is a collection of control parameters of the system.
This is a reasonable setting in e.g. the production of quantum devices, which have controls ($C$), but whose actual performance (dependant on $\bm{\theta}$) needs to be confirmed. Further, since production devices are seldom identical, it is beneficial to even further generalize this setting, by allowing the unknown parameters $\bm{\theta}$ to be only probabilistically characterized. More precisely, they are probabilistically dependent on another set of hyperparameters $\bm{\zeta} = (\zeta_1, \ldots, \zeta_k)$, such that the parameters $\bm{\theta}$ are distributed according to a known conditional probability distribution $P(\bm{\theta}|\bm{\zeta})$. 
This generalized task of estimating the hyperparameters $\bm{\zeta}$ thus allows the treatment of systems with inherent stochastic noise, when the influence of noise is understood (given by $P(\bm{\theta}|\bm{\zeta})$).
Such very general scenarios are addressed in \cite{2012_Granade_WiebeCory}, relying on classical learning techniques of Bayesian experimental design (BED)\cite{2004_Loredo}, combined with Monte Carlo methods.
The details of this method are beyond the scope of this review, but, roughly speaking,
BED assumes a Bayesian perspective on the experiments of the type described above. The estimation methods of the general problem (ignoring the hyperparameters and noise, for simplicity, although the same techniques apply)
realize a conditional probability distribution $P(\bm{d} | \bm{\theta} ; C)$ where $\bm{d}$ corresponds to experimental data, i.e. measurement outcomes collected in the experiment.
Assuming some prior distribution over hidden parameters ($P(\bm{\theta})$), the posterior distribution, given experimental outcomes, is given via Bayes theorem by
\EQ{
P(\bm{\theta} |\bm{d} ; C ) = \dfrac{P(\bm{d} | \bm{\theta} ; C ) P(\bm{\theta})}{P(\bm{d} | C)}.
}
The evaluation of above is already non trivial, predominantly as the normalization factor $P(\bm{d} | C)$ includes an integration over the parameter space. Further, of particular interest are scenarios where an experiment is iterated many times. In this case, analogously to the adaptive setting for metrology discussed above, it is beneficial to tune the control parameters $C$ dependent on the outcomes. 

BED \cite{2004_Loredo}, tackles such adaptive settings, by selecting the subsequent control parameters $C$ as to maximize a utility function\footnote{The utility function is an object stemming from decision theory and, in the case of BED it measures how well the experiment improves our inferences. It is typically defined by the prior-posterior gain of information as measured by the Shannon entropy, although there are other possibilities.}, for each update step. 
The Bayes updates consist of the computing of  $P(\bm{\theta} |\bm{d}_1, \ldots, \bm{d}_{l-1} \bm{d}_k) \propto  P(\bm{d_k} | \bm{\theta}) P (\bm{\theta} | \bm{d}_1, \ldots, \bm{d}_{l-1})$ at each step.
The evaluation of the normalization factor $P(\bm{d} | C)$ is, however, also non-trivial, as it includes an integration over the parameter space.
In \cite{2012_Granade_WiebeCory} this integration is tackled via numerical integration techniques, namely sequential Monte Carlo, yielding a novel technique for robust Hamiltonian estimation.

The robust Hamiltonian estimation method was subsequently expanded to use access to trusted quantum simulators, which forms a more powerful and efficient estimation scheme  \cite{2014_Wiebe_Cory_B}\footnote{This addition partially circumvents the computation of the \textit{likelihood function} $P(\bm{d} | \bm{\theta} ; C )$ which requires the simulation of the quantum system, and is in fact, in general intractable.}, which was also shown to be robust to moderate noise and imperfections in the trusted simulators \cite{2014_Wiebe_Cory_A}. A restricted version of the method of estimation with simulators was experimentally realized in \cite{2017_Wang}.
More recently, connected to the methods of robust Hamiltonian estimation, Bayesian and sequential Monte Carlo based estimation have further been combined with particle swarm optimization techniques \cite{2016_Stenberg_Wilhelm}. There the goal was to achieve reliable coupling strength and frequency estimation in simple decohering systems, corresponding to realistic physical models. More specifically, the studied problem is the estimation of field-atom coupling terms, and the mode frequency term, in  the Jaynes-Cummings model. The controlled parameters are the local qubit field strength, measurements are done via swap spectroscopy.

Aside from using ML to  perform partial process tomography of controlled quantum systems, ML can also help in the genuine problems of quantum control, specifically, the design of target quantum gates. This forms the subsequent topic. 

\subsection{Design of target evolutions}

\boxTLDR{
One of the main tasks quantum information is the \textbf{design of target quantum evolutions}, including quantum gate design. This task can be tackled by \textbf{quantum control} which studies controlled physical systems where certain parameters can be adjusted during system evolution, or by using extended systems, and unmodulated dynamics.
 Here, the underlying problem is an \textbf{optimization problem}, that is, the problem of finding optimal control functions or extended system parameters, of a system which is otherwise fully specified.  Under realistic constraints these optimization tasks are often \textbf{non-convex}, thus hard for conventional optimizers, yet \textbf{amenable to advanced ML technologies}.
Target evolution design problems can also be tackled by using feed-back from the actual experimental system, leading to the use of \textbf{on-line optimization} methods and \textbf{RL}.
}


From a QIP perspective, one of the most important tasks is the design of elementary quantum gates, needed for quantum computation. The paradigmatic approach to this is via quantum control, which aims to identify how control fields of physical systems need to be adapted in time, to achieve desired evolutions. The designing of target evolutions can also be achieved in other settings, e.g. by using larger systems, and unmodulated dynamics. In both cases, ML optimization techniques can be used to design optimal strategies, off line. However, target evolutions can also be achieved in run-time, by interacting with a tunable physical system, and without the need for the complete description of the system. We first consider off-line settings, and briefly comment on the latter on-line settings thereafter.

\subsubsection{Off-line design}
The paradigmatic setting in quantum control considers a Hamiltonian with a controllable ($c$) and a drift part ($dr$), e.g. $H(C(t)) = H_{dr} + C(t) H_{c}$. The free part is modulated via a (real-valued) control field $C(t)$. The resulting time-integrated operator  $U = U[C(t)] \propto \exp \left( - i \int_{0}^T dt  H(C(t)) \right)$, over some finite time $T$, is a function of the chosen field function $C(t)$. The typical goal is to specify the control field $C(t)$ which maximizes the transition probability from some initial state $\ket{0}$ to a final state $\ket{\phi},$ thus find $\textup{argmax}_{C} |\bra{\phi} U[C(t)] \ket{0}|$ \footnote{An example of such additional fields would be controlled laser fields in ion trap experiments, and the field function $C$ specifies how the laser field strengths are modulated over time.}. Generically, the mappings $C(t) \mapsto U[C(t)]$ are highly involved, but nonetheless, empirically it was shown that greedy optimization approaches provide optimal solutions (which is the reason why greedy approaches dominate in practice). This empirical observation was later elucidated theoretically \cite{2004_Rabitz_Carey}, suggesting that in generic systems local minima do not exist, which leads to easy optimization (see also \cite{2017_RusselRabitz} for a more up-to-date account). 
 This is good news for experiments, but also suggests that quantum control has no need for advanced ML techniques. However, as is often the case with claims of such generality, the underlying subtle assumptions are fragile which can often be broken. In particular, greedy algorithms for optimizing the control problem as above can fail, even in the low dimensional case, if we simply place rather reasonable constraints on the control function and parameters. Already for 3-level and 2-qubit systems with constraints on the allowed evolution time $t,$ and the precision of the linearization of the time-dependent control parameters\footnote{It is assumed that the field function $C(t)$ describing parameter values as functions of time is step-wise constant, split in $K$ segments. The larger the value $K$ is, the better is the approximation of a smooth function which would arguably be better suited for greedy approaches.}, it is possible to construct examples where greedy approaches fail, yet global (derivative-free) approaches, in particular differential evolution, succeed \cite{2014_Zahedinejad_Sanders}. 
 
Another example of hard off-line control concerns the design of high fidelity single-shot three-qubit gates\footnote{This includes the Toffoli (and Fredkin) gate which is of particular interest as it forms a universal gate set together with the simple single-qubit Hadamard transform \cite{2002_Shi} (if ancillas qubits are used).}, which is in \cite{2015_Zahedinejad_Sanders_B,2016_Zahedinejad_Sanders} addressed using a specialized novel optimization algorithm the authors called subspace-selective self-adaptive differential evolution (SuSSADE) . 

An interesting alternative approach to gate design is by utilizing larger systems. Specifically designed larger systems can naturally implement desired evolutions on a subsystem, without the need of time-dependent control (c.f. QC with always-on interaction \cite{2003_Benjamin}). In other words, local gates are realized despite the fact that the global dynamics is unmodulated. The non-trivial task of constructing such global dynamics, for the Toffoli gate, is in \cite{2016_Banchi_Bose} tackled by a method which relies stochastic gradient descent, and draws from supervised learning techniques.

\subsubsection{On-line design}
Complementary to off-line methods, here we assume access to an actual quantum experiment, and the identification of optimal strategies relies on on-line feedback. In these cases, the quantum experiment need not be fully specified beforehand. Further, the required methodologies lean towards on-line planning and RL, rather than optimization. In the case optimization is required,  the parameters of optimization are different due to experimental constraints, see \cite{2012_Shir_Rabitz} for an extensive treatment of the topic. 

The connections between on-line methods which use feedback from experiments to ``steer'' systems to desired evolutions, have been connected to ML in early works \cite{2008_Bang,2009_GammelmarkMolmer}. These exploratory works deal with generic control problems via experimental feedback, and have, especially at the time, remained mostly unnoticed by the community.  In more recent times, feedback-based learning and optimization has received more attention.
For instance in \cite{2014_Chen_Tarn} the authors have explored the applicability of a modified Q-learning algorithm for RL (see section \ref{RL-section}) on canonical control problems. Further, the potential of RL methods had been discussed in the context of optimal parameter estimation, but also typical optimal control scenarios in \cite{2016_WittekSanders}. In the latter work, the authors also provide a concise yet extensive overview of related topics, and outline a perspective which unifies various aspects of ML and RL in an approach to resolve hard quantum measurement and control problems. In \cite{2016_Clausen}, RL based on PS updates was analyzed in the context of general control-and-feedback problems. Finally, ideas of unified computational platforms for quantum control, albeit without explicit emphasis on ML techniques had been previously provided in \cite{2011_Machnes}.

In the next section, we further coarse-grain our perspective, and consider scenarios where ML techniques control various gates, and more complex processes, and even help us learn how to do interesting experiments.

 \subsection{Controlling quantum experiments, and machine-assisted research} 
 
 \boxTLDR{ML and RL techniques can help us \textbf{control complex quantum systems, devices, and even quantum laboratories}. Furthermore, almost as a by-product, they may also \textbf{help us to learn more about the physical systems and processes studied in an experiment}. 
Examples include adaptive control systems (agents) which learn how to control quantum devices, e.g. how to \textbf{preserve the memory} of a quantum computer, \textbf{combat noise} processes, \textbf{generate entangled quantum states}, and \textbf{target evolutions} of interest. In the process of learning such optimal behaviours even simple artificial agents also \textbf{learn, in an implicit, embodied embodied, sense, about the underlying physics}, which can be used by us to obtain \textbf{novel insights}. In other words \textbf{artificial learning agents can genuinely help us do research}.
}
  \label{control-in-lab}

The prospects of utilizing ML and AI in quantum experiments have been investigated also for ``higher-level" experimental design problems. Here one considers automated machines that control complex processes which e.g. specify the execution of longer sequences of simple gates, or the execution of quantum computations.
Moreover, it has been suggested that learning machines can be used for, and integrated into, the very design of quantum experiments, thereby helping us in conducting genuine research.
We first present two results where ML and RL methods have been utilized to control more complex processes (e.g. generate sequences of quantum gates to preserve memory), and consider the perspectives of machines genuinely helping in research thereafter.
 
 \subsubsection{Controlling complex processes}
The simplest example of involved ML machinery used to generate control of slightly more complex systems was done in the context of is the problem of \textit{dynamical decoupling} for quantum memories. In this scenario, a quantum memory is modelled as a system coupled to a bath (with a local Hamiltonian for the system ($H_S$) and the bath $H_B$), and decoherence is realized by a coupling term $H_{SB}$;  the local unitary errors are captured by $H_S$. The evolution of the total Hamiltonian  $H_{noise}= H_S + H_B + H_{SB}$ would destroy {the contents of the memory}, but this can be mitigated by adding a controllable local term $H_C$ acting on the system alone\footnote{For the sake of intuition, a frequent application of $X$ gates, referred to as bang-bang control, on a system which is freely evolving with respect to $\sigma_z$ effectively flips the direction of rotation of the system Hamiltonian, effectively undoing its action.  }. 
Certain optimal choices of the control Hamiltonian $H_C$ are known. For instance, we can consider the scenario where $H_C$ is modulated such that it implements instantaneous\footnote{By instantaneous we mean that it is assumed that the implementation requires no evolution time, e.g. by using infinite field strengths.} Pauli-$X$ and Pauli-$Y$ unitary operations, sequentially, at intervals $\Delta t$.  {As this interval, which is also the time of the decoherence-causing free evolution, approaches zero, so $\Delta t \rightarrow 0$, this process is known to ensure perfect memory.} However, the moment the setting is made more realistic, allowing finite $\Delta t$ times, the space of optimal sequences becomes complicated. In particular, optimal sequences start depending on $\Delta t,$ the form of the noise Hamiltonian, and total evolution time. 

   \begin{wrapfigure}{l}{0.4\textwidth}
 \includegraphics[width=0.4\textwidth,clip=true,trim =300 80 130 50]{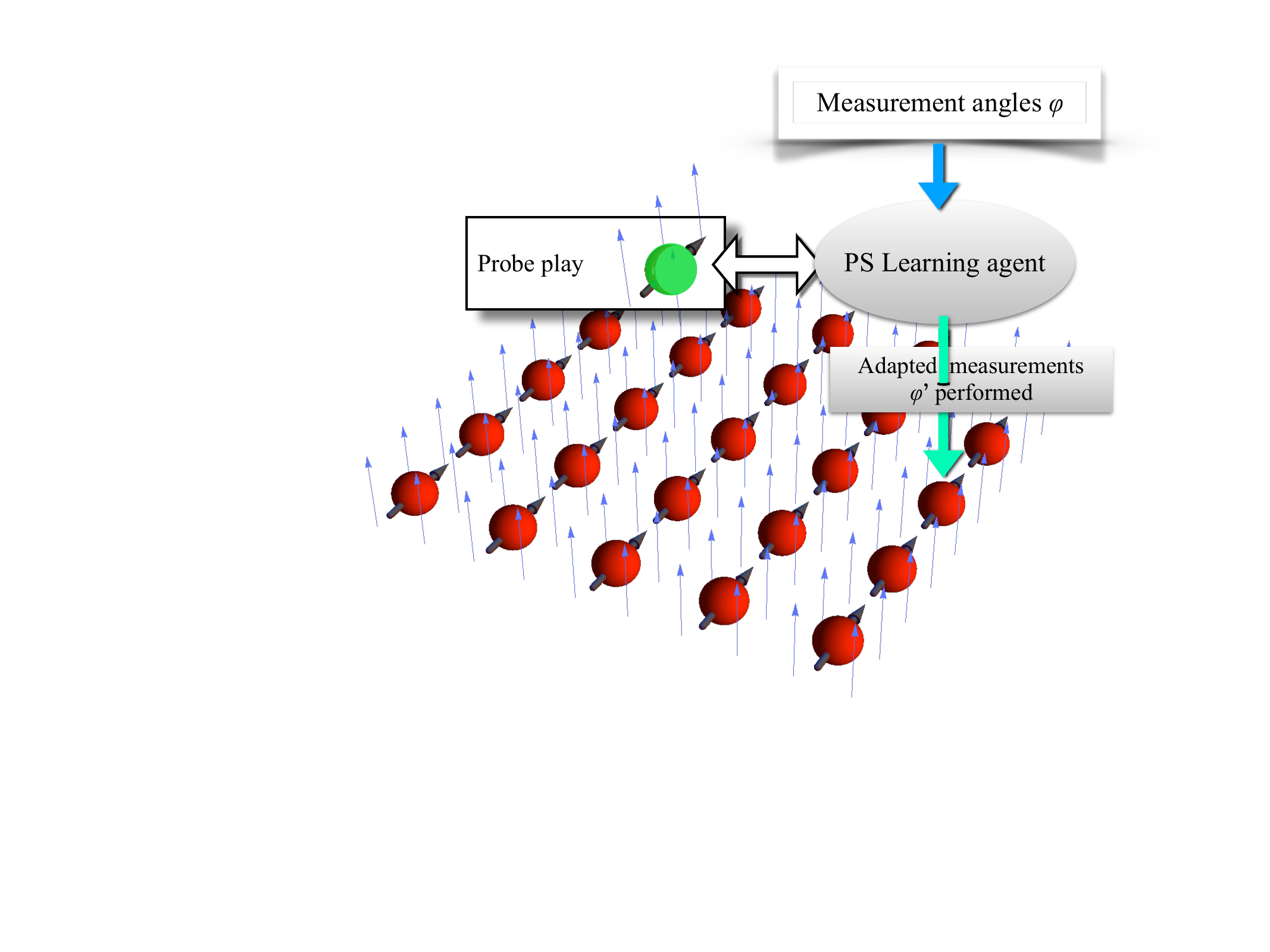}
 \vspace{-2.0cm}
\caption{\label{fig:PS-MBQC}{The learning agent learns how to correctly perform MBQC measurements in an unknown field.}  \vspace{-0.cm}}
 \end{wrapfigure}
To identify optimal sequences, in \cite{2017_AugustNi}, the authors employ recurrent NNs, which are trained as a generative model -- meaning they are trained to generate sequences which minimize final noise. The entire sequences of pulses (Pauli gates) which the networks generated were shown to outperform well-known sequences.

In a substantially different setting, where interaction necessarily arises, the authors studied how AI/ML techniques can be used to make quantum protocols themselves adaptive. Specifically, the authors applied RL methods based on PS \cite{2012_Briegel} (see section \ref{QLI})
to the task of protecting quantum computation from local stray fields 
\cite{2014_Tiersch}.
In MBQC \cite{2001_Raussendorf,2009_Briegel}, the computation is driven by performing adaptive single-qubit projective measurements on a large entangled resource state, such as the cluster state \cite{2001_Raussendorf}. In a scenario where the resource state is exposed to a stray field, each qubit undergoes a local rotation. To mitigate this,
in \cite{2014_Tiersch}, the authors introduce learning agent, which ``plays'' with a local probe qubit, initialized in say the $+1$ eigenstate of $\sigma_x$,  denoted  $\ket{+}$, learning how to compensate for the unknown field. In essence, given a measurement, the agent chooses a different measurement, obtaining a reward whenever a $+1$ outcome is observed. The agent is thus trained to compensate for the unknown field, and serves as an ``interpreter'' between desired measurements and the measurements which should be performed in the given setting (i.e. in the given field with given frequency of measurements $(\Delta t)$), see Fig. \ref{fig:PS-MBQC}.
The problem of mitigating such fixed stray fields could naturally be solved with non-adaptive methods where we use the knowledge about the system to solve our problem, by e.g. measuring the field and adapting accordingly, or by using fault-tolerant constructions.
From a learning perspective, such direct methods have a few shortcomings which may be worth presenting for didactic purposes. Fault tolerant methods are clearly wasteful, as they fail to gain utilize any knowledge about the noise processes. In contrast, field estimation methods \textbf{learn too much}, and \textbf{assume a model of the world}. To clarify the latter, to compensate the measured field, we need to use quantum mechanics, specifically the Born rule. In contrast, RL approach is model-free: the Born rule plays no part, and ``correct behavior'' is learned, and established exclusively based on experience. This is conceptually different, but also operatively critical, as model-free approaches allow for more autonomy and flexibility (i.e. the same machinery can be used in more settings without intervention)\footnote{Indeed, the authors also show that correct behavior can be established when additional unknown parameters are introduced, like time-and-space dependent fields (see \cite{2014_Tiersch} for results), where hand-crafted methods would fail.}.
Regarding learning too much, one of the basic principles of statistical learning posits that ``when solving a problem of interest, one should not solve a more general problem as an intermediate step" \cite{1995_Vapnik}, which is intuitive. The problem of the presented setting is ``how to adapt the measurement settings,'' and not ``characterize the stray fields''. While in the present context, the information-theoretic content of the two questions may be the same, it should easy to imagine that if more complex fields are considered, full process characterization contains a lot more information than needed to optimally adapt the local measurements. 
The approaches of \cite{2014_Tiersch} can further be generalized to utilize information from stabilizer measurements \cite{2016_Orsucci}, or similarly outcomes of syndrome measurements when codes are utilized \cite{2014_Combes}, (instead of probe states) to similar ends. Addressing somewhat related problems, but using supervised learning methods, the authors in \cite{2017_Mavadia} have shown how to compensate for qubit decoherence (stochastic evolution) also in experiments .  
 
 \subsubsection{Learning how to experiment}
 
One of the first examples of applications of RL in QIP appears in the context of experimental photonics, where one of the current challenges lies in the generation of highly entangled, high dimensional, multi-party states. Such states are generated on optical tables, the configuration of which, to generate complex quantum states, can be counter-intuitive and unsystematic. The searching for configurations which are interesting can be mapped to a RL problem, where a learning agent is rewarded whenever it generates an interesting state (in a simulation). In a precursor work \cite{2016_Krenn}, the authors used a feedback-assisted search algorithm to identify previously unknown configurations which generate novel highly entangled states. This demonstrated that the design of novel quantum experiments can also be automatized, which can significantly aid in research.
This idea given in the context of optical tables, has subsequently been combined with earlier proposals to employ AI agents in quantum information protocols and as "lab robots" in future quantum laboratories \cite{Briegel}. 
This led to the application of more advanced RL techniques, based on the PS framework, for the tasks of understanding the Hilbert space accessible with optical tables, and the autonomous machine-discovery of useful optical gadgets \cite{2017_Melnikov}. 
Related to the topic of learning new insight from experimenting machines, in \cite{2017_Bukov} the authors consider the problem of preparing target states by means of chosen pulses implementing (a restricted set) of rotations. This is a standard control task, and authors show that RL achieves respectable and sometimes near-optimal results. However, for our purposes, the most relevant aspects of this work pertain to the fact that the authors also illustrate how of ML/RL techniques can be used to obtain new insights in quantum experiments, and non-equilibrium physics, by circumventing human intuition which can be flawed.
Interestingly, the authors also demonstrate the reverse, i.e. how physics insights can help elucidate learning problems\footnote{For instance, the authors investigate the strategies explored by the learning agent, and identify spin-glass like phase transition in the space of protocols as a function of the protocol duration. This highlights the difficulty of the learning problem. }.

\subsection{Machine learning in condensed-matter and many-body physics}
\label{QMB}
\boxTLDR{One of the quintessential problems of many-body physics is the \textbf{identification of phases of matter}. A popular overlap between ML and this branch of physics demonstrates that supervised and unsupervised systems can be trained to classify different phases. More interestingly, unsupervised learning can be used to \textbf{detect phases, and even discover order parameters} -- possibly genuinely leading to novel physical insights. Another important overlap considers the \textbf{representational power of (generalized) neural networks}, to characterize \textbf{interesting families of quantum systems}. Both suggest a \textbf{deeper link} between certain learning models, on the one side, and physical systems, on the other side, the scope of which is currently an important research topic.}

ML techniques have, over the course of last 20 years, become an indispensable toolset of many natural sciences which deal with highly complex systems. These include biology (specifically genetics, genomics, proteomics, and the general field of computational biology) \cite{2015_Libbrecht}, medicine (e.g. in epidemiology, disease development, etc.) \cite{2015_Cleophas}, chemistry \cite{2007_Cartwright}, high energy and particle physics \cite{2015_Castelvecchi}. Unsurprisingly, they have also permeated various aspects of condensed matter and many-body physics. Early examples of this were proposed in the context of quantum chemistry and density functional theory \cite{2003_Curtarolo,2012_Snyder, 2012_Rupp, 2015_Li_DeVita}, or for the approximation of the Green's function of the single-site Anderson impurity
model \cite{2014_Arsenault}. The interest in connections between NNs and many-body and condensed matter physics has undergone immense growth since. 
Some of the results which we cover next deviate from the primary topic of this review, those concerning the overlaps of QIP and ML. However, since QIP, condensed matter, and many-body physics share significant overlaps we feel it is important to at least briefly flesh out the basic ideas. 

One of the basic lines of research in this area deals with the learning of phases of matter, and the detection of phase transitions in physical systems. A canonical example is the discrimination of samples of configurations stemming from different phases of matter, e.g. Ising model configurations of thermal states below, or above the critical temperature. This problem has been tackled using principal component analysis and nearest neighbour unsupervised learning techniques \cite{2016_Wang} (see also \cite{2017_Wenjian}). 
Such methods also have the potential to, beyond just detecting phases, actually identify order parameters \cite{2016_Wang} -- in the above case, magnetization.
More complicated discrimination problems, e.g. discriminating Coulomb phases, have been resolved using basic feed-forward networks, and convolutional NNs were trained to detect topological phases, \cite{2017_Carrasquilla_Melko}, but also phases in fermionic systems on cubic lattices \cite{2016_Chng_Khatami}. Neural networks have also been combined with quantum Monte Carlo methods \cite{2016_Broecker}, and with unsupervised methods \cite{2017_vanNieuwenburg} (applied also in \cite{2016_Wang}), in both cases to improve classification performance in various systems. It is notable that all these methods prove quite successful in ``learning'' phases, without any information of the system Hamiltonian.
While the focus in this field had mostly been on neural network architectures, other supervised methods, specifically kernel methods (e.g. SVMs) had been used for the same purpose \cite{2017_Ponte}. Kernel methods may be in some cases advantageous as they can have a higher interpretability: it is often easier to understand the reason behind the optimal model in the cases of kernel methods, rather than NNs, which also means that learning about the underlying physics may be easier in the cases of kernel methods. Note that this will most likely be challenged by deep NN approaches in years to come.

A partial explanation behind the success of neuronal approaches for classifying phases of matter may lie in their form. Specifically, they may have the capacity to encode important properties of physical systems both in the classical in quantum case. This motivates the second line of research we mention in this context. 
BMs, even in their restricted variant, are known to have the capacity to encode complicated distributions. In the same sense, restricted BMs, extended to accept complex weights (i.e. the weights $w_{ij}$ in   Eqs. (\ref{ising}) and  (\ref{boltz})) encode quantum states, and the hidden layer captures correlations, both classical and quantum (entanglement). In \cite{2017_Carleo_Troyer} it was shown that  this approach describes equilibrium and dynamical properties of many prototypical systems accurately: that is, restricted BMs form a useful ansatz for interesting quantum states (called \textit{neural-network quantum states} (NQS)), where the number of neurons in the hidden layer controls the size of the representable subset of the Hilbert space. {This is analogous to how, for instance, the bond dimension controls the scope of the matrix product state ansatz \cite{2008_Verstraete}}. This property can also be exploited in order to achieve efficient quantum state tomography\footnote{This method can be thought of as effectively by assigning a prior stating that the analyzed state is well approximated by a NQS.}\cite{2017_Torlai_Troyer_Carleo}. In subsequent works, the authors have also analyzed the structure of entanglement of NQS states \cite{2017_Deng}, and have provided analytic proofs of the representation power of deep restricted BMs, proving they can e.g. represent ground states of any k-local Hamiltonians with polynomial-size gaps \cite{2017_Gao}. 
It is worthwhile to note that representational powers of standard variational representations (e.g. that of the variational renormalization group) had previously been contrasted to those of deep NNs \cite{2014_Mehta}, with the goal of elucidating the success of deep networks. Related to 
this, the Tensor Network \cite{1995_Ostlund,2004_Verstraete} formalism has been used for the efficient description of deep convolutional arithmetic circuits, establishing also a formal connection between quantum many-body states and deep learning \cite{2017_Levine}.
Very recently, the intersection between ML and many-body quantum physics have also inspired research into ML-motivated entanglement witnesses and classifiers \cite{2017_Yue-Chi, 2017_Lu2}, and also into furthering the connections between ML and many-body physics, specifically, entanglement theory. These recent results have positioned NNs as one of the most exciting new techniques to be applied in the context of both condensed-matter and many-body physics. Additionally, they also show the potential of the converse direction of influence -- the application of mathematical formalism of many-body physics for the deepening of our understanding of complex learning models.

\section{Quantum generalizations of machine learning concepts}
\label{QgenML}

The onset of quantum theory necessitated a change in how we describe physical systems, but also a change in our understanding of what information is\footnote{Arguably, in the light of the physicalistic viewpoint on the nature of information, which posits that ``Information is [ultimately] physical''.}.
Quantum information is a more general concept, and QIP exploits the genuine quantum features
for more efficient processing (using quantum computers) and more efficient communication. Such quintessential quantum properties, such as the fact that even pure states cannot be perfectly copied \cite{1982_Wootters}, are often argued to be at the heart of many quantum applications, such as cryptography. 
Similarly, quintessential information processing operations are more general in the quantum world: closed quantum systems can undergo arbitrary unitary evolutions, whereas the corresponding classical closed-system evolutions correspond to the (finite) group of permutations\footnote{Classical evolutions are guaranteed to transform computational basis states (the ``classical states'') to computational basis states, and closed-system implies the dynamics must be reversible, leaving only permutations.}. 
The majority of  ML literature deals with learning from, and about data -- that is, classical information. This section examines the question of what ML looks like, when the data (and perhaps its processing) is fundamentally quantum.
We will first explore quantum generalizations of supervised learning, where the ``data-points'' are now genuine quantum states. This generates a plethora of scenarios which are indistinguishable in the classical case (e.g. having one or two copies of the same example is not the same!). Next, we will consider another quantum generalization of learning, where quantum states are used to represent the generalizations of unknown concepts in CLT -- thus we talk about the \textbf{learning of quantum states}. Following this we will present some results on quantum generalizations of POMDP's which could lead to quantum-generalized reinforcement learning (although this actually just generalizes the mathematical structure).

\subsection{Quantum generalizations: machine learning of quantum data}

\boxTLDR{A significant fraction of the field of ML deals with data analysis, classification, clustering, etc. QIP generalizes standard notions of data, to include quantum states. The processing of quantum information comes with restrictions (e.g. no-cloning or no-deleting), but also new processing options. This section addresses the question of \textbf{how conventional ML concepts can be extended to the quantum domain}, mostly focusing on  aspects of supervised learning and learnability of quantum systems, but also concepts underlying RL. }

One of the basic problems of ML is that of supervised learning, where a training set $D = \{ (\mathbf{x}_i, y_i)\}_{i}$  is used to infer a labeling rule mapping data points to labels $\mathbf{x}_i \stackrel{rule}{\rightarrow} y_i$ (see section \ref{AIMLC} for more details). More generally, supervised learning deals with classification of classical data. In the tradition of QIP, data can also be quantum -- that is, all quantum states carry, or rather represent, (quantum) information. What can be done with datasets of the type 
$\{ (\rho_i, y_i)\}_{i}$,§§ where $\rho_i$ is a quantum state?
 Colloquially it is often said that one of the critical distinction between classical and quantum data is that quantum data cannot be copied. In other words, having one instance of an example, by notation abuse denoted $(\rho_i \otimes y_i)$, is not generally as useful as having two copies $(\rho_i \otimes y_i)^{\otimes 2}.$ In contrast in the case of classification with functional labeling rules, this is the same.
The closest classical analog of dealing with quantum data is the case where labelings are not deterministic, or equivalently, where the conditional distribution $P(\textup{label} | \textup{datapoint})$ is not extremal (Dirac). This is the case of classification (or learning) of random variables, or probabilistic concepts, where the task is to produce the best guess label, specifying the random process which ``most likely'' produced the datapoint\footnote{Note that in this setting we do not have the descriptions of the stochastic processes given a-priory -- they are to be inferred from the training examples. }. In this case, having access to two examples in the training phase which are independently sampled from the same distribution is not the same as having two copies of one and the same individual sample--these are perfectly correlated and carry no new information\footnote{In this sense, no-cloning theorem also applies to classical information: an unknown random variable cannot be cloned. In QIP language this simply means that no-cloning theorem applies to diagonal density matrices, i.e. $\rho \not \rightarrow \rho\otimes\rho$, even when $\rho$ is promised to be diagonal.}. 
To obtain full information about a distribution, or random variable, one in principle needs infinitely many samples. Similarly, in the quantum case, having infinitely many copies of the same quantum state $\rho$ is operatively equivalent to having a classical description of the given state. 

Despite similarities, quantum information is still different from mere stochastic data. The precursors of ML-type classification tasks can be identified in the theories of quantum state discrimination, which we briefly comment on first. Next, we review some early works dealing with ``quantum pattern matching'' which spans various generalizations of supervised settings, and first works which explicitly propose the study of \textit{quantum-generalized machine learning}. Next, we discuss more general results, which characterize inductive learning in quantum settings.
Finally, we present a CLT perspective on learning with quantum data, which addresses the learnability of quantum states. 
\subsubsection{State discrimination, state classification, and machine learning of quantum data}
\label{QGML}
\paragraph{State discrimination}
The entry point to this topic can again be traced to seminal works of Helstrom and Holevo \cite{1969_Helstrom,1982_Holevo} as the problems of state discrimination can be rephrased as variants of supervised learning problems. In typical state discrimination settings, the task is the identifying of a given quantum state (given as an instance of a quantum system prepared in that state), under the promise that it belongs to a (typically finite) set $\{ \rho_i\}_i$, where the set is fully classically specified. Recall, state estimation, in contrast, typically assumes continuous parametrized families, and the task is the estimation of the parameter. In this sense, discrimination is a discretized estimation problem\footnote{Intuitively, estimation is to discrimination, what regression is to classification in the ML world.}, and the problems of identifying optimal measurements (under various figures of merit), and success bounds have been considered extensively and continuously throughout the history of QIP \cite{1969_Helstrom,2008_Croke,2017_Slussarenko}.

\underline{\textbf{Remark:}} Traditional quantum state discrimination can be rephrased as degenerate supervised learning setting for quantum states. Here, the space of ``data-points'' is restricted to a finite (or parametrized) family $\{ \rho_i\}_i$, and the training set contains an effective infinite number of examples $D = \{  (\rho_i, i)^{\otimes \infty} \}$; naturally, this notation is just a short-hand for having the complete classical description of the quantum states \footnote{From an operative, and information content perspective, having infinitely many copies is equivalent to having a full classical description: infinite copies are sufficient and necessary for perfect tomography -- yielding the exact classical description -- whereas having an exact classical description is sufficient and necessary for generating an unbounded copy number.  }. In what follows we will sometimes write $\rho^{\otimes \infty}$ to denote a quantum system containing the classical description of the density matrix $\rho$.

\paragraph{Quantum template matching -- classical templates}
A variant of discrimination, or class assignment task, which is one of the first instances of works which establish explicit connections with ML and discrimination-type problems, is ``template matching'' \cite{2001_Sasaki_Jozsa}.
In this pioneering work, the authors consider discrimination problems where the input states $\psi$ may not correspond to the (known) \textit{template states} $\{ \rho_i\}_i$, and the correct matching label is determined by the largest the Uhlmann fidelity. More precisely, the task is defined as follows: given a classically specified family of template states $\{ \rho_i\}_i$, given $M$ copies of a quantum input $\psi^{\otimes M}$, output the label $i_{corr}$ defined with $i_{corr} = \textup{argmax}_{i} \textup{Tr} \left [ \sqrt{ \sqrt{\psi} \rho_i \sqrt{\psi} }\right]^{2}.$
In this original work, the authors focused on two-class cases, with pure state inputs, and identify fully quantum, and semi-classical strategies for this problem. ``Fully quantum strategies'' identify the optimal POVM. Semi-classical strategies impose a restriction of measurement strategies to separable measurements, or perform state estimation on the input, a type of ``quantum feature extraction''.

\paragraph{Quantum template matching -- quantum templates.}

In a generalization of the work in \cite{2001_Sasaki_Jozsa}, the authors in \cite{2002_Sasaki_Carlini} consider the case where instead of having access to the classical descriptions of the template states 
$\{ \rho_i\}_i$, we are given access to a certain number $K$ of copies.
In other words, we are given access to a quantum system in the state $\bigotimes_{i} \rho_i^{\otimes K}.$.
Setting $K \rightarrow \infty,$ recovers the case with classical templates.
This generalized setting introduces many complications, which do not exist in the ``more classical'' case with classical templates. For instance, classifying measurements now must ``use up'' copies of template states, as they too cannot be cloned. 
The authors identify various flavors of semi-classical strategies for this problem. For instance, if the template states are first estimated, we are facing the scenario of classical templates (albeit with error). The classical template setting itself allows semiclassical strategies, where all systems are first estimated, and it allows coherent strategies. 
The authors find optimal solutions for $K=1,$ and show that there  exists a fully quantum  procedure that is strictly superior to  straightforward semiclassical extensions.

\underline{\textbf{Remark:}} Quantum template matching problems can be understood as quantum-generalized supervised learning, where the training set is of the form $\{ (\rho^{\otimes K}_i, i)_i \},$ data beyond the training set comes from the family $ \left\{ \psi^{\otimes M} \right\}$ (number of copies is known), and the classes are defined via minimal distance, as  measured by the Uhlmann fidelity. The case $K \rightarrow \infty$ approaches the special case of classical templates. Restricting the states $\psi$ to the set of template states (\textbf{restricted template matching}), and setting $M=1$ recovers standard state discrimination.
 
\paragraph{Other known optimality results for (restricted) template matching}

For the restricted matching case, where the input is promised to be from the template set, the optimal solutions for the two-class setting, minimum error figure of merit, and uniform priors of inputs, have been found in \cite{2005_Bergou, 2005_Hayashi} for the qubit case. In  \cite{2006_Hayashi} the authors found optimal solutions for the unambiguous discrimination case\footnote{In unambiguous discrimination, the device is allowed to output an ambiguous ``I do not know'' outcome, but is not allowed to err in the case it does output an outcome. The goal is to minimize the probability of the ambiguous outcome.}.  An asymptotically optimal strategy restricted matching with finite templates $K<\infty,$ for arbitrary priors, and mixed qubit states was later found in \cite{2010_Guta}. This work also provides a solid introduction to the topic, a review of quantum analogies for statistical learning, and emphasizes connections to ML methodologies and concepts. 

Later, in \cite{2012_Sentis} the authors introduced and compared all three strategies: classical estimate-and-discriminate, classical optimal, and quantum strategy, for the restricted template matching case with finite templates. Recall, the adjective ``classical'' here denotes that the training states are fully measured out as the first step -- the quantum set is converted to \textit{classical information}, meaning that no quantum memory is further required, and that the learning can be truly inductive.
 A surprising result is that the intuitive estimate-and-discriminate strategy, which reduces supervised classification to optimal estimation coupled with a (standard) quantum state discrimination problem, is not optimal for learning.  Another measurement provides not only better performance, but matches the optimal quantum strategy exactly (as opposed to asymptotically). Interestingly, the results of \cite{2010_Guta} and \cite{2012_Sentis} opposite claims for \textit{essentially} the same setting:  no separation, vs. separation between coherent (fully quantum) and semi-classical strategies, respectively.  This discrepancy is caused by differences in the chosen figures of merit, and a different definition of asymptotic optimality \cite{2017_Sentis}, and serves as an effective reminder of the subtle nature of quantum learning. Optimal strategies had been subsequently explored in other settings as well, e.g. when the data-set comprises coherent states  \cite{2015_Sentis}, and or in the cases where an error margin is in an otherwise unambiguous setting \cite{2013_Sentis}.

\paragraph{Quantum generalizations of (un)supervised learning}
The works of the previous paragraph consider particular families of generalizations of supervised learning problems. The first attempts to classify and characterize \textbf{what ML could look like in a quantum world} from a more general perspective was, however, first explicitly done in \cite{2006_Aimeur_Gambs}. 
There, the basic object introduced is the database of labeled quantum or classical objects,  i.e.  $D_n^{K} = \{ (\ket{\psi_i}^{\otimes _i}, y_i)\}_{i=1}^{n}$\footnote{Such a dataset can be stored in, or instantiated by, a 2-$n$ partite quantum system, prepared in the state $\bigotimes_{i=1}^{n} \ket{\psi_i}^{\otimes K_i}\ket{y_i}$.}, which may come in copies. Such a database can, in general then be processed to solve various types of tasks, using classical or quantum processing.
The authors propose to characterize quantum learning scenarios in terms of classes,  denoted  $L^{context}_{goal}.$ Here  \textit{context} may denote we are dealing with classical or quantum data and whether the learning algorithm is relying on quantum capabilities or not. The \textit{goal} specifies the learning task or goal (perhaps in very broad terms). Examples include $L^{c}_c,$ which corresponds to standard classical ML, and $L^{q}_c,$ which could mean we use a quantum computer to analyze classical data. The example of template matching classical templates ($K=\infty$)  \cite{2001_Sasaki_Jozsa}  considered earlier in this section would be denoted $L^{c}_q,$ and the generalization with finite template numbers $K<\infty$ would fit in $L^{\otimes K}_q.$
While the formalism above suggests focus on supervised settings, the authors also suggest that datasets could be inputs for (unsupervised) clustering.  The authors further study quantum algorithms for determining closeness of quantum states\footnote{These are based on the SWAP-test (see section \ref{qcircuit}), in terms of Uhlmann fidelity}, which could be the basic building block of quantum clustering algorithms, and also compute certain error bounds for special cases of classification (state discrimination) using well known results of Helstrom \cite{1969_Helstrom}. 
Similar ideas were used in \cite{2014_Lu} for the purpose of definition of a quantum decision tree algorithm for data classification in the quantum regime.

The strong connection between quantum-generalized learning theory sketched out in \cite{2006_Aimeur_Gambs} and the classical\footnote{Here we mean classical in the sense of ``being a classic'', rather than pertaining to classical systems.} theory of Helstrom \cite{1969_Helstrom} was more deeply explored in \cite{2008_Gambs}. There, the author computed the lower bounds of sample complexity -- in this case the minimal number of copies $K$ -- needed to solve a few types of classification problems. For this purpose the author introduced a few techniques which reduce ML-type classification problems to the settings where theory \cite{1969_Helstrom}  of could be directly applied. 
These types of results contribute to the establishing of a deeper connection between problems of ML and techniques of QIP.

\paragraph{Quantum inductive learning}
Recall that inductive, eager learning, produces a best guess classifier which can be applied to the entire domain of data-points, based on the training set. But, already the results of 
\cite{2002_Sasaki_Carlini} discussed in paragraph on template matching with quantum templates, point to problems with this concept in the quantum realm -- the optimal classifier may require a copy of the quantum data-points to perform classification, which seemingly prohibits unlimited use.
The perspectives of such quantum generalizations of supervised learning in its inductive form, were recently addressed from a broad perspective \cite{2017_Monras}.
Recall that inductive learning algorithms, intuitively, use only the training set to specify a hypothesis (the estimation of the true labeling function). In contrast, in \textit{transductive} learning, the learner is also given the data points the labels of which are unknown.
These unlabeled points may correspond to the cross-validation test set, or the actual target data.
Even though the labels are unknown, they carry additional information of the complete dataset  which can be helpful in identifying the correct labeling rule \footnote{For instance, a transductive algorithm may use unsupervised clustering techniques to assign labels, as the whole set is given in advance.}. Another distinction is that transductive algorithms need only label the given points, whereas inductive algorithms need to specify a classifier, i.e., a labeling function, defined on the entire space of possible points.
In  \cite{2017_Monras}, the authors notice that the property of an algorithm being inductive corresponds to a non-signaling property\footnote{The outcome of the entire learning and evaluation process can be viewed as a probability distribution $P(\mathbf{y})=P(y_1 \ldots y_k | x_1 \ldots x_k ; A )$, where $A$ is the training set, $x_1, \ldots x_k$ are the points of the test state and $y_1 \ldots y_k$ the respective labels the algorithm assigns with the probability $P(\mathbf{y})$. No signaling implies that the marginal distribution for the $k^{th}$ test element $P(y_k)$ only depends on $x_k$ and the training set, but not on other test points $\{ x_l\}_{l\not=k}.$ }, using which they can prove that ``being inductive'' (i.e. being ``no signalling'') is equivalent to having an algorithm which outputs a classifier $h$ based on the training set alone, which is then applied to every training instance. A third equivalent characterization of inductive learning is that the training and testing cleanly separate as phases.
While these observations are quite intuitive in the classical case, they are in fact problematic in the quantum world. Specifically, if the training examples are quantum objects, quantum no-cloning, in general, prohibits the applying of a hypothesis function (candidate labeling function) $h$ arbitrarily many times.
This is easy to see since each instance of $h$ must depend on the quantum data in \textit{some non-trivial way}, if we are dealing with a learning algorithm. Multiple copies of $h$ would then require multiple copies of (at least parts of) the quantum data.

A possible implication of this would be that, in the quantum realm, inductive learning cannot be cleanly separated into training and testing. Nonetheless, the authors show that the no-signalling criterion, for certain symmetric measures of performance, implies that a separation is, asymptotically, possible.
Specifically, the authors show that for any quantum inductive no-signalling algorithm $A$ there exists another,  perhaps different algorithm $A'$ which does separate in a training and testing phase and which, asymptotically, attains the same performance \cite{2017_Monras}. 
Such a protocol $A'$, essentially, utilizes a semi-classical strategy.
In other words, for inductive settings, classical intuition survives, despite no-cloning theorems.

\subsubsection{Computational learning perspectives: quantum states as concepts}
\label{quantum-concepts}
The previous subsections addressed the topics of classification of quantum states, based on quantum database examples. The overall theory, however, relies on the assumption that there exists a labeling rule, which generates such examples, and \textbf{what is learned is the labeling rule}. 
This rule is also known as concept, in CLT (e.g. \textit{PAC} learning, see section \ref{CPAC} for details). A reasonable sufficient criterion is, if one can predict the probabilities of outcomes of any two-outcome measurements on this state, as this already suffices for a full tomographic reconstruction.
What would ``the learning of quantum states'' mean, from this perspective? What does it mean to ``know a quantum state''? A natural criterion is that one ``knows'' a quantum state, if one can predict the measurement outcome probabilities of any given measurement. 
In \cite{2007_Aaronson}, the author addressed the question of the learnability of quantum states in the sense above, where the role of a concept is played by a given quantum state,
and ``knowing'' the concept then equates to the possibility of predicting the outcome probability of a given measurement and its outcome.
One immediate distinction from conventional CLT, discussed in \ref{CPAC},  is that the concept range is  no longer binary. However, as as we clarified, classical CLT theory has generalizations with continuous ranges. In particular, so called \textit{p-concepts} have range in $[0,1]$ \cite{1994_Kearns}, and  quantities which are analogs of the VC-dimension, and analogous theorems relating this to generalization performance, exist for the $p$-concept case as well (see \cite{2007_Aaronson}).
Explicitly, the basic elements of such the generalized theory are: domain of concepts $\mathbf{X},$ a sample $\mathbf{x} \in \mathbf{X}$ and the p-concept $f: \mathbf{X} \rightarrow [0,1].$
These abstract objects are mapped to central objects of quantum information theory \cite{2007_Aaronson} as follows:
the domain of concepts is the set of two-outcome quantum measurement, and a sample is a POVM element $\Pi$\footnote{More precisely $\Pi $ is a positive-semidefinite operator such that $\mathbbmss{1} - \Pi$ is positive-semidefinite as well.} (in short: $\mathbf{x} \leftrightarrow \Pi$); the $p$-concept to be learned is a quantum state $\psi$
and the evaluation of the concept/hypothesis on the sample corresponds to the probability $\textup{Tr}[\Pi \psi]\in [0,1]$ of observing the measurement outcome associated with $\Pi$ when the state $\psi$ is measured. 

To connect the data classification-based perspectives of supervised learning to the CLT perspective above, note that in the given quantum state CLT this framework, the quantum concept -- quantum state -- ``classifies'' quantum POVM elements (the \textit{effects}) according to the probability of observing that effect. 
The training set elements for this model are of the form $(\Pi, \textup{Tr}(\rho \Pi)),$ with $0 \leq \Pi \leq \mathbbmss{1}$.

In the spirit of CLT, the concept class  ``quantum states'', is said to be learnable under some distribution $\mathcal{D}$ over two-outcome generalized measurement elements $(\Pi)$, if for every concept --quantum state $\rho$ -- there exists an algorithm with access to examples of the form $(\Pi, \textup{Tr}(\rho \Pi))$, where $\Pi$ is drawn according to $\mathcal{D}$, which outputs a hypothesis $h$ which (approximately) correctly predicts the label $ \textup{Tr}(\rho \Pi')$ with high probability, when $\Pi'$ is drawn from $\mathcal{D}$. Note that the role of a hypothesis here can simply be played by a ``best guess'' \textit{classical description} of the \textbf{quantum state} $\rho$.
The key result of \cite{2007_Aaronson} is that quantum states are learnable with sample complexity scaling only \textit{linearly in the number of qubits}\footnote{The dependencies on the allowed inverse error and inverse allowed failure probability are polynomial and polylogarithmic, respectively.}, that is logarithmically in the dimension of the density matrix. 
In operative terms, if Alice wishes to send an $n$ qubit quantum state to Bob who will perform on it a two-outcome measurement (and Alice does not know which), she can achieve near-ideal performance by sending ($O(n)$) \textit{classical} bits\footnote{Here we assume Alice can locally generate her states at will. A classical strategy (using classical channels) is thus always possible, by having Alice send the outcomes of full state tomography (or equiv. the classical description of the state), but this requires the using of $O(2^{n})$ bits already for pure states.}, which has clear practical but also theoretical importance. 
In some sense, these results can also be thought of as a generalized variant of Holevo bound theorems \cite{1982_Holevo}, limiting how much information can be stored and retrieved in the case of quantum systems.
This latter result has thus far been more influential in the contexts of tomography than quantum machine learning, despite being quite a fundamental result in quantum learning theory.
However, for fully practical purposes. The results above come with a caveat. The learning of quantum state is efficient in \textbf{sample complexity} (e.g. number of measurements one needs to perform), however, the \textbf{computational complexity} of the reconstruction of the hypothesis is, in fact, likely exponential in the qubit number. 
Very recently, the efficiency of also the reconstruction algorithms for the learning of stabilizer states  was shown in \cite{2017_Rocchetto}.

\subsection{(Quantum) learning and quantum processes }
\boxTLDR{The notion of quantum learning has been used in literature to refer to the studying of various aspects of ``learning about'' quantum systems. Beyond the learning of quantum states, one can also consider the \textbf{learning of quantum evolutions}. Here ``knowing'' is operatively defined as having the capacity to implement the given unitary at a later point -- this is similar to how ``knowning'' in computational learning theory implies we can apply the concept function at a later point.
Finally, as learning can pertain to learning in interactive environments -- RL -- one can consider the quantum generalizations of such settings. One of the first results in this direction formulates a \textbf{quantum generalization of POMDPs}. Note as POMDPs form the mathematical basis of RL, the quantum-generalized mathematical object -- quantum POMDP, may form a basis of quantum-generalized RL.}

\paragraph{Learning of quantum processes}

The concept of learning is quite diffuse and ``quantum learning'' has been used in literature quite often, and not every instance corresponds to generalizations of ``classical learning'' in a machine or statistical learning sense. Nonetheless, some such works further illustrate the distinctions between the approaches one can employ with access to classical (quantum) tools, while learning about classical or quantum objects. 
\vspace{0.2cm}\\
\underline{{Learning unitaries}}\ 
For instance ``quantum learning of unitary operations" has been used to refer to the task of optimal storing and retrieval of unknown unitary operations, which is a two stage process. In the storing phase, one is given access to a few uses of some unitary $U$. In the retrieval phase, one is asked to approximate the state $U \ket{\psi},$ given one or few instances of a (previously fully unknown) state $ \ket{\psi}$. Like in the case of quantum template states (see section \ref{QGML}), we can distinguish semi-classical prepare-and-measure strategies (where $U$ is estimated and represented as classical information), and quantum strategies, where the unitaries are applied on some resource state, which is used together with the input state $ \ket{\psi}$ in the retrieval stage. There is no simple universal answer to the question of optimal strategies. In \cite{2010_Bisio}, the authors have shown that, under reasonable assumptions, the surprising result that optimal strategies are semi-classical. In contrast, in \cite{2011_Bisio} the same question was asked for generalized measurements, and the opposite was shown: optimal strategies require quantum memory. See e.g. \cite{2017_Sedlak} for some recent results on probabilistic unitary storage and retrieval, which can be understood as genuinely quantum learning \footnote{Quantum in that that which is learned is encoded in a quantum state.} of quantum operations.
\vspace{0.2cm}\\
\underline{{Learning measurements}}\ 
The problem of identifying which measurement apparatus one is facing has first been in comparatively fewer works, see e.g. \cite{2014_Sedlak} for a more recent example.
Related to this, we encounter a more learning-theoretical perspective on the topic of learning measurements. In the comprehensive paper \cite{2016_Cheng} (which can serve as a review of parts of quantum ML in its own right), the authors explore the question of the learnability of quantum measurements. This can be thought of as the dual of the task of learning quantum states discussed previously in this section. Here, the examples are of the form $(\rho, Tr(\rho E)),$ and it is the measurement that is fixed. In this work, the authors compute a number of complexity measures, which are closely related to the VC dimension (see section \ref{CPAC}), for which sample complexity bounds are known.
From such complexity bounds one can, for instance, rigorously answer various relevant operative questions, such as, how many random quantum probe states we need to prepare on average, to accurately estimate a quantum measurement.
Complementing the standard estimation problems, here we do not compute the optimal strategy, but effectively gauge the information gain of a randomized strategy.
 These measures are computed for the family of hypotheses/concepts which can be obtained by either fixing the POVM element (thus learning the quantum measurement), or by fixing the state (which is the setting of \cite{2007_Aaronson}), and clearly illustrate the power of ML theory when applied in QIP context.

\paragraph{Foundations of quantum-generalized RL}

The majority of quantum generalizations of machine learning concepts fit neatly in the domain of supervised learning, however, with few notable exceptions. In particular, in \cite{2014_Barry_Aaronson}, the authors introduce a quantum generalization of partially observable Markov decision processes (POMDP), discussed in section \ref{RL-section}.  For convenience of the reader we give a brief recap of these objects. A fully observable MDP is a formalization of task environments: the environment can be in any number of states $\mathcal{S}$ the agent can observe. An action $a\in\mathcal{A}$ of the agent triggers a transition of the state of the environment -- the transition can be stochastic, and is specified by a Markov transition matrix $P^a$.\footnote{In other words, for any environment state $s$, producing an action $a$ causes a transition to some state $s'$ with probability $\vec{s'}^{\tau}  P^{a} \vec{s}$, where states are represented as canonical vectors.} Additionally, beyond the dynamics, each MDP comes with a reward function $R:\mathcal{S}\times \mathcal{A}\times\mathcal{S} \rightarrow \Lambda$, which rewards certain state-action-state transitions. In POMDP, the agent does not see the actual state of the environment, but rather just \textit{observations} $o\in\mathcal{O}$, which are (stochastic) functions of the environmental state\footnote{In general, the observations output can also depend on the previous action of the agent.}. 
Although the exact environmental state of the environment is not directly accessible to the agent, given the full specification of the system, the agent can still assign a probability distribution over the state space given an interaction history. This is called \textbf{a belief state}, and, can be represented as a mixed state (mixing the ``classical" actual environmental states), which is diagonal in the POMDP state basis. The quantum generalization promotes the environment belief state to any quantum state defined on the Hilbert space spanned by the orthonormal basis $\{ \ket{s} | s\in \mathcal{S}\}$. The dynamic of the quantum POMDP are defined by actions which correspond to quantum instruments (superoperators) the agent can apply: to each action $a$, we associate the set of Krauss operators $\{ K^a_o \}_{o \in \mathcal{O}},$ which satisfy $\sum_{o} {K^a}_o^\dagger K_{o}^a = \mathbbmss{1}$. If the agent performs the action $a$, and observes the observation $o$, the state of the environment is mapped as $\rho \rightarrow  K_o^a \rho {K_o^a}^{\dagger}/Tr[K_o^a \rho {K_o^a}^{\dagger}],$ where $Tr[K_o^a \rho {K_o^a}^{\dagger}]$ is the probability of observing that outcome. Finally, rewards are defined via the expected values of action-specific positive operators $R_a$, so $Tr[R_a \rho],$ given the state $\rho$.
In \cite{2014_Barry_Aaronson}, the authors have studied this model from the computational perspective of the hardness of identifying the best strategies for the agent, contrasting this setting to classical settings, and proving separations. In particular, the complexity of deciding policy existence for finite horizons\footnote{That is, given a full specification of the setting, decide whether there exist a policy for the agent which achieves a cumulative reward above some value, in a certain number of states.}, are the same for the quantum and classical cases\footnote{This decision problem is undecidable in the infinite horizon case, already for the classical problem, and thus trivially undecidable in the quantum case as well.}.
However, a separation can be found with respect to the goal reachability problem, which asks whether there exists a policy (of any length) which, with probability 1, reaches some target state. This separation is maximal -- this problem is decidable in the classical case, yet undecidable in the quantum case. While this particular separation may not have immediate consequences for quantum learning, it suggests that there may be other (dramatic) separations, with more immediate relevance.

\section{Quantum enhancements for machine learning}
\label{QeML}
One of the most advertised aspects of quantum ML deals with the question of whether quantum effects can help us solve classical learning tasks more efficiently, ideally mirroring the successes of quantum computation.
The very first attempts to apply quantum information techniques to ML problems were made even before the seminal works of Shor and Grover \cite{1997_Shor,1996_Grover}. Notable examples include the pioneering research into quantum neural networks and quantum perceptrons \cite{1994_Lewenstein,1995_Kak}, and also in the potential of quantum computational learning theory \cite{1998_Bshouty}. The topic of quantum neural networks (quantum NNs) has had sustained growth and development since these early days, exploring various types of questions regarding the interplay of quantum mechanics and neural networks. Most of the research in this area is not directly targeted at algorithmic improvements, hence will be only briefly mentioned here. 
A fraction of the research into quantum NNs, which was disproportionately more active in the early days, considered the speculative topics of the function of quantum effects in neural networks, both artificial and biological \cite{1989_Penrose,1995_Kak}. Parts of this research line has focused concrete models, such as the effect of transverse fields in HNs  \cite{1996_Nishimori}, and decoherence in models of biological nets \cite{2000_Tegmark}, which, it is argued, would destroy any potential quantum effect.
A second topic which permeates the research in quantum NNs is concerned with the fundamental question of a meaningful quantization of standard feed-forward neural networks. The key question here is finding the best way to reconcile the linear nature of quantum theory,  and the necessity for non-linearities in the activation function of a neural network (see section \ref{ANNs}), and identifying suitable physical systems to implement such a scheme. Early ideas here included giving up on non-linearities \textit{per se}, and considering networks of unitaries which substitute layers of neurons \cite{1994_Lewenstein}. Another approach exploits non-linearities which stem from measurements and post-selection (arguably first suggested in \cite{1995_Kak}).
The same issue is addressed by Behrman et al. \cite{1996_Behrman} by using a continuous mechanical system where the non-linearity is achieved by coupling the system with an environment \footnote{Similar ideas were also discussed by Peru\v{s} in \cite{2000_Perus}.}, in the model system of quantum dots. The purely foundational research into implementations of such networks, and analysis of their quantum mechanical features, has been and is continuing to be an active field of research (see e.g. \cite{2017_Altaisky}). For more information on this topic we refer the reader to more specialized reviews \cite{2014_Schuld,2011_Garman}.

Unlike the research into quantum NNs, which has a foundational flavor, majority of works studying quantum effects for classical ML problems are specifically focused on identifying improvements.. First examples of quantum advantages in this context were provided in the context of \textbf{quantum computational learning theory}, which is the topic of the first subsection below.
In the second subsection we will survey research suggesting the possibilities of improvement of \textbf{the capacity of associative memories}. 
The last subsection deals with proposals which address computational \textbf{run-time improvements} of classical learning algorithms, the first of which came out already in the early 2000s. Here we will differentiate approaches which focus on quantum improvements in the training phase of a classifier by means of \textbf{quantum optimization} (mostly focused on exploiting near-term technologies, and restricted devices), and approaches which build algorithms based on, roughly speaking, quantum parallelism and ``quantum linear algebra'' -- which typically assume universal quantum computers, and often ``pre-filled'' database. It should be noted that the majority of research in quantum ML is focused precisely on this last aspect, and the results here are already quite numerous. We can thus afford to present only a chosen selection of results.

\subsection{Learning efficiency improvements: sample complexity}
\boxTLDR{The first results showing the separation between quantum and classical computers were obtained in the context of oracles, and for sample complexity -- even the famous Grover's search algorithm constitutes such a result. Similarly, CLT deals with the learning, i.e., the identification or the approximation of \textbf{concepts}, which are also nothing but oracles. Thus, quantum oracular computation settings and learning theory \textbf{share the same underlying framework}, which is investigated and exploited in this formal topic. To talk about quantum CLT, and improvements, or bounds, on sample complexity, the classical concept oracles are thus upgraded to quantum concept oracles, which \textbf{output quantum states}, and/or \textbf{allow access in superposition}.}
\label{qPAC}

As elaborated in section \ref{CPAC}, CLT deals with the problem of learning \textit{concepts}, typically abstracted as boolean functions of bit-strings of length $n$ so $c: \{0,1 \}^n \rightarrow \{0,1\}$, from input-output relations alone. For intuitive purposes it is helpful to think of the task of optical character recognition (OCR), where we are given a bitmap image (black-and-white scan) of some size $n = N\times M$, and a concept may be say ``everything which represents the letter A'', more precisely, \textit{the concept}, specifying which bitmaps correspond to the bitmaps of the letter ``A''. 
Further, we are most often interesting in a learning performance for a set of concepts: a concept class $\mathcal{C} = \{ c| c: \{0,1 \}^n\rightarrow \{0,1 \}\}$ -- in the context of the  running example of OCR, we care about algorithms which are capable of recognising all letters, and not just ``A''.
 
 The three typical settings studied in literature are the PAC model, exact learning from membership queries, and the agnostic model, see section \ref{CPAC}. These models differ in the type of access to the concept oracle which is allowed. In the PAC model, the oracle outputs labeled examples according to some specified distribution, analogous to basic supervised learning. In the membership queries model, the learner gets to choose the examples, and this is similar to active supervised learning. In the agnostic model, the concept is ``noisy", i.e. forms a stochastic function, which is natural in supervised settings (the joint datapoint-label distribution $P(x,y)$ need not be functional), for details we refer the reader to section \ref{CPAC}. 
 
 All three models have been treated from a quantum perspective, and whether or not quantum advantages are obtainable greatly depends on the details of the settings. 
Here we give a very succinct overview of the main results, partially following the structure of the recent survey on the topic by Arunachalam and de Wolf \cite{2017_deWolf}.
\subsubsection{Quantum PAC learning}
\label{qqpac}

The first quantum generalization of PAC learning was presented in \cite{1998_Bshouty}, where the quantum example oracle was defined to output coherent superpositions
\EQ{
  \sum_\mathbf{x} \sqrt{p_D(\mathbf{x})} \ket{x, c(x)} ,
}
for a given distribution $D$ over the data points $x,$ for a concept $c$.
Recall, classical PAC oracles output a sample pair $(x,c(x))$, where $x$ is drawn from $D$, which can be understood as copies of the mixed state  $\sum_\mathbf{x} p_D(\mathbf{x}) \dm{x, c(x)} $, with 
$p_D(\mathbf{x}) = P(D=\mathbf{x})$.  The quantum oracle reduces to the standard oracle if the quantum example is measured in the standard (computational) basis. 
This first pioneering work showed that quantum algorithms, with access to such a quantum-generalized oracle can provide more efficient learning of certain concept classes.  
The authors have considered the concept class of DNF formulas, under the uniform distribution: here the concepts are $s$-term formulae in disjunctive normal form. In other words, each concept $c$  is 
of the form $c(\mathbf{x})=
\bigvee_I \bigwedge_j (\textbf{x}_I)'_j,
$
where $\textbf{x}_I$ is a substring of $\textbf x$ associated to $I$, which is a subset of the indices of cardinality at most $s$,  and $(\textbf{x}_I)'_j$ is a variable or its negation (a literal). An example of a DNF is of the form
$(x_1 \wedge x_3 \wedge \neg x_6) \vee (x_4 \wedge \neg x_8 \wedge  x_1) \cdots$, where parentheses (terms) only contain variables or their negations in conjunction (ANDs, $\wedge$), whereas all the parentheses  are in disjunction (ORs, $\vee$).

The uniform DNF learning problem (for $n$ variables, and $poly(n)$ terms) is not known to be efficiently PAC learnable, but, in \cite{1998_Bshouty} it was proven to be efficiently quantum PAC learnable. The choice of this learning problem was not accidental: DNF learning is known to be learnable in the membership query model, which is described in detail in the next section. 
The corresponding classical algorithm which learns DNF in the membership query model directly inspired the quantum variant in the PAC case \footnote{To provide the minimal amount of intuition, the best classical algorithm for the membership query model, heavily depends on Fourier transforms (DFT) of certain sets -- the authors then use the fact that FT can be efficiently implemented on the amplitudes of the states generated by the quantum oracle using quantum computers. We refer the reader to \cite{1998_Bshouty} for further details.}.
If the underlying distribution over the concept domain is uniform, other concept classes can be learned with a quantum speed-up as well, specifically, so called $k$-juntas: $n$-bit binary functions which depend only on $k<n$ bits.
 In  \cite{2007_Atici}, At\i c\i\ and Servedio have shown that there exists a quantum algorithm for learning $k$-juntas using $O(k \log(k)/\epsilon)$ uniform quantum examples, $O(2^k)$
uniform classical examples, and $O(n\ k\ log(k)/\epsilon + 2^k
log(1/\epsilon))$ time. Note the improvement in this case is not in query complexity, but rather in the classical processing, which, for the best known classical algorithm has complexity at least $O(n^{2k/3})$ (see \cite{2007_Atici,2017_deWolf} for further details).
Diverging from perfect PAC settings, in \cite{2015_Cross}, the authors considered the learning of 
linear boolean functions\footnote{The learning of such functions is in QIP circles also known as the (non-recursive) Bernstein-Vazirani problem defined first in \cite{1997_Bernstein}.} under the uniform distribution over the examples. The twist in this work is the assumption of noise\footnote{However, the meaning of noise is not exactly the same in the classical and quantum case.} which allows for  evidence of a classical quantum learnability separation.

\paragraph{Distribution-free PAC} While the assumption of the uniform distribution $D$ constitutes a convenient theoretical setting, in reality, most often we have few guarantees on the underlying distribution of the examples. For this reason PAC learning often refers to distribution-free learning, meaning, learning under the worst case distribution $D$. Perhaps surprisingly, it was recently shown that the quantum PAC learning model offers no advantages, in terms of sample complexity, over the classical model.
Specifically, in \cite{2016_deWolf} the authors show that if $C$ is a concept class of VC dimension $d+1,$ then for every (non-negative) $\delta \leq 1/2$ and $\epsilon \leq 1/20$, every $(\epsilon, \delta)$-quantum PAC learner requires $\Omega(d/\epsilon+ \log(d^{-1})/\epsilon)$ samples.
The same number of samples, however, is also known to suffice for a classical PAC learner (for any $\epsilon$ and $\delta$).

A similar result, showing no separation between quantum and classical \textit{agnostic learning} was also proven in \cite{2016_deWolf}\footnote{ 
The notions of efficiency and sample complexity in the agnostic model are analogous to those in the PAC model, as is the quantum oracle which provides the coherent samples $
 \sum_\mathbf{x,y} \sqrt{p_D(\mathbf{x},y)} \ket{\mathbf{x}, y}$. See section \ref{CPAC} for more details.
} .

\paragraph{Quantum predictive PAC learning}
Standard PAC learning settings do not allow exponential separations between classical and quantum sample complexity of learning, and consequently the notion of learnable concepts is the same in the classical and the quantum case. This changes if we consider weaker learning settings, or rather, a weaker meaning of what it means to learn. 
The PAC learning setting assumes that the learning algorithm outputs a hypothesis $h$ with a low error with high confidence. In the classical case, there is no distinction between expecting that the hypothesis $h$ can be applied once, or any arbitrary number of times. However, in the quantum case, where the examples from the oracle may be quantum states, this changes, and inductive learning in general may not be possible in all settings, see section \ref{QgenML}. In \cite{2012_Gavinsky}, the author considers a quantum PAC settings where only one (or polynomially few) evaluations of the hypothesis are required, called the Predictive Quantum (PQ) model\footnote{In a manner of speaking, to learn a concept, in the PAC sense, implies we can apply what we have learned arbitrarily many times. In PQ it suffices that the learner be capable of applying what it had learned just once once, to be considered successful. It however follows that if the number of examples is polynomial, PQ learnability also implies that the verification of learning can be successfully executed polynomially many times as well.}. 
In this setting the author identifies a relational concept class (i.e. each data point may have many correct labels) which is not (polynomially) learnable in the classical case, but is PQ learnable under a standard quantum oracle, under the uniform distribution. The basic idea is to use quantum states, obtained by processing quantum examples, for each of the testing instances -- in other words, the ``implementation'' of the hypothesis contains a quantum state obtained from the oracle. This quantum state cannot be efficiently estimated, but can be efficiently obtained using the PQ oracle. The concept class, and the labeling process are inspired by a distributed computation problem for which an exponential classical-quantum separation had been identified earlier in \cite{2008_BarYossef}. This work provides another noteworthy example of the intimate connection between various aspects of QIP -- in this case, quantum communication complexity theory -- and quantum learning.

\subsubsection{Learning from membership queries}
\label{LMQ}
In the model of exact learning from membership queries, the learner can choose the elements from the concept domain it wishes labeled (similar to active learning), however, the task is to identify the concept exactly (no error), except with probability $\delta<1/3$\footnote{As usual, success probability which is polynomially bounded away from 1/2 would also do.}
Learning from membership queries has, in the quantum domain, usually been called \textit{oracle identification}.
While quantum improvements in this context are possible, in 
\cite{2004_Servedio}, the authors show that they are at most low-degree polynomial improvements in the most general cases.
More precisely, if a concept class $C$ over $n-bits$ has classical and quantum membership query complexities
$D(C)$ and $Q(C)$, respectively, then $D(C) = O(n Q(C)^3)$\footnote{This simple formulation of the claim of  \cite{2004_Servedio} was presented in \cite{2017_deWolf}.} -- in other words, improvements in sample complexity can be at most polynomial. Polynomial relationships have also been established for worst-case exact learning sample complexitites (so-called $(N,M)$-query complexity), see \cite{2013_Kothari} and \cite{2017_deWolf}.
The above result  is in spirit similar to earlier results in \cite{2001_Beals}, where it was shown quantum query complexity cannot provide a better than polynomial improvement over classical results, unless structural promises on the oracle are imposed.

The results so far considered are standard, comparatively simple generalizations of classical learning settings, leading to somewhat restricted improvements in sample complexity. More dramatic improvements are possible if computational (time) complexity is taken into account, or if slightly non-standard generalizations of the learning model are considered. Note, we are not explicitly bringing computational complexity separations into the picture. Rather, under the assumption that certain computation problems are hard for the learner, we obtain a sample complexity separation.

In particular, already in \cite{1994b_Kearns} the authors have constructed several classes of Boolean functions in the distribution-free model whose efficient learning (in the sample complexity sense) implies the capacity of factoring of so-called Blum integers - a task not known to be solvable classically, but solvable on a quantum computer\footnote{
These ideas exploit the connections between asymmetric cryptography and learning. In asymmetric cryptography, a message can be decrypted easily using a public key, but the decryption is computationally hard, unless one has a private key. To exemplify public key can be a Blum integer, whereas the private key one of the factors. The data-points are essentially the encryptions of integers $k$ $E(k,N)$, for a public key $N$. The concept is defined by the least significant bit of $k$, which, provably, is not easier to obtain with bounded error than the decryption itself -- which is computationally hard. A successful efficient learner of such a concept could factor Blum integers.The full proposal has further details we omit for simplicity. 
}. Using this observations, Servedio and Gortler have demonstrated classes which are efficiently quantum PAC learnable, and classes which are efficiently learnable in the quantum membership query model, but which not efficiently learnable in the corresponding classical models, unless Blum integers \footnote{The integer $n$ is a Blum integer if it is a product of two distinct prime numbers $p$ and $q$, which are congruent to 3 mod 4 (i.e. both can be written in the form $4 t + 3,$ for a non-negative integer $t$.).} can be efficiently factored on a classical computer \cite{2004_Servedio}.

.

\subsection{Improvements in learning capacity}
\label{LearningCap}
\boxTLDR{The observation that a complete description of quantum systems typically requires the specification of exponentially many complex-valued amplitudes has lead to the idea that those same amplitudes could be used to store data using only logarithmically few systems. While this idea fails for most applications, it has inspired some of the first proposals to use quantum systems for the \textbf{dramatic improvement of the capacities of associative, or content-addressable memories}.
More likely quantum upgrades of CAM memories, however, may come from a substantially different direction -- which explores methods of extracting information from HNs -- used as CAM memories -- and which is inspired by quantum adiabatic computing to realize a recall process which is similar yet different from standard recall methods. The quantum methods may yield advantages by \textbf{outputting superpositions of data}, and it has been suggested they also utilize the memory more efficiently, \textbf{leading to increased capacities}.
}
The pioneering investigations in the areas between CLT, NNs and QIP, have challenged the classical sample complexity bounds. 
Soon thereafter (and likely independently), the first proposals suggesting quantum improvements in the context of space complexity emerged -- specifically the efficiency of associative memories. Recall, associative, or content-addressable memory (abbreviated CAM) is a storage device which can be loaded with \textit{patterns,} typically a subset of $n$-bit bit-strings $P = \{\mathbf{x}_i \}_i,$ $\mathbf{x}_i \in \{0,1 \}^n,$ which are then, unlike in the case of standard RAM-type memories, not recovered by address but by content similarity: given an input string  $\mathbf{y} \in \{0,1 \}^n,$ the memory should return $\mathbf{y}$ if it is one of the stored patterns (i.e. $\mathbf{y} \in P$), or a stored pattern which is ``closest'' to $\mathbf{y}$, with respect to some distance, typically the Hamming distance. Deterministic perfect storage of any set of patterns clearly requires $O(n \times 2^n)$ bits (there are in total $2^n$ distinct patterns each requiring $n$ bits), and the interesting aspects of CAMs begin when the requirements are somewhat relaxed.
We can identify roughly two basic groups of ideas which were suggested to lead to improved capacities. The first group, sketched next, relies directly on the structure of the Hilbert space, whereas the second group of ideas stems from the quantization of a well-understood architecture for a CAM memory system: the Hopfield network.

\subsubsection{Capacity from amplitude encoding}
\label{QCAM}
 In some of the first works \cite{2000_Ventura, 2001_Trugenberger} it was suggested that the proverbial ``exponential-sized'' Hilbert space describing systems of qubits may allow exponential improvements: intuitively even exponentially numerous pattern sets $P$ can be ``stored'' in a quantum state of only $n$ qubits:
$\ket{\psi_{P}} = |P|^{-\frac{1}{2}} \sum_{\mathbf{x}\in P} \ket{x}.$ These early works suggested creative ideas on how such a memory could be used to recover patterns (e.g. via modified amplitude amplification), albeit, often suffering from lack of scalability, and other quite fundamental issues to yield complete proposals\footnote{For a discussion on some of the shortcomings see e.g. \cite{2003_Brun,2003_Trugenberger}, and we also refer the reader to more recent reviews \cite{2014_Schuld, 2014_Schuld2} for further details and analysis of the potential application of such memories to pattern recognition problems.}, and thus we will not dig into details. We will, however, point out that these works may be interpreted to propose some of the first examples of ``amplitude encoding'' of classical data, which is heavily used in modern approaches to quantum ML. In particular, the stored memory of a CAM can always be represented as a single bit-string $(b_{(0\cdots 0)},b_{(0\cdots 1)} \ldots, b_{(1\ldots1)})$ of length $2^n$ (each bit in the bit-string is indexed by a  pattern, and its value encodes if it is stored or not). This data-vector (in this case binary, but this is not critical) is thus encoded into amplitudes of a quantum state of an exponentially smaller number of qubits: $\mathbf{b} = (b_{(0\cdots 0)},b_{(0\cdots 1)} \ldots, b_{(1\ldots1)}) \rightarrow  \sum_{\mathbf{x}\in \{ 0,1 \}^n} b_{\mathbf{x}} \ket{\mathbf{x}}$ (up to normalization).

\subsubsection{Capacity via quantized Hopfield networks}

A different approach to increasing the capacities of CAM memories arises from the ``quantization'' of different aspects of classical HNs, which constitute well-understood classical CAM systems.
\paragraph{Hopfield networks as a content-addressable memory}
Recall, a HN is a recurrent NN characterized by a set of $n$ neurons, whose connectivity is given by a (typically symmetric) real matrix of weights $W = (w_{ij})_{ij}$ and a vector of (real) local thresholds $\{\theta_i \}_{i=1}^n$.
In the context of CAM memories, the matrix $W$ encodes the stored patterns, which are in this setting best represented as sequences of signs, so $\mathbf{x} \in \{1,-1\}^{n}$. 
The retrieval, given an input pattern $\mathbf{y} \in \{1,-1\}^{n}$, is realized by setting the $k^{th}$ neuron $s_k$ to the $k^{th}$ value of the input pattern $y_k$, followed by the ``running of the network'' according to standard perceptron rules: each neuron $k$ computes its subsequent value by checking if its inbound weighted sum is above the local threshold: $s_k \leftarrow \textup{sign}(\sum_{l} w_{kl} s_l - \theta_k)$ (assuming $\textup{sign}(0)=+1$)\footnote{The updates can be synchronous, meaning all neurons update their values at the same time, or asynchronous, in which case usually a random order is assigned. In most analyses, and here, asynchronous updates are assumed.}.
As discussed previously, under moderate assumptions the described dynamical system converges to local attractive points, which also correspond to the energy minima of the Ising functional 
\EQ{
E(\mathbf{s}) = -\dfrac{1}{2} \sum_{ij} w_{ij} s_i s_j + \sum_{i} \theta_i s_i.}
Such a system still allows significant freedom in the rule specifying the matrix $W,$ given a set of patterns to be stored: intuitively, we need to ``program'' the minima of $E$ (choosing the appropriate $W$ will suffice, as the local thresholds can be set to zero) to be the target patterns, ideally without storing too many unwanted, so-called spurious, patterns. 
This, and other properties of a useful storing rule, that is, rule which specifies $W$ given the patterns, are given as follows \cite{1997_Storkey}: 
$a)$ \textit{locality:} an update of a particular connection should depend only on the information
available to the neurons on either side of the connection\footnote{Locality matters as the lack of it prohibits parallelizable architectures.}; $b)$ \textit{incrementality}: the rule should allow the updating of the matrix $W$ to store an additional pattern based only on the new pattern and $W$ itself \footnote{In particular, it should not be necessary to have external memory storing e.g. all stored patters, which would render HN-based CAM memories undesirably non-adaptive and inflexible.} $c)$ \textit{immediateness:} the rule should not require a limiting computational process for the evaluation of the weight matrix (rather, it should be a simple computation of few steps). The most critical property of a useful rule is that it $d)$ results in a CAM with a non-trivial capacity: it should be capable of storing and retrieving some number of patters, with controllable error (which includes few spurious patterns, for instance).

The historically first rule, the Hebbian rule, satisfies all the conditions above and is given by a simple recurrence relation: for the set of patterns $\{ \mathbf{x}^k\}_k$ the weight matrix is given with $w_{ij} = \sum_{k}\mathbf{x}_i^k \mathbf{x}_j^k / M$  (where $\mathbf{x}_j^k$ is the $j^{th}$ sign of the $k^{th}$ pattern, and $M$ is the number of patterns). 
The capacity of HN's under standard recall and Hebbian updates has been investigated from various perspectives, and in the context of absolute capacity (the asymptotic ratio of the number of patterns that can be stored without error to the number of neurons, as the network size tends to infinity), it is known to scale as $O(\frac{n} { 2 ln(n)} ).$ A well known result in the field improves on this to the capacity of $O(\frac{n} { \sqrt{2 ln(n)}} )$, and is achieved by a different rule introduced by Storkey \cite{1997_Storkey}, while maintaining all the desired properties. Here, we should emphasize that, in broad terms, the capacity is typically (sub)-linear in $n$. Better results, however, can be achieved in the classical settings if some of the assumptions $a) - c)$ are dropped, but this is undesirable.

\paragraph{Quantization of Hopfield-based CAMs} In early works \cite{2006_Rigatos, 2007_Rigatos}, the authors have considered fuzzy and probabilistic learning rules, and have broadly argued that a) such probabilistic rules correspond to a quantum deliberation process and that b) the resulting CAMs can have significantly larger capacities.
However, more rigorous (and fully worked out) results were shown more recently, by combining HNs with ideas from adiabatic QC. 

The first idea, presented in \cite{2009_Neigovzen} connects HNs and quantum annealing. Recall that the HN can be characterized by the Ising functional $E(\mathbf{s}) = -\dfrac{1}{2} \sum_{ij} w_{ij} s_i s_j $ (see  Eq. \ref{ising}), where the stored patterns correspond to local minima, and where we have, without the loss of generality, assumed that the local thresholds are zero. The classical recall corresponds to the problem of finding local minima closest to the input pattern $\mathbf{y}$. However, an alternative system, with similar features, is obtained if the input pattern is added in place of the local thresholds:
$E(\mathbf{s}, \mathbf{y}) = -\dfrac{1}{2} \sum_{ij} w_{ij} s_i s_j  - \Gamma  \sum_{i} y_{i} s_i.$ Intuitively, this lowers the energy landscape of the system specifically around the input pattern configuration. But then, the stored pattern (previous local minimum) which is closest to the input pattern is the most likely candidate for a global minimum.
Further, the problem of finding such configurations can now be tackled via quantum annealing: 
we define the quantum ``memory Hamiltonian'' naturally as $H_{mem} = -\dfrac{1}{2} \sum_{ij} w_{ij} \sigma_i^z \sigma_j^z,$ and the HN Hamiltonian, given input $\mathbf{y}$ with  $H_{p} = H_{mem} + \Gamma H_{inp},$ where the input Hamiltonian is given with $ H_{inp}=   - \sum_{i} y_{i} \sigma_i^z$.
The quantum recall is obtained by the adiabatic evolution via the Hamiltonian trajectory $H(t) = \Lambda(t) H_{init} +H_{p},$ where $\Lambda(0)$ is large enough that  $ H_{init}$ dominates, and $\Lambda(1) =0$. The system is initialized in the ground state of the (arbitrary and simple) Hamiltonian $H_{init},$ and if the evolution in $t$ is slow enough to satisfy the criteria of the adiabatic theorem, the system ends in the ground state of $H_{p}$. This proposal exchanged local optimization (classical retrieval) for global optimization. While this is generally a bad idea\footnote{Generically, local optimization is easier than global, and in the context of the Ising system, global optimization is known to be NP-hard.}, what is gained is a quantum formulation of the problem which can be run on adiabatic architectures, and also the fact that this system can return quantum superpositions of recalled patterns, if multiple stored patterns are approximately equally close to the input, which can be an advantage \cite{2009_Neigovzen}.
However, the system above does not behave \textit{exactly} the same as the classical recall network, 
 which was further investigated in subsequent work \cite{2014_Seddiqi} analysing the sensitivity of the quantum recall under various classical learning rules. Further, in \cite{2016_Santra} the authors have provided an extensive analysis of the capacity of the Hebb-based HN, but under quantum annealing recall as proposed in \cite{2009_Neigovzen} showing, surprisingly, that this model yields \textit{exponential} storage capacity, under the assumption of random memories. This result stands in apparent stark contrast to standard classical capacities reported in textbooks\footnote{At this point it should be mentioned that recently exponential capacities of HNs have been proposed for fully classical systems, by considering different learning rules \cite{2014_Hillar, 2014_Karbasi}, which also tolerate moderate noise. The relationship and potential advantages of the quantum proposals remains to be elucidated.}.
 
 Regarding near-term implementability,  in \cite{2016_Santra} the authors have investigated the suitability of the Chimera graph-based architectures of D-Wave programmable quantum annealing device for quantum recall HN tasks, showing potential for demonstrable quantum improvements in near-term devices.




\subsection{Run-time improvements: computational complexity}
\boxTLDR{The theory of quantum algorithms has provided examples of \textbf{computational speed ups} for decision problems, various functional problems, oracular problems, sampling tasks, and optimization problems. This section presents quantum algorithms which provide \textbf{speed-ups for learning-type problems}. The two main classes of approaches differ in the underlying computational architecture -- a large class of algorithms relies on \textbf{quantum annealers}, which may not be universal for QC, but may natively solve certain sub-tasks important in the context of ML. These approaches then have an increased likelihood of being \textbf{realizable with near-term devices}. In contrast, the second class of approaches assumes universal quantum computers, and often data prepared and accessible in quantum database, \textbf{but offers up to exponential improvements}. Here we distinguish between \textbf{quantum amplitude amplification} and \textbf{amplitude encoding approaches}, which, with very few exceptions, cover all quantum algorithms for supervised and unsupervised learning.}

The most prolific research area within quantum ML in the last few years has focused on identifying ML algorithms, or their computationally intensive subroutines, which may be sped up using quantum computers. While there are multiple natural ways to classify the performed research, an appealing first-order delineation follows the types of quantum computational architectures assumed\footnote{Other classification criteria could be according to tasks, i.e. supervised vs. unsupervised vs. generative models etc., or depending on the underlying quantum algorithms used, e.g. amplitude amplification, or equation solving.}. Here we can identify research which is focused on using quantum annealing architectures, which are experimentally well justified and even commercially available in recent times (mostly in terms of the D-Wave system set-ups). In most of such research, the annealing architecture will be utilized to perform a classically hard optimization problem usually emerging in the training phases of many classical algorithms. 
An involved part of such approaches will often be a meaningful rephrasing of such ML optimization to a form which an annealing architecture can (likely) handle. 
While the overall supervised task comprises multiple computational elements, it is only the optimization that will be treated by a quantum system in these proposals. 

The second approach to speeding up ML algorithms assumes universal quantum computation capabilities. Here, the obtained algorithms are typically expressed in terms of quantum circuits. For most proposals in this research line, to guarantee actual speed-ups, there will be additional assumptions. For instance, most proposals can only guarantee improvements if the data, which is to be analyzed, is already present in a type of a quantum oracle or a quantum memory, and, more generally, that certain quantum states, which depend on the data, can be prepared efficiently. The overhead of initializing such a memory in the first place is not counted, but this may not unreasonable as in practice, the same database is most often used for a great number of analyses. 
Other assumptions may also be placed on the structure of the dataset itself, such as low condition numbers of certain matrices containing the data \cite{2015_Aaronson}.

\subsubsection{Speed-up via adiabatic optimization}
\label{Adiabatic}
Quantum optimization techniques play an increasingly important role in quantum ML. Here, we can roughly distinguish two flavours of approaches, which differ in what computationally difficult aspect of training of a classical model is tackled by adiabatic methods. In the (historically) first approach, we deal with clear-cut optimization in the context of binary classifiers, and more specifically, boosting (see \ref{othermodels}). Since, it has been shown that annealers can also help by generating samples from hard-to-simulate distributions. We will mostly focus on the historically first approaches, and only briefly mention the other more recent results.
\paragraph{Optimization for boosing}
The representative line of research, which also initiated the development of this topic of quantum-enhanced ML based on adiabatic quantum computation, focuses on a particular family of optimization problems called \textit{quadratic unconstrained optimization} (QUBO) problems of the form
 \EQ{\mathbf{x}^\ast = (x_1^\ast,\ldots,x_n^\ast)  = \textup{argmin}_{(x_1,\ldots,x_n)}\sum_{i<j} J_{ij} x_i x_j,\ x_k \in\{0,1 \} }
specified by a real matrix $J$. QUBO problems are equivalent to the problem of identifying lowest energy states of the Ising functional\footnote{More precisely, an efficient algorithm which solves general QUBO problems can also efficiently solve arbitrary Ising ground state problems. One direction is trivial as QUBO optimization is a special case of ground state finding, where the local fields are zero. In the converse, given an Ising ground state problem over $n$ variables, we can construct a QUBO over $n+1$ variables, which can be used to encode the local terms.} $E(\mathbf{s}) = -\dfrac{1}{2} \sum_{ij} J_{ij} s_i s_j + \sum_{i} \theta_i s_i$, provided we make no assumptions on the underlying lattice.  Modern annealing architectures provide means for tackling the problem of finding such ground states using adiabatic quantum computation. Typically we are dealing with systems which can implement the tunable Hamiltonian of the form 
\EQ{
H(t) = -A(t)\underbrace{\sum_{i} \sigma^x}_{H_{initial}} + B(t)\underbrace{\sum_{ij} J_{ij} \sigma^z_i \sigma^z_j}_{H_{target}},
}
where $A,B$ are smooth positive functions such that $A(0)\gg B(0)$ and $B(1)\gg A(1),$ that is, by tuning $t$ sufficiently slowly, we can perform adiabatic preparation of the ground state of the Ising Hamiltonian $H_{target}$, thereby solving the optimization problem. In practice, the parameters $J_{ij}$ cannot be chosen fully freely (e.g. the connectivity is restricted to the so-called Chimera graph \cite{2015_Hen} in D-Wave architectures), and also the realized interaction strenght values have a limited precision and accuracy \cite{2009_Neven,2010_Bian}, but we will ignore this for the moment.
In general, finding ground states of the Ising model is functional NP-hard\footnote{Finding ground states is not a decision problem, so, technically it is not correct to state it is NP-hard. The class functional NP (FNP) is the extension of the NP class to functional (relational) problems. }, which is likely beyond the reach of quantum computers. However, annealing architectures still may have many advantages, for instance it is believed they may still provide speed ups in all, or at least average instances, and/or that they may provide good heuristic methods, and hopefully near optimal solutions\footnote{Indeed, one of the features of adiabatic models in general is that they provide an elegant means for (generically) providing approximate solutions, by simply performing the annealing process faster than prescribed by the adiabatic theorem.}.

 In other words, any aspect of optimization occurring in ML algorithms which has an efficient mapping to (non-trivial) instances of QUBO problems, specifically those which can be realized by experimental set-ups, is a valid candidate for quantum improvements. 
Such optimization problems have been identified in a number of contexts, mostly dealing with training binary classifiers, thus belong to the class of supervised learning problems. 
The first setting considers the problem of building optimal classifiers from linear combinations of simple hypothesis functions, which minimize empirical error, while controlling the model complexity through a so-called regularization term. This is the common optimization setting of {boosting} (see \ref{othermodels}), and, with appropriate mathematical gymnastics and few assumptions, it can be reduced to a QUBO problem.

The overarching setting of this line of works can be expressed in the context of training a binary classifier by combining weaker hypotheses. 
For this setting, consider a dataset $D=\{\mathbf{x}_i,y_i \}_{i=1}^{M}$, $\mathbf{x}_i \in \mathbbmss{R}^{n}$, $y_i \in\{-1,1 \}$, and a set of hypotheses $\{h_j\}_{j=1}^{K}, h_j:  \mathbbmss{R}^{n}\rightarrow \{-1,1\}.$ 
For a given weight vector $\mathbf{w}  \in \mathbbmss{R}^{n}$ we define the composite classifier of the form 
$hc_{{\mathbf{w}}}(\mathbf{x}) = \textup{sign}(\sum_k {w}_k h_k(\mathbf{x})).$ 

The training of the composite classifier is achieved by the optimization of the vector $\mathbf{w}$ as to minimize misclassification on the training set, and as to decrease the risk of overtraining.
The misclassification cost is specified via a loss function $L$, which depends on the dataset, and the hypothesis set in the boosting context. The overtraining risk, which tames the complexity of the model, is controlled by a so-called regularization term $R$. Formally we are solving
\EQ{
 \textup{argmin}_{\mathbf{w}}\ L(\mathbf{w}; D) + R(\mathbf{w}).
}
This constitutes the standard boosting frameworks exactly, but is also closely related to the training of certain SVMs, i.e. hyperplane classifiers\footnote{If we allow the hypotheses $h_j$ to attain continuous real values, then by setting $h_j$ to be the projection on the $j^{th}$ component of the input vector, so $h_j(\mathbf{x}) = x_j,$ then the combined classifier attains  attains the inner-product-threshold form
 $hc_{\mathbf{w}}(\mathbf{x}) = \textup{sign}(\mathbf{w}^\tau \mathbf{x} )$ which contains hyperplane classifiers -- the only component missing is the hyperplane offset $b$ which incorporated into the weight vector by increasing the dimension by 1. }. In other words, quantum optimization techniques which work for boosting setting can also help for hyperplane classification.

There are a few well-justified choices for $L$ and $R$, leading to classifiers with different properties. Often, best choices (the definition of which depends on the context) lead to hard optimization\cite{2010_Long}, and some of those can be reduced to QUBOs, but not straightforwardly.

In the pioneering paper on the topic \cite{2008_Neven}, Neven and co-authors consider the boosting setting. The regularization term  is chosen to be proportional to the 0-norm, which counts the number of non-zero entries, that is, $R(\mathbf{w},\lambda) = \lambda \| \mathbf{w}\|_0.$ The parameter $\lambda$ controls the relative importance of regularization in the overall optimization task. 
A common choice for the loss function would be the 0-1 loss function $L_{0-1}$, optimal in some settings, given with $L_{0-1}(\mathbf{w}) = \sum_{j=1}^{M} \Theta \left(-y_j \sum_k {w}_k h_k(\mathbf{x}_j)  \right)$ (where $\Theta$ is the step function), which simply counts the number of misclassifications.
This choice is reasonably well motivated in terms of performance, and is likely to be computationally hard.
With appropriate discretization of the weights $\mathbf{w}$, which the authors argue likely does not hurt performance, the above forms a solid candidate for a general adiabatic approach.
However, it does not fit the QUBO structure (which has only quadratic terms), and hence cannot be tackled using existing architectures. To achieve the desired QUBO structure the authors impose two modifications: they opt for a quadratic loss function $L_{2}(\mathbf{w}) = \sum_{j=1}^{M} |y_j  - \sum_k {w}_k h_k(\mathbf{x}_j)  |^2$, and restrict the weights to binary (although this can be circumvented to an extent). Such a system is also tested using numerical experiments. In a follow-up paper \cite{2009_Neven}, the same team has generalized the initial proposal to accommodate another practical issue: problem size. Available architectures allow optimization over a few thousand variables, whereas in practice the number of hypotheses one optimizes over ($K$) may be significantly larger. To resolve this, the authors show how to break a large optimization problem into more manageable chunks while maintaining (experimentally verified) good performance. These ideas were also tested in an actual physical architecture \cite{2009_Neven2}, and combined and refined in a more general, iterative algorithm in \cite{2012_Neven}, tested also using actual quantum architectures.

While $L_{0-1}$ loss functions were known to be good choices, they were not the norm in practice as they lead to non-convex optimization -- so convex functions were preferred. 
However, in 2010 it became increasingly clear that convex functions are provably bad choices. For instance, in the seminal paper \cite{2010_Long} Long and Servedio\footnote{Servedio also, incidentally, provided some of the earliest results in quantum computational learning theory, discussed in previous sections.}, showed that boosting with convex optimization completely fails in noisy settings. Motivated by this in \cite{2012_Denchev}, the authors re-investigate D-Wave type architectures, and identify a reduction which allows a non-convex optimization. Expressed in the hyperplane classification setting (as explained, this is equivalent to the boosting setting in structure), they identify a reduction which (indirectly) implements a non-convex function  $l_{q}(x) = \min\{(1-q)^2, (\max (0, 1-x))^2 \}$. This function is called the \textit{q-loss} function, where $q$ is a real parameter. The implementation of the q-loss function allows for the realization of optimization relative to the total loss of the form $L_{q}(\mathbf{w},b ; D) =\sum_j  l_{q}( y_j(\mathbf{w}^\tau \mathbf{x} + b )).$ The resulting regularization term is in this case proportional to the 2-norm of $\mathbf{w},$ instead of the 0-norm as in the previous examples, which may be sub-optimal.
Nonetheless, the above forms a prime example where quantum architectures lead to ML settings which would not have been explored in the classical case (the loss $L_q$ is unlikely to appear naturally in many settings) yet are well motivated, as \textit{a)} the function is non-convex and thus has the potential to circumvent all the no-go results for convex functions, and \textit{b)} the optimization process can be realized in a physical system. The authors perform a number of numerical experiments demonstrating the advantages of this choice of a non-convex loss function when analysing noisy data, which is certainly promising. In later work \cite{2015_Denchev}, it was also suggested that combinations of loss-regularization which are realizable in quantum architectures can also be used for so-called totally corrective boosting with cardinality penalization, which is believed to be classically intractable.

 The details of this go beyond the scope of this review, but we can at least provide a flavour of the problem. In corrective boosting, the algorithm updates the weights $\mathbf{w}$ essentially one step at a time. In totally corrective boosting, at the $t^{th}$ step of the boosting algorithm optimization, $t$ entries of  $\mathbf{w}$ are updated simultaneously. This is known to lead to better regularized solutions, but the optimization is harder. Cardinality penalization pertains to using explicitly the $0$-norm for the regularization (discussed earlier), rather than the more common 1-norm. This, too, leads to harder optimization which may be treated using an annealing architecture.  
In \cite{2014_Babbush}, the authors significantly generalized the scope of loss functions which can be embedded into quantum architectures, by observing that any polynomial unconstrained binary optimization  can, with small overhead, be mapped onto a (slightly larger) QUBO problem. This, in particular, opens up the possibility of implementing odd-degree polynomials which are non-convex and can approximate the 0-1 loss function. This approach introduced new classes of unusual yet promising loss functions.

\paragraph{Applications of quantum boosting}

Building on the ``quantum boosting'' architecture described above, in \cite{2013_Pudenz},  the authors explore the possibility of (aside from boosting) realizing anomaly detection, specifically envisioned in the computationally challenging problem of software verification and validation\footnote{A software is represented as a map $P$ from input to output spaces, here specified as subset of the space of pairs $(x_{input},x_{output})$. An implemented map (software) $P$ is differentiated from the ideal software $\hat{P}$ by the mismatches in the defining pairs.}.
In the proposed learning step the authors use quantum optimization (boosting) to learn the characteristics of the program being tested. In the novel testing step the authors modify the target Hamiltonian as to lower the energy of the states which encode input-outputs where the real and ideal software differ.  These can then be prepared in superposition (i.e. they can prepare a state which is a superposition over the inputs where the $P$ will produce an erroneous output) similarly to the previously mentioned proposals in the context of adiabatic recall of superpositions in HN \cite{2009_Neigovzen}.
\paragraph{Beyond boosting}
Beyond the problems of boosting, annealers have been shown to be useful for the training of so-called Bayesian Network Structure Learning problems \cite{2015_OGorman}, as their training can also be reduced to QUBOs. 
Further, annealing architectures can also be used the training of deep neural networks, relying on sampling, rather than optimization. A notable approach to this is based on the fact that the training of deep networks usually relies on the use of a so-called generative \textbf{deep belief network,} which are, essentially, restricted BMs with multiple layers\footnote{In other words, they are slightly less restricted BMs, with multiple layers and no within-layer connectivity.}. The training of deep belief networks, in turn, is the computational bottleneck, as i requires the sampling of hard-to-generate distributions, which may be more efficiently prepared using annealing architectures, see e.g. \cite{2015_Adachi}. Further. novel ideas introducing fully quantum BM-like models have been proposed \cite{2016_Mohammad}. Further, in recent work \cite{2017_Sieberer} which builds on the flexible construction in \cite{2015_Lechner}, the authors have shown how to achieve programmable adiabatic architectures, which allows running algorithms where the weights themselves are in superposition. This possibility is also sure to inspire novel QML ideas.
Moving on from BMs, in recent work \cite{2017b_Wittek}, the authors have also shown how suitable annealing architectures may be useful to speed-up the performing of probabilistic inference in so-called Markov logic networks\footnote{Markov logic networks \cite{2006_Richardson} combine first-order logic as used for knowledge representation and reasoning, and statistical modelling -- essentially, the world is described via first-order sentences (a knowledge base), which gives rise to a graphical statistical model (a Markov random field), where correlations stem from the relations in the knowledge base.}.
This task involves the estimation of partition functions of arising from statistical models, concretely Markov random fields, which include the Ising model as a special case. Quantum annealing may speed up this sub-task. 

More generally, general, the ideas that restricted, even simple, quantum systems which may be realizable with current technologies, could implement information processing elements useful for supervised learning are beginning to be explored in setting beyond annealers. For instance, in \cite{2017_Schuld}, a simple interferometric circuit is used for the efficient evaluation of distances between data-vectors, useful for classification and clustering. A more complete account of these recent ideas is beyond the scope of this review.

\subsubsection{Speed-ups in circuit architectures}
\label{qcircuit}

One of the most important applications of ML in recent times has been in the context of data mining, and analyzing so-called \textit{big data}.  
The most impressive improvements in this context have been achieved by proposing specialized quantum algorithms which solve particular ML problems.
Such algorithms assume the availability of full-blown quantum computers, and have been tentatively probed since early 2000s. In recent times, however, we have witnessed a large influx of ideas.
Unlike the situation we have seen in the context of quantum annealing, where an optimization subroutine alone was run on a quantum system, in most of the approaches of this section, the entire algorithm, and even the dataset may be quantized.

The ideas for quantum-enhancements for ML can roughly be classified into two groups: a) approaches which rely on Grover's search and amplitude amplification to obtain up-to-quadratic speed-ups, and, b) approaches which encode relevant information into quantum amplitudes, and which have a potential for even exponential improvements. 
The second group of approaches forms perhaps the most developed research line in quantum ML, and collects a plethora quantum tools -- most notably quantum linear algebra, utilized in quantum ML proposals.

\paragraph{Speed-ups by amplitude amplification}

In \cite{2003_Anguita}, it was noticed that the training of support vector machines may be a hard optimization task, with no obviously better approaches than brute-force search. In turn, for such cases of optimization with no structure, QIP offers at least a quadratic relief, in the guise of variants of Grover's \cite{1996_Grover} search algorithm or its application to minimum finding \cite{1999_Durr}. This idea predates, and is, in spirit, similar to some of the early adiabatic-based proposals of the previous subsection, but the methodology is substantially different.
 The potential of quadratic improvements stemming from Grover-like search mechanisms was explored more extensively in \cite{2013_Aimeur}, in the context of unsupervised learning tasks. 
There the authors assume access to a black-box oracle which computes a distance measure between any two data-points. Using this, combined with amplitude amplification techniques (e.g. minimum finding in \cite{1999_Durr}), the authors achieve up to quadratic improvements in key subroutines used in clustering (unsupervised learning) tasks. Specifically, improvements are obtained in algorithms performing minimum spanning tree clustering, divisive clustering and $k$-medians clustering\footnote{
In minimum tree clustering, data is represented as a weighted graph (weight being the distance), and a minimum weight spanning tree is found. $k$ clusters are identified by simply removing the $k-1$- highest weight edges. Divisive clustering is an iterative method which splits sets into two subsets according to a chosen criterion, and this process is iterated. $k-$median clustering identifies clusters which minimize the cumulative within-cluster distances to the median point of the cluster. 
}. Additionally, the authors also show that quantum effects allow for a better parallelization of clustering tasks, by constructing a distributed version of Grover's search. This construction may be particularly relevant as large databases can often be distributed.

More recently, in \cite{2014_Wiebe} the author considers the problem of training deep (more than two-layered) BMs. As we mentioned earlier, one of the bottlenecks of exactly training BMs stems from the fact that it requires the estimation of probabilities of certain equilibrium distributions. Computing this analytically is typically not possible (it is as hard as computing partition functions), and sampling approaches are costly as it requires attaining the equilibrium distribution and many iterations to reliably estimate small values. This is often circumvented by using proxy solutions (e.g. relying on contrastive divergence) to train approximately, but it is known that these methods are inferior to exact training. In \cite{2014_Wiebe}, a quantum algorithm is devised which prepares coherent encodings of the target distributions, relying on quantum amplitude amplification, often attaining quadratic improvements in the number of training points, and even exponential improvements in the number of neurons, in some regimes. 
Quadratic improvements have also been obtained in pure data mining contexts, specifically in \textit{association rules mining} \cite{2016_Yu}, which, roughly speaking identifies correlations between objects in large databases \footnote{To exemplify the logic behind association rules mining, in the typical context of shopping, if shopping item (list element) B occurs in nearly every shopping list in which shopping item A occurs as well, one concludes that the person buying A is also likely to buy B. This is captured by the rule denoted $B \Rightarrow A$.}.
As our final example in the class of quantum algorithms relying on amplitude amplification we mention the algorithm for the training perceptrons \cite{2016_Wiebe}.
Here, quantum amplitude amplification was used to quadratically speed up training, but, interestingly, also to quadratically reduce the error probability. Since perceptrons constitute special cases of SVMs, this result is similar in motivation to the much older proposal \cite{2003_Anguita}, but relies on more modern and involved techniques.

\paragraph{Precursors of amplitude encoding}

In an early pioneering, and often overlooked, work \cite{2003_Schutzhold}, Sch\"{u}tzhold proposed an interesting application of QC on pattern recognition problems, which addresses many ideas which have only been investigated, and re-invented, by the community relatively recently. 
The author considers the problem of identifying ``patterns'' in images, specified by $N \times M$ black-and-white bitmaps, characterized by a function $f:\{1,\ldots,N  \}\times\{1,\ldots,M \}\rightarrow \{0,1 \}$ (which technically coincides with a \textit{concept} in CLT see \ref{CPAC}), specifying the color-value $f(x,y)\in \{ 0,1\}$ of a pixel at coordinate $(x,y)$. The function $f$ is given as a quantum oracle $\ket{x}\ket{y}\ket{b} \stackrel{U_f}{\rightarrow}\ket{x}\ket{y}\ket{b\oplus f(x,y)}$.
The oracle is used in quantum parallel (applied to a superposition of all coordinates), and conditioned on the bit-value function being $1$ (this process succeeds with constant probability, whenever the density of points is constant,) leading to the state $\ket{\psi} = \mathcal{N} \sum_{x,y\ s.t. f(x,y)=1} \ket{x}\ket{y}$, where $\mathcal{N} $ is a normalization factor. Note, this state is proportional to the vectorized bitmap image itself, when given in the computational basis. Next, the author points out that ``patterns'' -- repeating macroscopic features -- can often be detected by applying discrete Fourier transform to the image vector, which has classical complexity $O(NM \log (NM))$. However, the quantum Fourier transform (QFT) can be applied to the state $\ket{\psi}$ utilizing exponentially fewer gates.
The author proceeds to show that the measurements of the QFT transformed state may yield useful information, such as pattern localization.  
This work is innovative in a few aspects. First, the author utilized the encoding of data-points (here strings of binary values) into amplitudes by using a quantum memory, in a manner which is related to the applications in the context of content-addressable memories discussed in \ref{QCAM}. It should be pointed out, however, that in the present application of amplitude encoding, non-binary amplitudes have clear meaning (in say grayscale images), although this is not explicitly discussed by the author. 
Second, in contrast to all previous proposals, the author shows the potential for a quantifiable exponential computational complexity improvement for a family of tasks. However, this is all contingent on having access of the pre-filled database ($U_f$) the loading of which would nullify any advantage. Aside from the fact that this may be considered a one-off overhead, Sch\"{u}tzhold discusses physical means of loading data from optical images in a quantum-parallel approach, which may be effectively efficient.

\paragraph{Amplitude encoding: linear algebra tools}
The very basic idea of amplitude encoding is to treat states of $N-$level quantum systems, as data vectors themselves. More precisely given a data-vector $\mathbf{x} \in \mathbbmss{R}^n$, the amplitude encoding would constitute the normalized quantum state $\ket{x} = \sum_{i}x_i \ket{i}/ ||\mathbf{x}||$, where it is often also assumed that norm of the vector $\| \mathbf{x} \|$ can always be accessed.

Note that $N-$dimensional data-points are encoded into amplitudes of $n \in O(\log(N))$ qubits. Any polynomial circuit applied to the $n$-qubit register encoding the data thus constitutes only a polylogarithmic computation relative to the data-vector size, and this is at the basis of all exponential improvements (also in the case of \cite{2003_Schutzhold}, discussed in the previous section)\footnote{In a related work \cite{2015b_Wiebe}, the authors investigate the learning capacity, of ``small'' quantum systems, and identify certain limitations in the context of Bayesian learning, based on Grover optimality bounds. Here, ``small'' pertains to systems of logarithmic size, encoding information in amplitudes. This work thus probes the potential of space complexity improvements for quantum-enhanced learning, related to early ideas discussed in \ref{LearningCap}.}.  

 These ideas have lead to a research area which could be called ``quantum linear algebra'' (QLA), that is, a collection of algorithms which solve certain linear algebra problems, by directly encoding numerical vectors into state vectors.
  These quantum sub-routines have then been used to speed up numerous ML algorithms, some of which we describe later in this section. 
 QLA includes algorithms for matrix inversion and principal component analysis \cite{2009_Harrow,2014_Lloyd}, and many others. For didactic purposes, we will first give the simplest example which performs the estimation of inner products in logarithmic time. \vspace{0.2cm}
 
\underline{\textit{Tool 1:} inner product evaluation}\ 
Given access to boxes which prepare quantum states $\ket{\psi}$ and $\ket{\phi},$ the overlap $|\langle \phi \ket{\psi} |^2$ can be estimated to precision $\epsilon$ using $O(1/\epsilon)$ copies, using the so-called the swap-test. 

The swap test \cite{2001_Buhrman} applies a controlled-SWAP gate onto the state $\ket{\psi}\ket{\phi},$ where the control qubit is set to the uniform superposition $\ket{+}$. The probability of ``succeeding'', i.e. observing $\ket{+}$ on the control after the circuit is given with $(1+  | \langle \phi \ket{\psi} |^2)/2$, and this can be estimated by iteration (a more efficient option using quantum phase estimation is also possible). If the states $\ket{\psi}$ and $\ket{\phi}$ encode unit-length data vectors, the success value encodes their inner product up to sign. Norms, and phases can also be estimated by minor tweaks to this basic idea -- in particular, actual norms of the amplitude-encoded states will be accessible in a separate oracle, and used in algorithms. The sample complexity of this process depends only on precision, whereas the gate complexity is proportional to $O(\log(N))$ as that many qubits need to be control-swapped and measured.

The swap test also works as expected if the reduced states are mixed, and the overall state is product. This method of computing inner products, relative to classical vector multiplication, offers an exponential improvement with respect to $N$ (if calls to devices which generate $\ket{\psi}$ and $\ket{\phi}$ take $O(1)$), at the cost of significantly worse scaling with respect to errors, as classical algorithms have typical error scaling with the logarithm of inverse error, $O(\log(1/\epsilon))$. However, in context of ML problems, this is can constitute an excellent compromise.\vspace{0.2cm}

\underline{\textit{Tool 2:} quantum linear system solving}\ 
Perhaps the most influential technique for quantum enhanced algorithms for ML is based on one of the quintessential problems of linear algebra: solving systems of equations.
In their seminal paper \cite{2009_Harrow}, the authors have proposed the first algorithm for ``quantum linear system'' (QLS) solving, which performs the following. Consider an $N\times N$ linear system $A \mathbf{x} = \mathbf{b}$, where $\kappa$ and $d$ are the condition number\footnote{Here, the condition number of the matrix $A$ is given by the quotient of the largest and smallest singular value of $A$.}, and sparsity of the Hermitian system matrix $A$\footnote{The assumption that $A$ is Hermitian is non-restrictive, as an oracle for any sparse matrix $A$ can be modified to yield an oracle for the symmetrized matrix $A' = \ket{0}\bra{1}\otimes A^{\dagger} +  \ket{1}\bra{0}\otimes A$.}. Given (quantum) oracles giving positions and values of non-zero elements of $A,$ (that is, given standard oracles for $A$ as encountered in Hamiltonian simulation, cf. \cite{2015_Berry}) and an oracle which prepares the quantum state $\ket{\mathbf{b}}$ which is the amplitude encoding of $\mathbf{b}$ (up to norm), the algorithm in  \cite{2009_Harrow} prepares the quantum state $\ket{\mathbf{x}}$
 which is $\epsilon-$close to the amplitude encoding of the solution vector $\mathbf{x}$. The run-time of the first algorithm is $\tilde{O}(\kappa^2 d^2 \log(N)/\epsilon).$ Note, the complexity scales proportionally to the \textit{logarithm} of the system size. Note that any classical algorithm must scale at least with $N$, and this offers room for exponential improvements. 
 The original proposal in  \cite{2009_Harrow} relies on Hamiltonian simulation (implementing $exp(i A t)$,) upon which phase estimation is applied. Once phases are estimated, inversely proportional amplitudes -- that is, the inverses of the eigenvalues of $A$ --  are imprinted via a measurement. It has also been noted that certain standard matrix pre-conditioning techniques can also be applicable in the QLS scheme \cite{2013_Clader}. The linear scaling in the error in these proposals stems from the phase estimation subroutine. In more recent work \cite{2015_Childs}, the authors also rely on best Hamiltonian simulation techniques, but forego the expensive phase estimation. Roughly speaking, they (probabilistically) implement a linear combination of unitaries of the form $\sum_{k} \alpha_k exp(i  kA t)$ upon the input state. This constitutes a polynomial in the unitaries which can be made to approximate the inverse operator $A^{-1}$ (in a measurement-accessible subspace) more efficiently. This, combined with other numerous optimizations, yields a final algorithm with complexity $\tilde{O}(\kappa d \textup{polylog}(N/\epsilon)),$ which is essentially optimal. It is important to note that the apparently exponentially more efficient schemes above do not trivially imply provable computational improvements, even if we assume free access to all oracles. For instance, one of the issues is that the quantum algorithm outputs a quantum state, from which classical values can only be accessed by sampling. This process for the reconstruction of the complete output vector would kill any improvements. On the other hand, certain functions of the amplitudes can be computed efficiently, the computation of which may still require $O(N)$ steps classically, yielding the desired exponential improvement. Thus this algorithm will be most useful as a sub-routine, an intermediary step of bigger algorithms, such as those for quantum machine learning.  
\vspace{0.2cm}

\underline{\textit{Tool 3:} density matrix exponentiation}\ 
Density matrix exponentiation (DME) is a remarkably simple idea, with few subtleties, and, arguably, profound consequences. 
Consider an $N$-dimensional density matrix $\rho.$ Now, from a mathematics perspective, $\rho$ is nothing but a semidefinite positive matrix, although it is also commonly used to denote the quantum state of a quantum system -- and these two are subtly different concepts. In the first reading, where $\rho$ is a matrix (we will denote it $[\rho]$ to avoid confusion), $[\rho]$ is also a valid description of a physical Hamiltonian, with time-integrated unitary evolution $\exp(-i [\rho] t)$.
 Could one approximate $\exp(-i [\rho] t)$, having access to quantum systems prepared in the state $\rho$? Given sufficiently many copies ($\rho^{\otimes n}$), the obvious answer is yes -- one could use full state tomography to reconstruct $[\rho],$ to arbitrary precision, and then execute the unitary using say Hamiltonian simulation (efficiency notwithstanding). In \cite{2014_Lloyd}, the authors show a significantly simpler method: given any input state $\sigma$, and one copy of $\rho$, the quantum state
 \EQ{ \sigma' = Tr_{B}[ \exp(-i \Delta t S) ( \sigma_{A} \otimes \rho_B)  \exp(i \Delta t S)],} where $S$ is the Hermitian operator corresponding to the quantum SWAP gate, approximates the desired time evolution to first order, for small $ \Delta t$:
 $ \sigma' = \sigma - i \Delta t [\rho, \sigma ]  +  O(\Delta t^2 )$. If this process is iterated, by using fresh copies of $\rho$, we obtain that the target state $\sigma_\rho = \exp(-i \rho t) \sigma \exp(i \rho t)$ can be approximated to precision $\epsilon,$ by setting $\Delta t$ to $O(\epsilon/t)$ and using $O(t^2/\epsilon)$ copies of the state $\rho$.  DME is, in some sense, a generalization of the process of using SWAP-tests between two quantum states, to simulate aspects of a measurement specified by one of the quantum states.
One immediate consequence of this result is in the context of Hamiltonian simulation, which can now be efficiently realized (with no dependency on the sparsity of the Hamiltonian), whenever one can prepare quantum systems in a state which is represented by the matrix of the Hamiltonian. In particular, this can be realized using qRAM stored descriptions of the Hamiltonian, whenever the Hamiltonian itself is of low rank. More generally, this also implies, e.g. that QLS algorithms can also be efficiently executed when the system matrix is not sparse, but rather dominated by few principal components, i.e. close to a low rank matrix\footnote{{Since a density operator is normalized, the eigenvalues of data-matrices are rescaled by the dimension of the system. If the eigenvalues are close to uniform, they are rendered exponentially small in the qubit number. This then requires exponential precision in DME, which would off-set any speed-ups. However, if the spectrum is dominated by a constant number of terms, the precision required, and overall complexity, is again independent from the dimension, allowing overall efficient algorithms.}}. 
\vspace{0.2cm}\\

\textbf{Remark:} Algorithms for QLS, inner product evaluation, quantum PCA, and consequently, almost all quantum algorithms listed in the remainder of this section also assume ``pre-loaded databases'', which allow accessing of information in quantum parallel, and/or the accessing or efficient preparation of amplitude encoded states.
The problem of parallel access, or even the storing of quantum states has been addressed and mostly resolved using so-called  quantum random access memory (qRAM) architectures \cite{2008_Giovannetti} \footnote{qRAM realizes the following mapping: 
$\ket{addr}\ket{b} \stackrel{qRAM}{\longrightarrow} \ket{addr}\ket{b \oplus d_{addr}},\label{qram} $
 where $d_{addr}$ represents the data stored at the address $addr$ (the $\oplus$ represents modulo addition, as usual), which is the reversible variant of conventional RAM memories. In \cite{2008_Giovannetti}, it was shown a $qRAM$ can be constructed such that its internal processing scales logarithmically in the number of memory cells. }.
The same qRAM structures can be also used to realize oracles utilized in the approaches based on quantum search. However, having access to quantum databases pre-filled with classical data does a-priori not imply that quantum amplitude encoded states can also be generated efficiently, which is, at least implicilty, assumed in most works below. For a separate discussion on the cost of some of similar assumptions, we refer the reader to \cite{2015_Aaronson}.

\paragraph{Amplitude encoding: algorithms}
With all the quantum tools in place, we can now present a selection of quantum algorithms for various supervised and unsupervised learning tasks, grouped according to the class of problems they solve.
The majority of proposals of this section follow a clear paradigm: the authors investigate established ML approaches, and identify those where the computationally intensive parts can be reduced to linear algebra problems, most often, diagonalization and/or equation solving. In this sense, further improvements in quantum linear algebra approaches, are likely to lead to new results in quantum ML. 

As a final comment, all the algorithms below pertain to discrete-system implementations. Recently, in  \cite{2017_Lau}, the authors have also considered continuous variable variants of qRAM, QLS and DME, which immediately lead to continuous variables implementations of all the quantum tools and most quantum-enhanced ML algorithms listed below.
\vspace{0.2cm}

\underline{{Regression algorithms}}\
One of the first proposals for quantum enhancements tackled linear regression problems, specifically, least squares fitting, and relied on QLS. In least squares fitting, we are given $N$ M-dimensional real datapoints paired with real labels, so $(\mathbf{x}_i, y_i)_{i =1}^{N},$   $\mathbf{x}_i = (x^j_i)_j \in \mathbbmss{R}^{M}, \mathbf{y} = (y_i)_i \in \mathbbmss{R}^N.$ In regression $y$ is called the response variable (also regressant or dependant variable), whereas the datapoints $\mathbf{x}_i$ are called predictors (or regressors or explanatory variables), and the goal of least-squares linear regression is to establish the best linear model, that is $\bm{\beta} = (\beta_j)_{j} \in  \mathbbmss{R}^{M}$ given with \EQ{
\textup{argmin}_{\bm{\beta}} \|  \mathbf{X} \bm{\beta} - \mathbf{y} \|^2,
}
 where the data matrix $\mathbf{X}$ collects the data-points $\mathbf{x}_i$ as rows. In other words, linear regression assumes a linear relationship between the predictors and the response variables. 
 It is well-established that the solution to the above least-squares problem is given with $\bm{\beta} =  \mathbf{X}^+  \mathbf{y}$, where $\mathbf{X}^+ $ is the Moore-Penrose pseudoinverse of the data-matrix, which is, in the case that $\mathbf{X}^\dagger \mathbf{X}$ is invertible, given with 
 $\mathbf{X}^+  = (\mathbf{X}^\dagger \mathbf{X})^{-1} \mathbf{X}^\dagger$. The basic idea in \cite{2012_Wiebe} is to  apply $\mathbf{X}^\dagger$ onto the initial vector $\ket{\mathbf{y}}$ which amplitude-encodes the response variables, obtaining a state proportional to $\mathbf{X}^\dagger \ket{\mathbf{y}}$. This can be done e.g. by  modifying the original QLS algorithm \cite{2009_Harrow} to imprint not the inverses of eigenvalues but the eigenvalues themselves. Following this, the task of applying $(\mathbf{X}^\dagger \mathbf{X})^{-1}$ (onto the generated state proportional to $\mathbf{X}^\dagger \ket{\mathbf{y}}$) is interpreted as an equation-solving problem for the system $(\mathbf{X}^\dagger \mathbf{X}) \bm{\beta} = \mathbf{X}^\dagger \mathbf{y}$.
 
The end result is a quantum state $\ket{\bm{\beta}}$ proportional to the solution vector $\bm{\beta},$ in time $O(\kappa^4 d^3 \log(N)/\epsilon),$ where $\kappa, d$ and $\epsilon$ are the condition number, the sparsity of the ``symmetrized'' data matrix $\mathbf{X}^\dagger \mathbf{X}$, and the error, respectively. Again, we have in general few guarantees on the behaviour of $\kappa$, and an obvious restriction on the sparsity $d$ of the data-matrix. However, whenever both are $O(\textup{polylog}(N))$, we have a potential\footnote{In this section we often talk about the ``potential'' for exponential speed-ups because some of the algorithms as given do not solve classical computational problems for which classical lower bounds are known. Consider the conditions which have to be satisfied for the QLS algorithm to offer exponential speed-ups.
First, we need to be dealing with problems where the preparation of the initial state and qRAM memory can be done in $O(polylog(N))$. Next, the problem condition number must be $O(polylog(N))$ as well. Assuming all this is satisfied, we are still not done: the algorithm generates a quantum state. As classical algorithms do not output quantum states, we cannot talk about quantum speed-ups. The quantum state can be measured, outputting at most $O(polylog(N))$ (more would kill exponential speed-ups due to printout alone) bits which are functions of the quantum state. However, the hardness of computing these output bits, given all the initial assumptions is clearly not obvious, needs to be proven. } for exponential improvements. This algorithm is not obviously useful for actually finding the solution vector $\bm{\beta},$ as it is encoded in a quantum state. Nonetheless, it is useful for estimating the quality of fit: essentially by applying $\mathbf{X}$ onto $\ket{\bm{\beta}}$ we obtain the resulting prediction of $\mathbf{y},$ which can be compared to the actual response variable vector via a swap test efficiently\footnote{In the paper, the authors take care to appropriately symmetrize all the matrices in a manner we discussed in a previous footnote, but for clarity, we ignore this technical step.}.  

 These basic ideas for quantum linear regression have since been extended in a few works.  In an extensive, and complementary work \cite{2014_Wang}, the authors rely on the powerful technique of ``qubitization'' \cite{2016_Low}, and optimize the goal of actually producing the best-fit parameters $\bm{\beta}$. By necessity, the complexity of their algorithm is proportional to the number of data-points $M$, but is logarithmic in the data dimension $N$, and quite efficient in other relevant parameters.
  In \cite{2016_Schuld}, the authors follow the ideas of \cite{2012_Wiebe} more closely, and achieve the same results as in the original work also when the data matrix is not sparse, but rather low-rank. Further, they improve on the complexities by using other state-of-the-art methods. This latter work critically relies on the technique of DME. \vspace{0.2cm}

\underline{{Clustering algorithms}}
In \cite{2013_Lloyd}, amplitude encoding and inner product estimation are used to estimate the distance $\|\mathbf{u} -  \bar{\mathbf{v}} \|$ between a given data vector $\mathbf{u}$ and the average of a collection of data points (centroid) $\bar{\mathbf{v}}  = \sum_i  \mathbf{v}_i /M$ for $M$ datapoints $\{ \mathbf{v}_i\}_i$, in time which is logarithmic in both the vector length $N$, and number of points $M$.
Using this as a building block, the authors also show an algorithm for $k$-means classification/clustering (where the computing of the distances to the centroid is the main cost), achieving an overall complexity $O(M\log(MN)/\epsilon),$ which may even further be improved in some cases. Here, it is assumed that amplitude-encoded state vectors, and their normalization values, are accessible via an oracle, or that they can be efficiently implemented from a qRAM storing all the values.
Similar techniques, combined with coherent quantum phase estimation, and Grover-based optimization, have been also used for the problem of $k$-nearest neighbour algorithms for supervised and unsupervised learning
\cite{2015_Wiebe}.\vspace{0.2cm}

\underline{{Quantum Principal Component Analysis}} 
The ideas of DME were in the same paper \cite{2014_Lloyd} immediately applied to a quantum version of principal component analysis (PCA). PCA constitutes one of the most standard unsupervised learning techniques, useful for dimensionality reduction but, naturally, has a large scope of applications beyond ML.
 In quantum PCA, for a quantum state $\rho$ one applies quantum phase estimation of the unitary $\exp(-i [\rho])$ using DME, applied onto the state $\rho$ itself. In the ideal case of absolute precision, given the  spectral decomposition $\rho = \sum_i \lambda_i \dm{\lambda_i},$ this process  generates the state
$
 \sum_i \lambda_i \dm{\lambda_i} \otimes  | \tilde{\lambda_i}  \rangle \langle \tilde{\lambda_i}|,
 $
where $\tilde{\lambda_i}$ denotes the numerical estimation of the eigenvalue $\lambda_i,$ corresponding to the eigenvector $\ket{\lambda_i}.$ Sampling from this state recovers both the (larger) eigenvalues, and the corresponding quantum states, which are amplitude-encoding the eigenvectors, which may be used in further quantum algorithms. The recovery of high-value eigenvalues and eigenvectors constitutes the essence of classical PCA as well.\vspace{0.2cm}

\underline{Quantum Support Vector Machines}
One of the most influential papers in quantum-enhanced ML relies on QLS and DME for for the task of quantizing support vector machine algorithms. For the basic ideas behind SVMs see section \ref{SVMs}.

We focus our attention to the problem of training SVMs, as given by the optimization task in its dual form, in Eq. (\ref{inner}), repeated here for convenience:

\EQ{
(\alpha^\ast_1 ,\ldots \alpha^\ast_N) = \textup{argmin}_{\alpha_1 \ldots \alpha_N} \sum_{i} \alpha_i - \frac{1}{2} \sum_{i,j} \alpha_i  \alpha_j  y_i y_j\mathbf{x}_i.\mathbf{x}_j \label{inner2},\
\textup{such\ that}\ \alpha_i\geq 0\ \textup{and}\ \sum_{i}\alpha_iy_i =0.\ \nonumber
}
The solution of the desired SVM is then easily computed by $\mathbf{w}^\ast = \sum_{i}y_i\alpha_i \mathbf{x}_i.$

As a warm-up result, in \cite{2014_Rebentrost} the authors point out that using quantum evaluation of inner products, appearing in Eq. (\ref{inner2}), already can lead to exponential speed-ups, with respect to the data-vector dimension $N$. The quantum algorithm complexity is, however, still polynomial in the number of datapoints $M$, and the error dependence is now linear (as the error of the inner product estimation is linear). The authors proceed to show that full exponential improvements can be possible (with respect to $N$ and $M$ both), however for the special case of least-squares SVMs.
Given the background discussions we have already done with respect to DME and QLS, the basic idea is here easy to explain. Recall that the problem of training least-squares SVMs reduces to a linear program, specifically a least-squares minimization. As we have seen previously, such minimization reduces to equation solving, which was given by the system in Eq. (\ref{LSSVM:sys}), which we repeat here:
\EQ{
\left[{\begin{matrix}0&1_{}^{T}\\1_{N}&\Omega +\gamma ^{-1}I_{}\end{matrix}}\right]\left[{\begin{matrix}b\\\bm{\alpha} \end{matrix}}\right]=\left[{\begin{matrix}0\\Y\end{matrix}}\right].
}
Here, $1$ is an ``all ones'' vector, $Y$ is the vector of labels $y_i$, $\bm{\alpha}$ is the vector of the Lagrange multipliers yielding the solution, $b$ is the offset, $\gamma$ is a parameter depending on the hyperparameter $C$, and $\Omega$ is the matrix collecting the (mapped) inner products of the training vectors so $\Omega_{i,j} = \mathbf{x}_i.\mathbf{x}_j$. 
The key technical aspects of \cite{2014_Rebentrost} demonstrate how the system above is realized in a manner suitable for QLS. To give a flavour of the approach, we will simply point out that the system sub-matrix $\Omega$ is proportional to the reduced density matrix of the quantum state $\sum_i |\mathbf{x}_i| \ket{i}_{1}\ket{\mathbf{x}_i}_{2},$ obtained after tracing out the subsystem 2. This state can, under some constraints, be efficiently realized with access to qRAM encoding the data-points. Following this, DME enables the application of QLS where the system matrix has a block proportional to $\Omega$, up to technical details we omit for brevity. 
The overall quantum algorithm generates the quantum state proportional to $\ket{\psi_{out}} \propto b\ket{0} + \sum_{i=1}^M \alpha_i \ket{i},$ encoding the offset and the multipliers. The multipliers need not be extracted from this state by sampling. Instead 
any new point can be classified by (1) generating an amplitude-encoded state of the input, and (2) estimating the inner product between this state and $\ket{\psi_{out}'} \propto b\ket{0}\ket{0} + \sum_{i=1}^M \alpha_i |\mathbf{x}_i| \ket{i}\ket{\mathbf{x}_i},$ which is obtained by calling the quantum data oracle using $\ket{\psi_{out}}$. This process has an overall complexity of $O(\kappa_{eff}^3 \epsilon^{-3} \log (MN))$, where $\kappa_{eff}$ depends on the eigenstructure of the data matrix. Whenever this term is polylogarithmic in data size, we have a potential for exponential improvements.\vspace{0.2cm} 

\underline{Gaussian process regression}
In \cite{2015_Zhao} the authors demonstrate how QLS can be used to dramatically improve Gaussian process regression (GPR), a powerful supervised learning method. GPR can be thought of as a stochastic generalization of standard regression: given a training set $\{ \mathbf{x}_i, y_i\}$, it models the latent function (which assigns labels $y$ to data-points), assuming Gaussian noise on the labels $f(\mathbf{x}) = y + \epsilon$ where $\epsilon$ encodes independent and identically distributed 
More precisely, GPR is a process in which an initial distribution over possible latent functions is refined by taking into account the training set points, using Bayesian inference.
Consequently, the output of GPR is, roughly speaking, a distribution over models $f$ which are consistent with the observed data (the training set). While the descriptions of such a distribution may be large, in computational terms, to predict the value of a new point $\mathbf{x}_\ast,$ in GPR, one needs to compute two numbers: a \textit{linear predictor} (also referred to as the \emph{predictive mean}, or simply \emph{mean}), and the \textit{variance of the predictor}, which are specific to  $\mathbf{x}_\ast.$ These numbers characterize the distribution of the predicted value $y_\ast$ by the GPR model which is consistent with the training data.  Further, it turns out, both values can be computed using modified QLS algorithms. The fact that this final output size is independent from the dataset size, combined with QLS, provides possibilities for exponential speed-ups in terms of data size. This, naturally holds, provided the data is available in qRAM, as is the case in most algorithms of this section. It should be mentioned that the authors take meticulous care of listing out all the ``hidden costs'', (and the working out intermediary algorithms) in the final tally of the computational complexity.\vspace{0.2cm}

\underline{Geometric and topological data analysis}
All the algorithms we presented in this subsection thus far critically depend on having access to ``pre-loaded'' databases -- the loading itself would introduce a linear dependence on the database size, whereas the inner-product, QLS and DME algorithms provide potential for just logarithmic dependence. However, this can be circumvented in the cases where the data-points in the quantum database can be efficiently computed individually. This is reminiscent of the fact that most applications of Grover's algorithm have a step in which the Grover oracle is efficiently computed. In ML applications, this can occur if the classical algorithm requires, as a computational step, a combinatorial exploration of the (comparatively small) dataset. Then, the quantum algorithm can generate the combinatorially larger space in quantum parallel -- thereby efficiently computing the effective quantum database.
The first example where this was achieved was presented in \cite{2016_Lloyd}, in context of topological and geometric data analysis. These techniques are very promising in the context of ML, as topological features of data do not depend on the metric of choice, and thus capture the truly robust, features of the data. The notion of topological features (in the ML world of discrete data points) are given by those properties which exist when data is observed at different spatial resolutions. Such persistent features are thus robust and less likely to be artefacts of noise, or choice of parameters, and are mathematically formalized through so-called persistent homology. A particular family of features of interest are the number of connected components, holes, voids (or cavities). These numbers, which are defined for simplicial complexes (roughly, a closed set of simplices), are called Betti numbers. To extract such features from data, one must thus construct nested families of simplical complexes from the data, and compute the corresponding features captured by the Betti numbers. However, there are combinatorially many simplices one should consider, and which should be analyzed, and one can roughly think of each possible simplex as data-points which need further analysis.
However, they are efficiently generated from a small set -- essentially the collection of the pair-wise distances between datapoints. The authors show how to generate quantum states which encode the simplexes in logarithmically few qubits, and further show that from this representation, the Betti numbers can be efficiently estimated. Iterating this at various resolutions allows the identification of persistent features.
As usual, full exponential improvements happen under some assumptions on the data, and here they are manifest in the capacity to efficiently construct the simplical states -- in particular, having the total number of simplices in the complex be exponentially large would suffice, although it is not clear when this is the case, see \cite{2015_Aaronson}.
 This proposal provides evidence that  quantum ML methods based on amplitude encoding may, at least in some cases, yield exponential speed-ups even if data is not pre-stored in a qRAM or an analogous system.

As mentioned a large component of modern approaches to quantum-enhanced ML, relies on quantum linear algebra techniques, and any progress in this area may lead to new quantum ML algorithms. A promising recent examples of this were given in terms of algorithms for quantum gradient descent \cite{2016b_Rebentrost,2017_Kerenidis}, which could e.g. lead to novel quantum methods for training neural networks.

 \section{Quantum learning agents, and elements of quantum AI}

\label{sec:QAI}
The topics discussed thus far in this review, with few exceptions, deal with the relationship between physics, mostly QIP, and traditional ML techniques which allow us to better understand data, or the process which generates it.
In this section, we go one step beyond data analysis and optimization techniques and address the relationship between QIP and more general learning scenarios, or even between QIP and AI.
As mentioned, in more general learning or AI discussions, we typically talk about agents, interacting with their environments, which may be, or more often fail to be, intelligent. In our view, by far the most important aspect of any intelligent agent, is its capacity to learn from its interactions with its environment. However, general intelligent agents learn in environments which are complex and changeable. Further, the environments are susceptible to being changed by the agent itself, which is the crux of e.g. learning by experiments. All this delineates general learning frameworks, which begin with RL, from more restricted settings of data-driven ML. 

In this section, we will consider physics-oriented approaches to learning via interaction, specifically the PS model, and then focus on quantum-enhancements in the context of RL\footnote{Although RL is a particularly mathematically clean model for learning by interaction, it is worthwhile to note it is not fully general -- for instance learning in real environments always involves supervised and other learning paradigms to control the size of the exploration space, but also various other techniques which occur when we try to model settings in continuous, or otherwise not turn-based fashion.}. 
Following this, we will discuss an approach for considering the most general learning scenarios, where the agent, the environment and their interaction, are treated quantum-mechanically: this constitutes a quantum generalization of the broad AE framework, underlying modern AI. We will finish off briefly discussing other results from QIP which may play a role in the future of QAI, which do not directly deal with learning, but which may still play a role in the future of QAI.
\subsection{Quantum learning via interaction}
\boxTLDR{The first proposal which addressed the specification of learning agents, designed with the possibility of \textbf{quantum  processing of episodic memory} in mind, was the model of Projective Simulation PS. The results on quantum improvements of agents which learn by interacting with classical environments have mostly been given within this framework. The PS agent deliberates by effectively projecting itself into conceivable situations, using its memory, which organizes its episodic experiences in a stochastic network. Such an agent can solve basic RL problems, meta-learn, and solve problems with aspects of generalization. 
The deliberation is a stochastic diffusion process, allowing for a few routes for quantization. Using \textbf{quantum random walks}, quadratic speed-ups can be obtained. 
}
\label{QLI}
The applications of QIP to reinforcement and other interactive learning problems has been comparatively less studied, when compared to quantum enhancements in supervised and unsupervised problems. One of the first proposals which provides a coherent view on learning agents from a physics perspective was that of Projective Simulation (abbrv. PS) \cite{2012_Briegel}. 
We first provide a detailed description the PS model, and review the few other works related to this topic at the end of the section.
PS is a flexible framework for the design of learning agents motivated both from psychology and physics, and influenced by modern views on robotics.   
One of the principal reasons why we focus on this model is that it provides a natural route to quantization, which will be discussed presently. However already the classical features of the model reveal an underlying physical perspective which may be of interest for the reader, and which we briefly expose first.

The PS viewpoint on (quantum) agents is conceived around a few basic principles.
First, in the PS view, the agent is a physical, or rather, an \textbf{embodied entity}, existing relative to its environment, rather than a mathematical abstraction\footnote{For instance, the Q-learning algorithm (see section \ref{subsec:RL}) is typically defined without an embodied agent-environment context. Naturally, we can easily promote this particular abstract model to an agent, by defining an agent which internally runs the Q-learning algorithm.}. Note, this does not prohibit computer programs to be agents: while the print-out of the code is not an agent, the executed instantiation of the code, the running program, so to speak, has its own well-defined virtual interfaces, which delineate it from, and allow interaction with other programs in its virtual world -- in this sense, that program too is embodied. 
Second, the interfaces of the agent are given by its sensors, collecting the environmental input, and the actuators, enabling the agent to act on the environment. 
Third, the learning is \textit{learning from experience}, and, the interfaces of the agent constrain the elementary experiences of the agent to be collections from the sets of percepts $\mathcal{S} = \{ s_i\}_i$ which the agent can perceive and actions $\mathcal{A}= \{a_i \}_i$. At this point we remark that the basic model assumes discretized time, and sensory space, which is consistent with actual realizations, although this could be generalized.
Fourth, a (good) learning agent's behaviour -- that is, the choice of actions, given certain percepts  -- is based on its cumulative experience, accumulated in the agent's memory, which is structured. This brings us to the central concept of the PS framework, which is the memory of the agent: the \textit{episodic and compositional memory} (ECM).

 The ECM is a structured network of units of experience which are called \textit{clips} or episodes. A clip, denoted $c_i$, can represent\footnote{Representation means that we, strictly speaking, distinguish actual percepts, from the memorized percepts, and the same for actions. This distinction is however not crucial for the purposes of this exposition.} an individual percept or action, so $c_i \in \mathcal{S}\cup\mathcal{A} $ -- and indeed there is no other external \textit{type} appearing in the PS framework. However, experiences may be more complex (such as an autobiographical episodic memory, similar to short video-clips, where we remember a temporally extended sequence of actions and percepts that we experienced). This brings us to the following recursive definition: a clip is either a percept, an action, or a structure over clips.
 \begin{wrapfigure}{l}{0.6\textwidth}
 \includegraphics[width=0.62\textwidth,clip=true,trim =60 290 240 170]{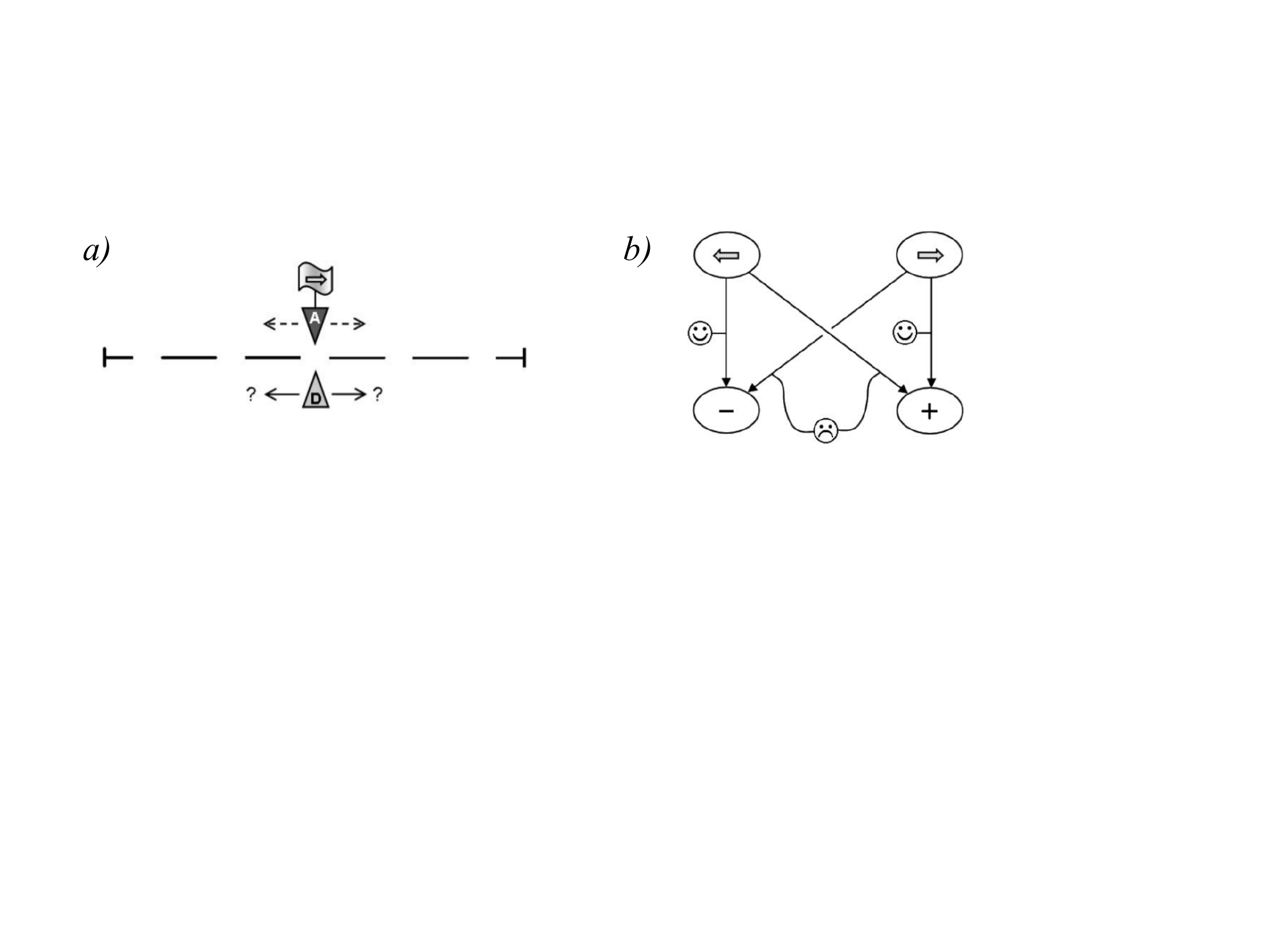}
 \vspace{-2.0cm}
\caption{\label{fig:inv}a) The agent learns to associate symbols to one of the two movements. b) the internal PS network requires only action and percept clips, arranged in two layers, with connections only from actions to percepts. The ``smiling'' edges are rewarded. Adapted from \cite{2012_Briegel}. \vspace{-0.3cm}}
 \end{wrapfigure}

Typical examples of structured clips are percept-action $(s_1, a_1, \ldots, s_k, a_k)$ sequences describing what happened, i.e. a $k-$length history of interaction between the agent and environment. Another example are simple sets of percepts  $(s_1\ \textup{or}\ s_2\ldots)$, which will be later used to generalize knowledge. The overall ECM is a network of clips (that is, a labeled directed graph, where the vertices are the clips), where the edges organize the agent's previous experiences, and has a functional purpose explained momentarily. 
Fifth, learning \textit{agent} must \textit{act}: that is, there has to be a defined deliberation mechanism, which given a current percept, the state of memory, i.e. the current ECM network, the agent, probabilistically decides on (or rather ``falls into'') the next action and performs it. Finally, sixth,  a \textit{learning} agent must learn, that is, the ECM network must change under experiences and this occurs in two modes: by (1) changing the weights of the edges, and (2) the topology of the network, through the addition of deletion of clips.
The above six principles describe the basic blueprint behind PS agents. The construction of a particular agent will require us to further specify certain components, which we will exemplify using the simplest example: a reinforcement learning PS agent, capable of solving the so-called \textit{invasion game}. In the invasion game, the agent Fig \ref{fig:inv} is facing an attacker, who must be blocked by appropriately moving to the left or right. These two options form the actions of the agent.

The attacker presents a symbol, say a left- or right- pointing arrow, to signal what its next move will be. Initially, the percepts have no meaning for the agent, and indeed the attacker can alter the meaning in time. The basic scenario here is, in RL terms a contextual two-armed bandit problem \cite{2008_Langford}, where the agent gets rewarded when it correctly couples the two percepts to the two actions.

The basic PS agent that can solve this is specified as follows. 
The action and percept spaces are the two moves, and two signals, so $\mathcal{A} = \{- , + \}$ (left and right move), and $\mathcal{S} = \{\leftarrow , \rightarrow \}$, respectively. The clips set is just the union of the two sets. The connections are directed edges from percepts to actions, weighted with real values, called $h-$values, $h_{ij}\geq 1$, which form the $h-$matrix. 
The deliberation is realized by a random walk in the memory space, governed proportionally to the $h-$matrix: that is the probability of transition from percept $s$ to action $a$ is given with $p(a | s) = \dfrac{h_{s,a}}{\sum_{a'}h_{s,a'}}.$ In other words, the column-wise normalized $h-$matrix specifies the stochastic transition matrix of the PS model, in the Markov chain sense. 
Finally, the learning is manifest in the tuning of the $h-$values, via an \textbf{update rule}, which is in its most basic form given with:
\EQ{h^{t+1}(c_j , c_i) = h^{t}(c_j , c_i) +\delta_{c_j,c_i}   \lambda,}
where $t,t+1$ denote consecutive time steps, $ \lambda$ denotes the reward received in the last step, and $\delta_{c_j,c_i}$ is 1 if and only if the $c_i$ to $c_j$ transition occurred in the previous step. 
Simply stated, used edges get rewards. The $h-$value $h^t(c_i, c_j)$ associated to the edges connecting clips $c_i, c_j$, when the time step $t$ is clear from context we will simply denote $h_{ij}$.

One can easily see that the above rule constitutes a simple RL mechanism, and that it will indeed over time lead to a winning strategy in the invasion game; since only the correctly paired transitions get rewards, they are taken more and more frequently. However, these $h-$values in this simple process diverge, which also makes re-learning, in the eventuality the rules of the game change, more difficult with time. To manage this, one typically introduces a decay, or dissipation, parameter $\gamma$ leading to the rule:

\EQ{h^{t+1}(c_j , c_i) = h^{t}(c_j , c_i) - \gamma(h^{t}(c_j , c_i)-1) + \delta_{c_j,c_i}   \lambda.}
The dissipation is applied at each time step.

  Note that since the dissipating term diminishes the values of $h^{t}(c_j , c_i)$ by an amount proportional to the deviation of these values from 1, where 1 is the initial value.
The above rule leads to the unit value $h=1$ when there are no rewards, and a limiting upper value of $1+ \lambda/\gamma$, when every move is rewarded.

  \begin{wrapfigure}{r}{0.5\textwidth}
 \includegraphics[width=0.51\textwidth,clip=true,trim =60 50 160 90]{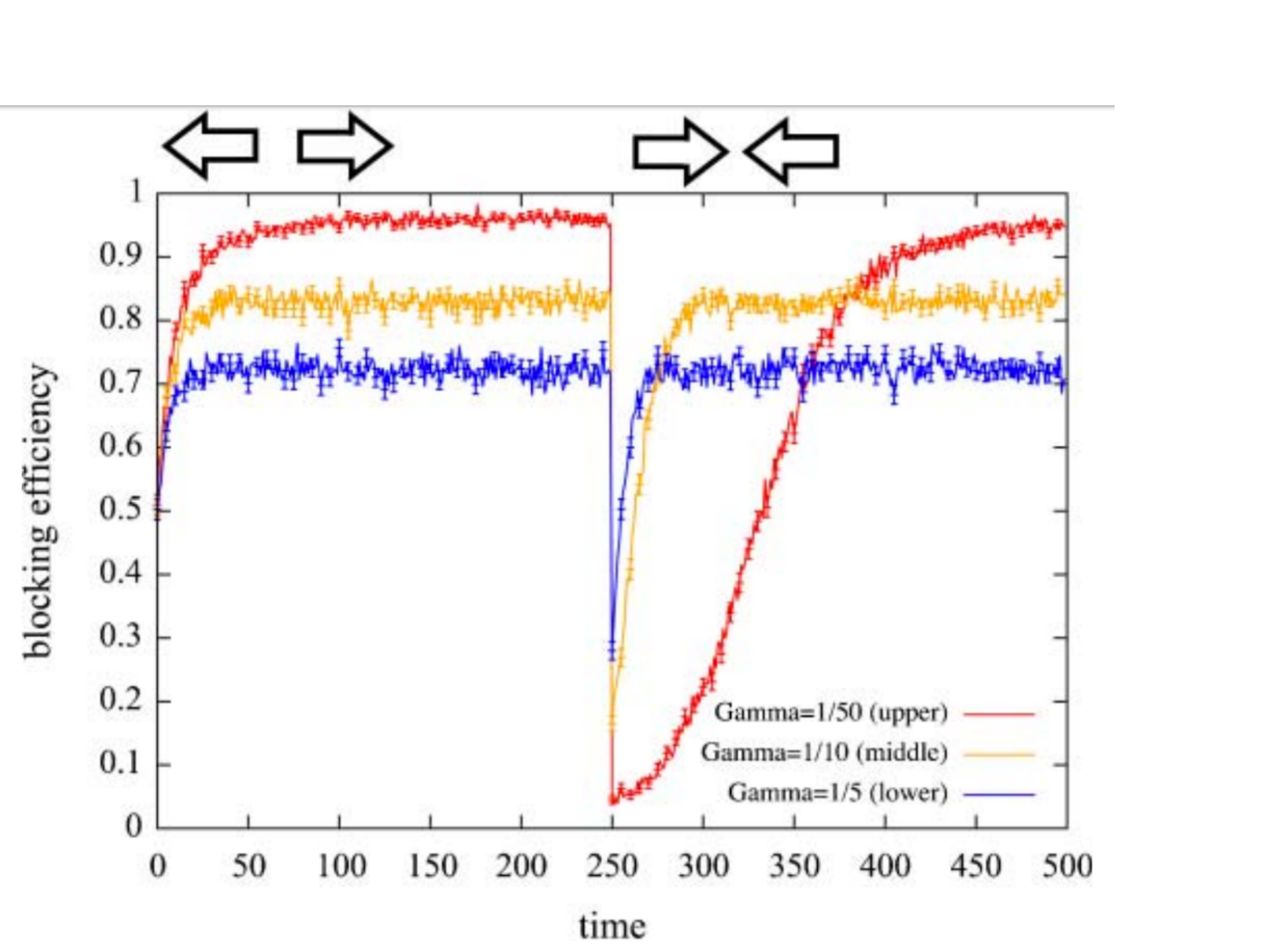}
 \vspace{-.5cm}
\caption{\label{fig:PsPlots}Basic learning curves for PS with non-zero $\gamma$  in the invasion game with a rules switch at time step 250. Adapted from \cite{2012_Briegel}. \vspace{-0.1cm}}
 \end{wrapfigure}
 
 This limits maximal efficiency to $1-(2+\lambda/\gamma)^{-1}$, but, as a trade-off, leads to much faster re-learning. 
This is illustrated in Fig. \ref{fig:PsPlots}.

\begin{wrapfigure}{r}{0.4\textwidth}
 \includegraphics[width=0.41\textwidth,clip=true,trim =130 100 100 100]{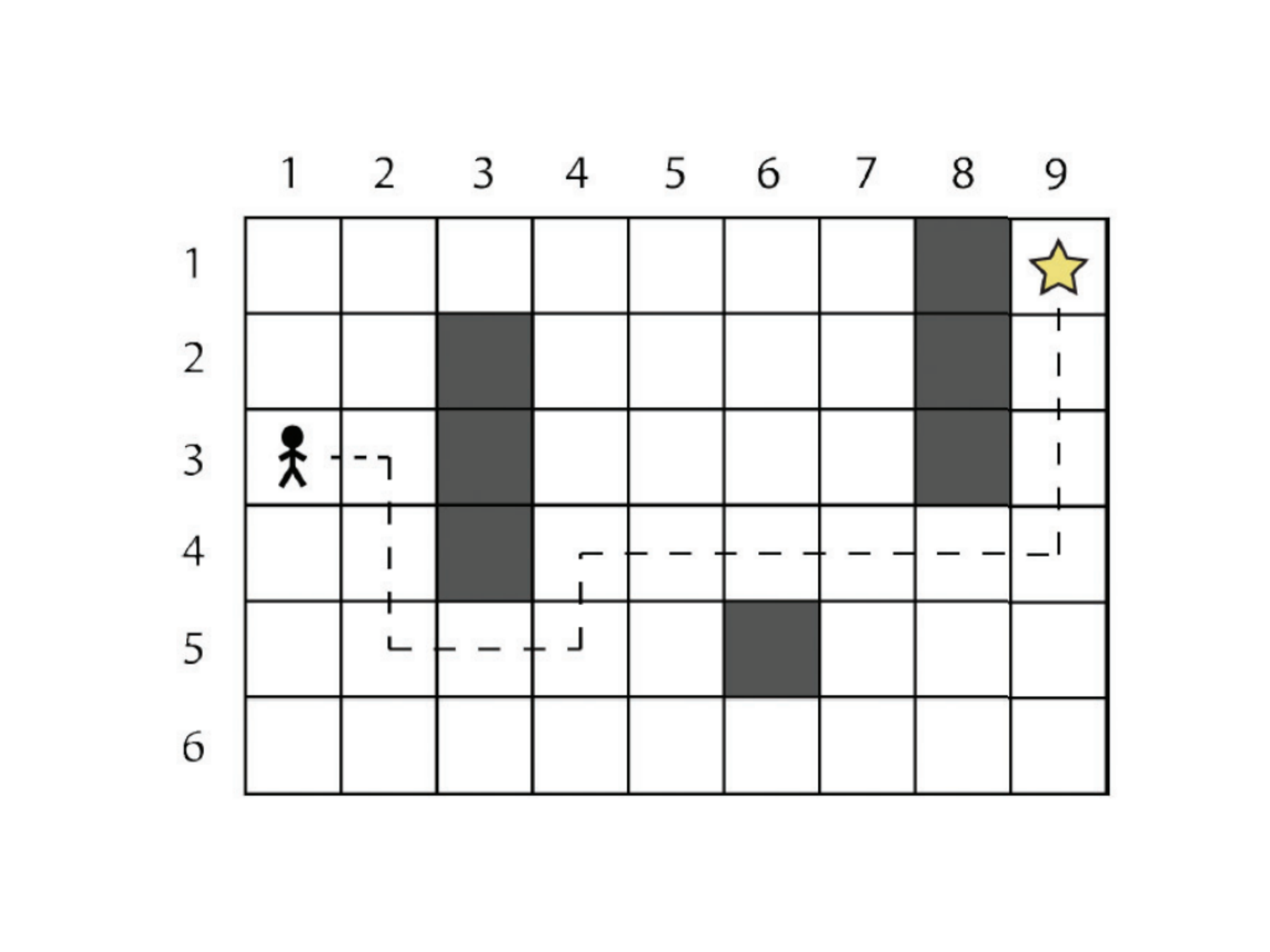}
 \vspace{-.5cm}
\caption{\label{fig:grid}The environment is essentially a grid, where each site has an individual percept, the moves dictate the movements of the agent (say up, down, left, right), and certain sites are blocked off -- walls. The agent explores this world looking for the rewarded site. When the exit is found, a reward is given and the agent is reset to the same initial position. Adapted from \cite{2014_Melnikov}.    \vspace{-0.3cm}}
 \end{wrapfigure}

The update rules can get a bit more involved, in the setting of delayed rewards. For instance, in a maze, or so called grid-world settings, illustrated in Fig. \ref{fig:grid}, it is a sequence of actions that leads to a reward. In other words, the final reward must ``propagate'' to all relevant percept-action edges which were involved in the winning move sequence.

In the basic PS model, this is done via a so-called \textit{glow} mechanism: to each edge in the ECM, a glow value $g_{ij}$ is assigned in addition to the $h_{ij}-$value. It is set to 1 whenever the edge is used, and decays with the rate $\eta \in [0,1]$, that is, $g^t_{ij} = (1-\eta)g^{t-1}_{ij}.$ The $h-$value update rule is appended to reward all ``glowing'' edges, proportional to the glow value, whenever a reward is issued: 

\EQ{
h^{t+1}(c_j , c_i) = h^{t}(c_j , c_i) - \gamma(h^{t}(c_j , c_i)-1) + g^t(c_j,c_i)\lambda.\label{PSall}}

In other words, all the edges which contributed to the final reward get a fraction, in proportion to how recently they were used. This parallels the intuition that the more recent actions relative to the rewarded move played a larger role in getting rewarded.

The expression in Eq. \ref{PSall} has functional similarities to the Q-learning action-value update rule in Eq. \ref{q-value}.
However, the learning dynamics is different, and the expressions are conceptually different -- Q-learning updates estimate bounded Q-values, whereas the PS is not a state-value estimation method, but rather a purely reward-driven system. 

\begin{wrapfigure}{l}{0.4\textwidth}
 \includegraphics[width=0.41\textwidth,clip=true,trim =100 250 170 150]{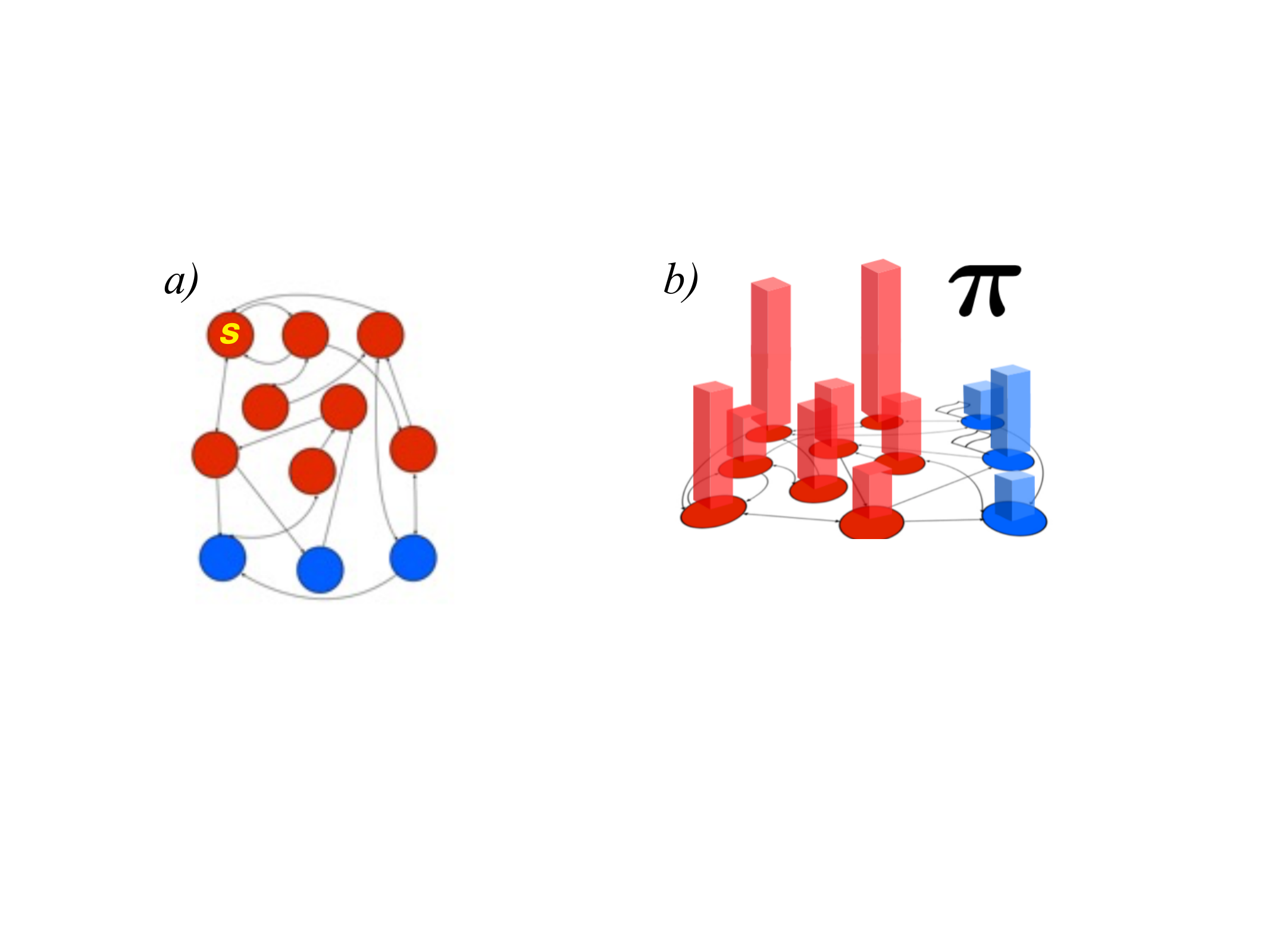}
 \vspace{-.5cm}
\caption{\label{fig:rPS} QrPS representation of network, and its steady state over non-action (red) and action (blue) clips.  \vspace{-0.cm}}
 \end{wrapfigure}

The PS framework allows other constructions as well. In \cite{2012_Briegel}, the authors also introduced emoticons -- edge-specific flags, which capture aspects of intuition. These can be used to speed-up re-learning via a \textit{reflection mechanism}, where a random walk can be iterated multiple times, until a desired -- flagged -- set of actions is hit, see \cite{2012_Briegel} for more detail. Further in this direction, the deliberation of the agent can be based not on a \textit{hitting process} -- where the agent performs the first action it hits -- but rather on a \textit{mixing process}. In the latter case, the ECM is a collection Markov chains, and the correct action is sampled from the stationary distribution over the ECM. This model is referred to as the reflective PS (rPS) model, see Fig. \ref{fig:rPS}. Common to all models, however, is that the deliberation process is governed by a stochastic walk, specified by the ECM.

Regarding performance, the basic PS structure, with a two-layered network encoding percepts and actions -- which matches standard tabular RL approaches --  was extensively analysed and benchmarked against other models \cite{2014_Melnikov, 2015_Mautner}. However, the questions that are emphasized in PS literature diverge from questions of performance in RL tasks, in two directions. First, the authors are interested in the capacities of the PS model beyond textbook RL.

For instance, in \cite{2015_Mautner} it was shown that the action composition aspects of the ECM allow the agent to perform better in some benchmarking scenarios, which had a natural application for example in the context of protecting MBQC from unitary noise \cite{2014_Tiersch}, and in the context of finding novel quantum experiments \cite{2017_Melnikov}, elaborated on in section \ref{control-in-lab}. Further, by utilizing the capacity of ECM to encode larger and multiple networks, we can also address problems which require generalization \cite{2015_Melnikov} -- inferring correct behaviour by percept similarity -- but also design agents which autonomously optimize their own meta-parameters, such as $\gamma$ and $\eta$ in the PS model. That is, the agents can meta-learn \cite{2016_Makmal}. These problems go beyond the basic RL framework, and the PS framework is flexible enough to also allow the incorporation of other learning models -- e.g. neural networks could be used to perform dimensionality reduction (which could allow for broader generalization capabilities), or even to directly optimize the ECM itself. The PS model has been combined with such additional learning machinery in an application to robotics and haptic skill learning \cite{2016_Hangl}. 

 However, there is an advantage into keeping the underlying PS dynamics homogenous, that is, essentially solely based on random walks over the PS network, in that if offers a few natural routes to quantization. This is the second direction of foundational research in PS.  
For instance, in \cite{2012_Briegel} the authors expressed the entire classical PS deliberation dynamics as a incoherent part of a Liouvillean dynamics (master equation for the quantum density operator), which also included some coherent part (Hamiltonian-driven unitary dynamics). This approach may yield advantages both in deliberation time and also expands the space of internal policies the agent can realize.

Another perspective on the quantization of the PS model was developed in the framerowk of discrete-time quantum walks.
In \cite{2014_Paparo}, the authors have exploited the paradigm of Szegedy-style quantum walks, to improve quadratically deliberation times of rPS agents. The Szegedy \cite{2004_Szegedy} approach to random walks can be used to specify a unitary random walk operator $U_P$, for a given transition matrix $P$\footnote{By transition matrix, we mean an entry-wise non-negative matrix, with columns adding to unity.}, whose spectral properties are intimately related to those of $P$ itself. We refer the reader to the original references for the exact specification of $U_{P}$, and just point out that $U_P$ can be efficiently constructed via a simple circuit depending on $P$, or given black-box access to entries of $P$.

Assume $P$ corresponds to an irreducible and aperiodic (guaranteeing a unique stationary distribution), and also time-reversible (meaning it satisfies detailed balance conditions) Markov chain. Let $\bm{\pi} = (\pi_i)_i$ be the unique stationary distribution of $P$, and $\delta$ the spectral gap of $P$\footnote{The spectral gap is defined with $\delta = 1 - |\lambda_2|$, where $\lambda_2$ is, in norm, the second largest eigenvalue.}, and $\ket{\bm{\pi}} = \sum_{i} \sqrt{\pi_i} \ket{i}$ be the coherent encoding of the distribution $\bm{\pi}.$ Then we have that \textit{a)} $U_{P} \ket{\bm{\pi}}  = \ket{\bm{\pi}} $, and \textit{b)} the eigenstates $\{ \lambda_i\}$ of $P$ and eigenphases $\theta_i$ of $U_P$ are related by $\lambda_i = cos(\theta_i)$\footnote{In full detail, these relations hold whenever the MC is lazy (all states transition back to themselves with probability at least 1/2 ), ensuring that all the eigenvalues are non-negative, which can be ensured by adding the identity transition with probability 1/2. This slows down mixing and hitting processes by an irrelevant factor of 2.  }. 
This is important as the spectral properties, specifically the spectral gap $\delta$ more-or-less tightly fixes the mixing time -- that is the number of applications of $P$ needed to obtain the stationary distribution -- to $\tilde{O}(1/\delta)$, by the famous Aldous bounds \cite{1982_Aldous}. This quantity will later bound the complexity of classical agents. In contrast, for $U_{P},$ we have that its non-zero eigenphases $\theta$ are not smaller than $\tilde{O}(1/\sqrt{\delta}).$ This quadratic difference between the inverse spectral eigenvalue gap in the classical case, and the eigenphase gap in the quantum case is at the crux of all speed-ups. 
 In \cite{2011_Magniez}, it was shown how the above properties of $U_P$ can be used to construct a quantum operator $R(\bm{\pi}) \approx \mathbbmss{1}- 2\dm{\bm{\pi}},$ which exponentially efficiently approximates the reflection over the encoding of the stationary distribution $\ket{\bm{\pi}}$. The basic idea in the construction of $R(\bm{\pi})$ is to apply phase estimation onto $U_P$ with precision high enough to detect non-zero phases, impose a global phase on all states with a non zero detected phase, and undo the process.

 Due to the quadratic relationship between the inverse spectral gap, and the smallest eigenphase, this can be achieved in time $\tilde{O}(1/\sqrt{\delta})$. That is, we can reflect over the (coherent encoding of the) stationary distribution, whereas obtaining it by classical mixing takes $\tilde{O}(1/\delta)$ applications of the classical walk operator.
 In \cite{2014_Paparo} this was used to obtain quadratically accelerated deliberation times for the rPS agent.
In the rPS model, the ECM network has a special structure, enforced by the update rules. In particular, for each percept $s$ we can consider the subnetwork ECM$_s$, which collects all the clips one can reach starting from $s$. By construction, it contains all the action clips, but also other, intermediary clips. The corresponding Markov chain $P_s$, governing the dynamics of  ECM$_s$, is, by construction, irreducible, aperiodic and time-reversible. In the deliberation process, given percept $s$, the deliberation process mixes the corresponding Markov chain $P_s$, and outputs the reached clip, provided it is an action clip, and repeats the process otherwise.

Computationally speaking, we are facing the problem of outputting a single sample, clip $c$, drawn according to the conditional probability distribution $p(c) = \pi_c/\epsilon$ if $c \in \mathcal{A}$ and $p(c)=0$ otherwise. Here $\epsilon$ is the total weight of all action clips in $\bm{\pi}.$ The classical computational complexity of this task is given by the product of $\tilde{O}(1/\delta)$ -- which is the mixing cost, and $O(1/\epsilon)$ which is the average time needed to actually hit an action clip.
Using the Szegedy quantum walk techniques, based on constructing the reflector $R(\bm{\pi})$, followed by an amplitude amplification algorithm to ``project'' onto the action space, we obtain a quadratically better complexity of 
$\tilde{O}(1/\sqrt{\delta})\times O(1/\sqrt{\epsilon})$. In full detail, this is achievable if we can generate one  copy of the coherent encoding of the stationary distribution efficiently at each step, and in the context of the rPS this can be done in many cases as was shown in  \cite{2014_Paparo}, and further generalized in \cite{2015_Dunjko} and \cite{2015bb_Dunjko}.

 The proposal in \cite{2014_Paparo} was the first example of a provable quantum speed-up in the context of RL \footnote{We point out that the first ideas suggesting that quantum effects could be useful had been previously suggested in \cite{2005_Dong}. }, and was followed up by a proposal for an experimental demonstration \cite{2015c_Dunjko}, which identified a possibility of a modular implementation based on coherent controlization -- the process of adding control to \textit{almost} unknown unitaries.

It is worth-while to note that further progress in algorithms for quantum walks and quantum Markov chain theory has the potential to lead to quantum improvements of the PS model. This to an extent mirrors the situation in quantum machine learning, where new algorithms for quantum linear algebra may lead to quantum speed-ups of other supervised and unsupervised algorithms.

Computational speed-ups of deliberation processes in learning scenarios are certainly  important, but in strict RL paradigm, such internal processing does not matter, and the learning efficiency depends only on the number of interaction steps needed to achieve high quality performance. Since the rPS and its quantum analog, the so-called quantum rPS agent are, by definition, behaviorally equivalent (i.e. they perform the same action with the same probability, given identical histories), their learning efficiency is the same.
The same, however, holds in the context of all the supervised learning algorithms we discussed in previous sections, where the speed-ups were in the context of computational complexity. In contrast, quantum CLT learning results did demonstrate improvements in sample complexity, as discussed in section \ref{qPAC}.

While formally distinct, computational and sample complexity can become more closely related the moment the learning settings are made more realistic.
For instance, if the training of a given SVM requires the solution of a BQP complete problem\footnote{BQP stands for \textit{bounded-error quantum polynomial}, and collects decision problems which can be solved with bounded error using a quantum computer. Complete problems of a given class are, in a sense, the hardest problems in that class, as all other are reducible to the complete instances using weaker reductions. In particular, it is not believed BQP complete problems are solvable on a classical computer, whereas all decision problems solvable by classical computers do belong to the class BQP.}, classical machines will most likely be able to run classification instances which are uselessly small. In contrast, a quantum computer could run such a quantum-enhanced learner. The same observation motivates most of research into quantum annealers for ML, see section \ref{Adiabatic}.

In  \cite{2014_Paparo}, similar ideas were more precisely formalized in the context of active reinforcement learning, where the interaction is occurring relative to some external real time. This is critical, for instance, in settings where the environment changes relative to this real time, which is always the case in reality. 
If the deliberation time is slow relative to this change, the agent perceives a ``blurred'', time-averaged environment where one cannot learn. In contrast, a faster agent will have time to learn before the environment changes -- and this makes a qualitative difference between the two agents. 
In the next section we will show how actual learning efficiency, in the rigid metronomic turn-based setting can also be improved, under stronger assumptions.

As mentioned, works which directly apply quantum techniques to RL, or other interactive modes of learning, are comparatively few in numbers, despite the ever growing importance of RL. These results still constitute quite isolated approaches, and we briefly review two recent papers.
In \cite{2016_Crawford} the authors design a RL algorithm based on a deep Boltzmann machine, and combine this with quantum annealing methods for training such machines to achieve a possible speed-up.This work combines multiple interesting ideas, and may be particularly relevant in the light of recent advances in quantum annealing architectures. In \cite{2017_Lamata}, the authors demonstrated certain building blocks of larger quantum RL agents in systems of superconducting qubits. 
 \subsection{Quantum agent-environment paradigm for reinforcement learning}
 \boxTLDR{To characterize the ultimate scope and limits of learning agents in quantum environments, one must first establish a framework for quantum agents, quantum environments and  their interaction: \textbf{a quantum AE paradigm}. Such a paradigm should maintain the correct classical limit, and preserve the critical conceptual components -- in particular the history of the agent-environment interaction, which is non-trivial in the quantum case. With such a paradigm in place the potential of quantum enhancements of classical agents is explored, and it is shown that quantum effects, under certain assumptions, can help \textbf{near-generically improve the learning efficiency of agents}. 
 A by-product of the quantum AE paradigm is a classification of learning settings, which is different and complementary to the classification stemming from a supervised learning perspective.}
  \label{QAE}
The topics of learning agents acting in quantum environments, and the more general questions of the how agent-environment interactions should be defined, have to this day only been broached in few works by the authors of this review and other co-authors. As these topics may form the general principles underlying the upcoming field of quantum AI, we take liberty to present them to substantial detail.

\begin{wrapfigure}{r}{0.5\textwidth}
\underline{{\small \textsf{RL:}}}\\
\includegraphics[width=0.47\textwidth, trim=0cm 12.9cm 0cm 12.4cm,clip=true]{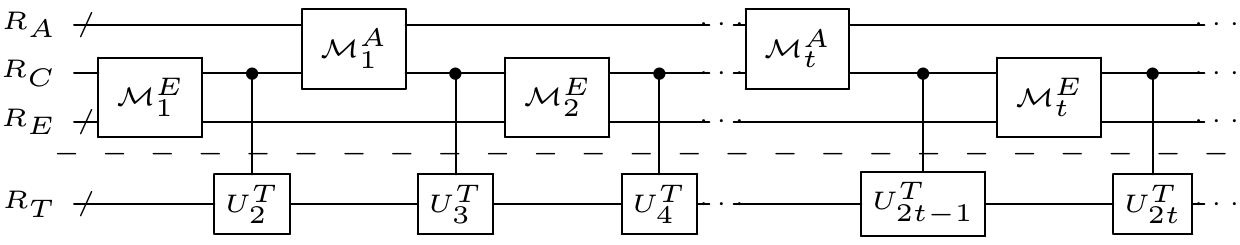}\\
\vspace{-1cm}\underline{{\small \textsf{DL:}}}\\
\includegraphics[width=0.47\textwidth, trim=0cm 12.9cm 0cm 12.4cm,clip=true]{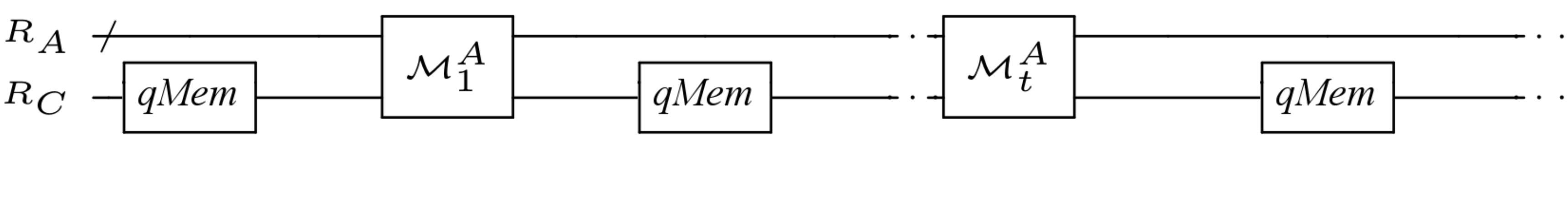}\\
\vspace{1cm}
\caption{\label{fig:QAE}
RL: Tested agent-environment interaction suitable for RL. In general, each map of the tester $U^T_k$ acts on a fresh subsystem of the register $R_T$, which is not under the control of the agent, nor of the environment. The crossed wires represent multiple systems.
DL: The simpler setting of standard quantum machine learning, where the environmental map is without internal memory, presented in the same framework. 
}
 \end{wrapfigure}

Motivated by the pragmatic question of the potential of quantum enhancements in general learning settings, in \cite{2016_Dunjko} it was suggested that the first step should be the identification of a quantum generalization of the AE paradigm, which underlies both RL and AI. This is comparatively easy to do in finite-sized, discrete space settings.
\paragraph{Quantum agent-environment paradigm}
The (abstract) AE paradigm, roughly illustrated in Fig. \ref{fig:AE}, can be understood as a two-party communication scenario, the quantum descriptions of which are well-understood in QIP. In particular, the two players -- here the agent, and the environment -- are modelled as (infinite) sequences of unitary maps
$\{ \mathcal{E}_A^i\}_i$, and $\{ \mathcal{E}_E^i\}_i$, respectively. They both have private memory registers $R_A$ and $R_E$, with matching Hilbert spaces $H_A,$ and $H_E$, and to enable precise specification of how they communicate (and to cleanly delineate the two players), the register of the communication channel, $R_C$, is introduced, and it is the register which is alone accessible to \textit{both} players -- that is, the maps of the agent act on $H_A\otimes H_C$ and of the environment on $H_E\otimes H_C$\footnote{Other delineations are possible, where the agent and environment have individually defined interfaces -- a part of E accesible to A and a part of A accessible to E -- leading to a four-partite system, but we will not be considering this here \cite{2015b_Dunjko}. }. 
The two players then interact by sequentially applying their respective maps in turn (see Fig. \ref{fig:QAE}).

To further tailor this fully general setting for the AE paradigm purposes, 
the percept and action sets are promoted to sets of orthonormal vectors $\{ \ket{s} | s \in \mathbf{S}\}$ and $\{ \ket{a} | a \in \mathbf{A} \}$, which are also mutually orthogonal. These are referred to as classical states. The Hilbert space of the channel is spanned by these two sets, so $H_C = \textup{span} \{ \ket{x} |\ x \in \mathbf{S} \cup  \mathbf{A}\}.$

This also captures the notion that the agent/environment only performs one action, or issues one percept, per turn. Without loss of generality, we can also assume that the state-spaces of the agent's and environment's registers are also spanned by sequences of percepts and actions. It is without loss of generality assumed that the reward status is encoded in the percept space.
It should be mentioned that the quantum AE paradigm also includes all other quantum ML settings as a special case. For instance, most quantum-enhanced ML algorithms assume access to  quantum database, a quantum memory, and this setting is illustrated in Fig. \ref{fig:QAE}, part DL. Since the quantum database is without loss of generality a unitary map, it requires no additional memory of its own, nor does it change over interaction steps.

At this point, the classical AE paradigm can be recovered when the maps of the agent and environment are restricted to ``classical maps'', which, roughly speaking do not generate superpositions of classical states, nor entanglement when applied to classical states.

Further, we now obtain a natural classification of generalized AE settings: CC, CQ, QC and QQ, depending on whether the agent or the environment are classical (C) or quantum (Q). {We will come back to this classification in section \ref{Taxonomy}.}

The performance of a learning agent, beyond internal processing time, is a function of the \textit{history of interaction}, which is a distribution over percept-action sequences (of a given finite length) which can occur between a given agent and environment. 
Any \textit{genuine learning-related} figure of merit, for instance, the probability of a reward at a given time-step (efficiency), or number of steps needed before the efficiency is above a threshold (learning speed) is a function of the interaction history. 
In the classical case, the history can simply be read out by a classical-basis measurement of the register $H_C$, as the local state of the communication register is diagonal in this basis, and not entangled to the other systems -- meaning the measurement does not perturb, i.e. commutes with the interaction.
In the quantum case this is not, in general, the case. To recover a robust notion of a history (needed for gauging of the learning), a more detailed description of measurement is used, which captures weaker measurements as well: an additional system, a \textit{tester} is added, which interchangeably couples to the $H_C$ register, and can copy full or partial information to a separate register. Formally, this is a sequence of controlled maps, relative to the classical basis, controlled by the states on $H_C$ and acting on a separate register, as illustrated in Fig. \ref{fig:QAE}.
The tester can copy the full information, when the maps are a generalized controlled-NOT gate -- in which case it is called a classical tester -- or even do nothing, in which case the interaction is untested. The restriction of the tester to maps which are controlled with respect to the classical basis guarantees that a classical interaction will never be perturbed by its presence. 
With this basic framework in place, the authors show a couple of basic theorems characterizing when any quantum separations in learning-related figures of merit of can be expected at all. The notion of quantum separations here are the same as in the context of oracular computation, or quantum PAC theory: a separation means no classical agent could achieve the same performance. 
The authors prove basic expected theorems: quantum improvements (separations) require a genuine quantum interaction, and, further, full classical testing prohibits this. 
Further, they show that for any specification of a classical environment, there exists a ``quantum implementation" -- a sequence of maps $\{ \mathcal{E}_E^i\}_i$ -- which is consistent with the classical specification, and prohibits any quantum improvements. 
\paragraph{Provable quantum improvements in RL}
 \begin{wrapfigure}{lh}{0.5\textwidth}
 \includegraphics[width=0.51\textwidth,clip=true,trim =0 0 0 0]{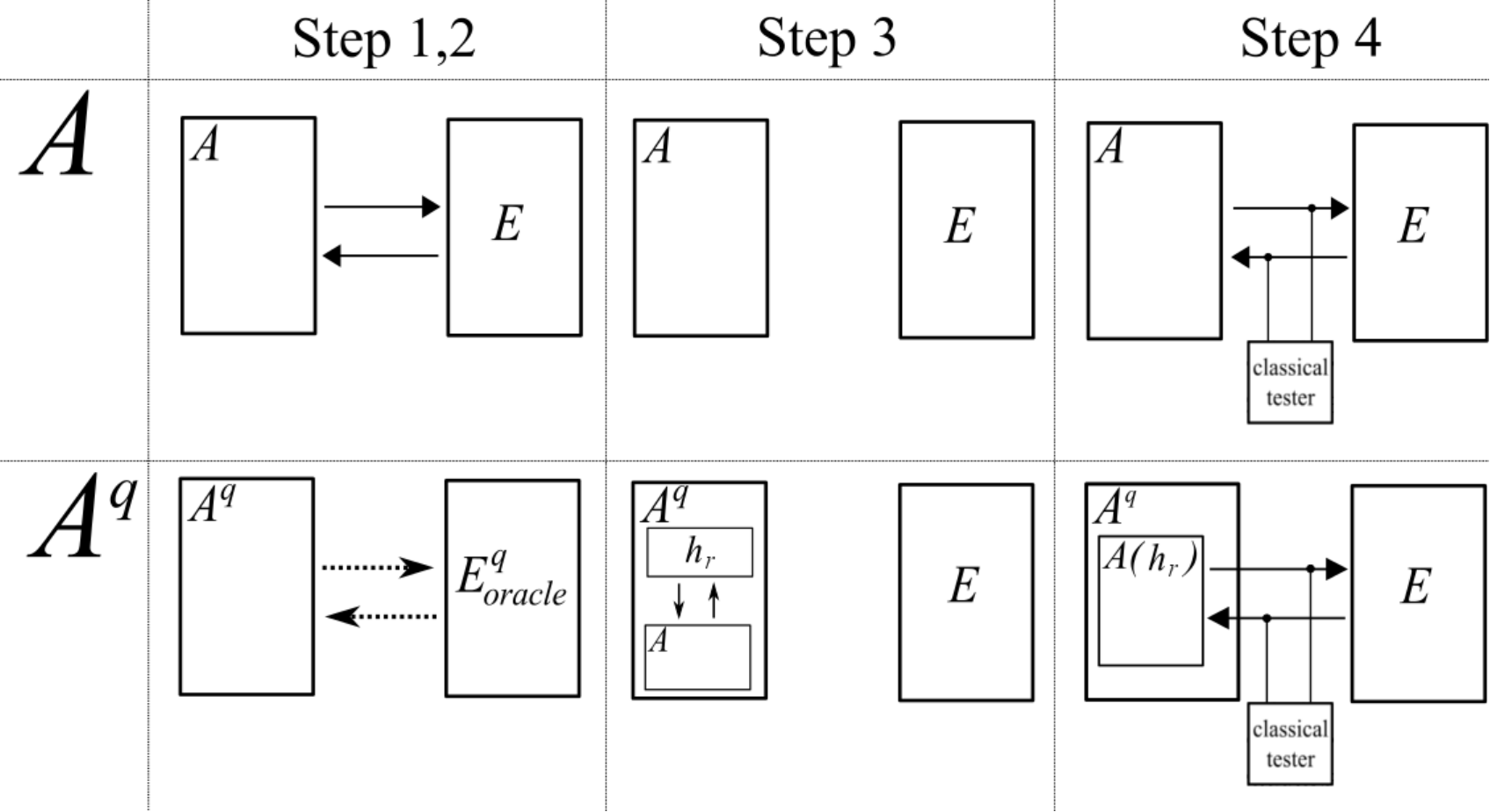}
 \vspace{-0.5cm}
\caption{\label{fig:aqc}
The interactions for the classical  ($A$) and quantum-enhanced classical agent ($A^q$). In Steps 1 and 2, $A^q$ uses quantum access to an oracularized environment $E^{q}_{oracle}$ to obtain a rewarding sequence $h_r$. Step 3: $A^q$ simulates the
agent $A$, and `trains' the simulation to produce the rewarding
sequence. In Step 4, $A^q$ uses the pre-trained agent for the remainder of the
now classically tested interaction, with the classical environment
$E$. Adapted from \cite{2016_Dunjko}.\vspace{0cm}}
 \end{wrapfigure}
 However, if the above no-go scenarios are relaxed, much can be achieved. The authors provide a structure of task environments (roughly speaking, maze-type problems), specification of quantum-accessible realizations of these environments, and a sporadic tester (which leaves a part of the interaction untested), for which classical learning agents can often be quantum-enhanced. 
The idea has a few steps, which we only very briefly sketch out. As a first step, the environments considered are deterministic and strictly episodic -- this means the task is reset after some $M$ steps. 
Since the environments are deterministic, whether or not rewards are given depends \textit{only} on the sequence of actions, as the interlacing percepts are uniquely specified. Since everything is reset after $M$ steps there are no correlations in the memory of the environment between the blocks, i.e. episodes. This allows for the specification of a quantum version of the same environment, which can be accessed in superpositions and which takes blocks of actions and returns the same sequence plus a reward status -- moreover, it can be realized such that it is self-inverse\footnote{This realization is possible under a couple of technical assumptions, for details see \cite{2015b_Dunjko}.}. With access to such an object, a quantum agent can actually Grover-search for an example of a winning sequence.
To convert this exploration advantage to a learning advantage, the set of agents and environments is restricted to pairs which are ``luck-favoring'', i.e. those where better performance in the past implies improved performance in the future, relative to a desired figure of merit. 
Under these conditions, any learning agent which is luck-favoring relative to a given environment can be quantum enhanced by first using quantum access to quadratically faster find the first winning instance, which is then used to ``pre-train'' the agent in question. The overall quantum-enhanced agent, provably outperforms the basic classical agent. 
The construction is illustrated in Fig. \ref{fig:aqc}.
These results can be generalized to a broader class of environments. 
Although these results form the first examples of quantum improvements in learning figures of merit in RL contexts, the assumptions of having access to ``quantized'' environments of the type used--in essence, the amount of quantum control the agent is assumed to have-- are quite restrictive from a practical perspective. 
The questions of minimal requirements, and the questions of the scope of improvements possible are still unresolved. 
\subsubsection{AE-based classification of quantum ML}
\label{Taxonomy}
The  AE paradigm is typically encountered in the contexts of RL, robotics, and more general AI settings, while it is less common in ML communities. 
Nonetheless, conventional ML scenarios can naturally be embedded in this paradigm, since it is, ultimately, mostly unrestrictive. For instance, supervised learning can be thought of as an interaction with an environment which is, for a certain number of steps, an effective database (or the underlying process, generating the data), providing training examples. After a certain number of steps, the environment starts providing unlabeled data-points, and the agent responds with the labels. If we further assume the environment additionally responds with the correct label to whatever the agent sent, when the data-point/percept was from the training set, we can straightforwardly read out the empirical risk (training set error)  from the history.
Since the quantization of the AE paradigm naturally leads to four settings -- CC, CQ, QC and QQ -- depending on whether the agent, or environment, or both are fully quantum systems, we can classify all of the results in quantum ML into one of the four groups. Such coarse grained division places standard ML in CC, results on using ML to control quantum systems in CQ, quantum speed ups in ML algorithms (without a quantum database, as is the case in annealing approaches) in QC, and quantum ML/RL where the environments, databases or oracles are quantum-accessible are QQ. 
This classification is closely related to the classification introduced in \cite{2006_Aimeur_Gambs}, which uses the $L^{context}_{goal},$ notation, where ``context'' may denote we are dealing with classical or quantum data and/or learner, and ``goal'' specifies the learning task (see section \ref{QGML} for more details). 

The QAE-based separation is not, however, identical to it: for instance classical learning tasks may require quantum or classical access -- this distinguishes the examples of quantum speed-ups in internal processing in ML which require a quantum database, and those which do not. In operative terms, this separation makes sense as the database must be pre-filled at some point, and if this is included we obtain a QC setting (which now may fail to be efficient in terms of communication complexity).
On the other hand, the $L^{context}_{goal}$ systematics does a nice job separating classical ML, from quantum generalizations of the same, discussed in   
section \ref{QgenML}. This mismatch also illustrates the difficulties one encounters if a sufficiently coarse-grained classification of the quantum ML field is required. 
 The classification criteria of this field, and also aspects of QAI in this review have been inspired by both the AE-induced criteria (perhaps natural from a physics perspective), and the $L^{context}_{goal}$ classification (which is more objective driven, and natural from a computer science perspective).

\subsection{Towards quantum artificial intelligence}
\boxTLDR{Can quantum computers help us build (quantum) artificial intelligence? The answer to this question cannot be simpler than the answer to the to the deep, and largely open, question of what intelligence is in the first place. Nonetheless, at least for very pragmatic readings of AI, \textbf{early research directions into what QAI may be} in the future can be identified. We have seen that quantum machine learning enhancements and generalizations cover data analysis and pattern matching aspects. Quantum reinforcement learning demonstrates how interactive learning can be quantum-enhanced. General QC can help with various planning, reasoning, and similar symbol manipulation tasks intelligent agents seem to be good at. Finally, the quantum AE paradigm provides a framework for the design and evaluation of whole quantum agents, built also from quantum-enhanced subroutines.
These conceptual components form a \textbf{basis for a behaviour-based theory of quantum-enhanced intelligent agents}.
}
\label{TQO}

AI is quite a loaded concept, in a manner in which ML is not. The question of how genuine AI can be realized is likely to be as difficult as the more basic question of what intelligence is at all, which has been puzzling philosophers and scientists for centuries. Starting a broad discussion of when \textit{quantum} AI will be reached, and what will be like, is thus clearly ill-advised.
We can nonetheless provide a few less controversial observations.
The first observation is the fact that the overall concept of quantum AI might have multiple meanings. First, it may pertain to a generalization of the very notions of intelligence, in the sense section \ref{QgenML} discusses how classical learning concepts generalize to include genuinely quantum extensions. 
A second, and a perhaps more pragmatic reading of quantum AI, may ask whether quantum effects can be utilized to generate more intelligent agents, where the notion of intelligence itself is not generalized: \textbf{quantum-enhanced artificial intelligence}. We will focus on this latter reading for the remainder of this review, as quantum generalization of basic learning concepts on its own, just as the notion of intelligence on its own, seem complicated enough.

To comment on the question of quantum-enhanced AI, we first remind the reader that the conceptual debates in AI often have two perspectives. The ultimately pragmatic perspective is concerned only with behavior in relevant situations. This is perhaps best captured by Alan Turing, who suggested that it may be irrelevant what intelligence is, if it can be recognized, by virtue of similarity to a ``prototype'' of intelligence -- a human \cite{1950_Turing} \footnote{Interestingly, the Turing test assumes that humans are good supervised learners of the concept of ``intelligent agents'', all the while being incapable of specifying the classifier -- the definition of intelligence -- explicitly. }. 
Another perspective tends to try to capture cognitive architectures, such as \textit{SOAR} developed by John Laird, Allen Newell, and Paul Rosenbloom \cite{2012_Soar}. Cognitive architectures try to identify the components needed to build intelligent agents, capable of many tasks, and thus also care about \textit{how} the intelligence is implemented. They often also serve as models of human cognition, and are both theories of what cognition is, and how to implement it.
A third perspective comes from the practitioners of AI who often believe that AI will be a complicated combination of various methods and techniques including learning and specialized algorithms, but are also sympathetic to the Turing test as the definitional method.
A simple reading of this third perspective is particularly appearing, as it allows us to all but equate computation, ML and AI. Consequently all quantum machine learning algorithms, and even broader, even most quantum algorithms already constitute progress in quantum AI. Aspects of such reading can be found in a few works on the topic \cite{2007_Sgabas,2014_Wichert,2015_Moret-Bonillo}\footnote{It should be mentioned that some of the early discussions on quantum AI also consider the possibilities that human brains utilize some form of quantum processing, which may be at the crux of human intelligence. Such claims are still highly hypothetical, and not reviewed in this work.}.

The current status of the broad field of quantum ML and related research is showing signs of activity with respect to all of the three aspects mentioned. The substantial activity in the context of ML improvements, in all aspects presented, is certainly filling the toolbox of methods which one day may play a role in the complicated designs of quantum AI practitioners. In this category, a relevant role may also be played by various algorithms which may help in planning, pruning, reasoning via symbol manipulation, and other tasks AI practice and theory encounters. Many possible quantum algorithms which may be relevant come to mind.  Examples include the algorithm for performing Bayesian inference \cite{2014_Low},  algorithms for quadratic and super-polynomial improvements in NAND- and boolean-tree evaluations, which are important in evaluation of optimal strategies in two-player games \footnote{See http://www.scottaaronson.com/blog/?p=207 for a simple explanation.} \cite{2008_Farhi,2009_Childs,2012_Zhan}. Further, even more exotic ideas, such as quantum game theory \cite{1999_Eisert}, may be relevant.
Regarding approaches to quantum artificial general intelligence, and, related, to quantum cognitive architectures, while no proposals explicitly address this possibility, the framework of PS offers sufficient flexibility and structure that it may be considered a good starting point. Further, this framework is intended to keep a homogenous structure, which may lead to more straightforward global quantization, in comparison to models which are built out of inhomogeneous blocks -- already in classical systems, the performance of system combined out of inhomogeneous units may lead to difficult-to-control behaviour, and it stands to reason that quantum devices may have a more difficult time to be synchronized. It should be mentioned that recently there have been works providing a broad framework describing how composite large quantum systems can be precisely treated \cite{2017_Portmann}.
Finally, from the ultimate pragmatic perspective, the quantum AE paradigm presented can offer a starting point for a quantum-generalized Turing test for QAI, as the Turing test itself fits in the paradigm: the environment is the administrator of the test, and the agent is the machine trying to convince the environment it is intelligent. Although, momentarily, the only suitable referees for such a test are classical devices -- humans -- it may be conceivable they, too, may find quantum gadgets useful to better ascertain the nature of the candidate \footnote{This is reminiscent to the problem of quantum verification, where quantum Turing test is a term used for the test which efficiently decides whether the Agent is a genuine quantum device/computer \cite{QTT}}. 
However, at this point it is prudent to remind ourselves and the reader, that all the above considerations are still highly speculative, and that the research into genuine AI has barely broken ground.

\label{QAI}

\section{Outlook}
\label{sec:discussion}
In this review, we have presented overviews of various lines of research that connect the fields of quantum information and quantum computation, on the one side, and machine learning and artificial intelligence, on the other side. 
Most of the work in this new area of research is still largely theoretical and conceptual, and there are, for example, hardly any dedicated experiments demonstrating how quantum mechanics can be exploited for ML and AI. However, there are a number of theoretical proposals \cite{2015_Friis, 2017_Lamata, 2015c_Dunjko} and also first experimental works showing how these ideas can be implemented in the laboratory  \cite{2009_Neigovzen,2015_Zhaokai,2015_Cai,2017_Riste}\footnote{These complement the experimental work based on superconducting quantum annealers \cite{2009_Neven2, 2015_Adachi}, which is closely related to one of the approaches to QML.}.
 At the same time it is clear that certain quantum technologies, which have been developed in the context in QIP and QC, can be readily applied to quantum learning, to the extent that learning agents or algorithms employ elements of quantum information processing in their very design. Similarly, it is clear, and there are by now several examples, how techniques from classical machine learning can be fruitfully employed in data analysis and the design of experiments in quantum many-body physics (see section \ref{QMB}). 
One may ask about the long-term impact of the exchange of concepts and techniques between QM and ML/AI. Which implications will this exchange have on the development of the individual fields, and what is the broader perspective of these individual activities leading towards a new field of research, with its own questions and promises?
Indeed, returning the focus back to the topics of this review, we can highlight
one overarching question encapsulating the collective effort of the presented research:
\begin{itemize}
\item[\hspace{-1cm}$\Rightarrow$] \textit{What are \textbf{the potential}, and \textbf{the limitations} of an interaction between quantum physics, and ML and AI?}
\end{itemize}

From a purely theoretical perspective, we can learn from analogies with the fields of communication, computation, or sensing. QIP has shown that to understand the limits of such information processing disciplines, both in pragmatic and conceptual sense, one must consider the full extent of quantum theory. Consequently, we should expect that the limits of learning, and of intelligence can also only be fully answered in this broader context. In this sense, the topics discussed in sections \ref{QgenML} already point to the rich and complex theory describing what learning may be, when even information itself is a quantum object, and aspects of the section \ref{QAI} point to how a general theory of quantum learning may be phrased\footnote{The question of whether information may be quantum, and whether we can talk about ``quantum knowledge'' as an outside observer broaches the completely fundamental questions of interpretations of quantum mechanics: for instance a Quantum Bayesianist would likely reject such a third-person perspective on learning. }. 
The motivation of phrasing such a general theory may be fundamental, but it also may have more pragmatic consequences. 
In fact, arguments can be made that the field of quantum machine learning and the future field of quantum AI may \textbf{constitute one of the most important research fields to emerge in recent times}. A part of the reason behind such a bold claim stems from the obvious potential of both directions of influence between the two constituent sides of quantum learning (and quantum AI). 
For instance, the potential of quantum enhancements for ML is profound. In a society where data is generated at geometric rate\footnote{ https://insidebigdata.com/2017/02/16/the-exponential-growth-of-data/ (accessed July 2017) }, and where its understanding may help us combat global problems, the potential of faster, better analyses cannot be overestimated. 
In contrast, ML and AI technologies are becoming indispensable tools in all high technologies, but  they are also showing potential to help us do research in a novel, better way. 
A more subtle reason supporting optimism  
 lies in positive feedback loops between ML, AI and QIP which are becoming apparent, and which is moreover, specific to these two disciplines.
 To begin with, we can claim that QC, once realized, will play an integral part in future AI systems, on general grounds. This can be deduced from even a cursory overview of the history of AI, which reveals that qualitative improvements in computing and information technologies result in progress in AI tasks, which is also intuitive. In simple terms, state-of-the-art in AI will always rely on state-of-the-art in computing. In contrast, ML and AI technologies are becoming indispensable tools in all high technologies. 
 
 The \textbf{perfect match between ML, AI and QIP}, however may have deeper foundations. In particular,\\
 
  \textit{$\rightarrow$advancements in ML/AI may help with critical steps in the building of quantum computers}. \\

In recent times, it has become ever more apparent that learning methods may make the difference between a given technology being realizable or being effectively impossible -- beyond obvious examples, for instance direct computational approaches to build a human-level Go-playing software had failed, whereas AlphaGo \cite{2016_Silver}, a fundamentally learning AI technology, achieved this complex goal. QC may in fact end up being such a technology, where exquisite fast, and adaptive control -- realized by an autonomous smart laboratory perhaps, helps mitigate the hurdles towards quantum computers.  
However, cutting edge research discussed in sections \ref{control-in-lab} and \ref{QMB} suggest that ML and AI techniques could help at an even deeper level, by helping us discover novel physics which may be the missing link for full blown quantum technologies.
Thus ML and AI may be what we need to build quantum computers. 

Another observation, which is hinted at increasing frequency in the community, and which fully entwines ML, AI and QIP, is that\\ 

\textit{$\rightarrow$ AI/ML applications may be the best reasons to build quantum computers.}\\

 Quantum computers have been proven to dramatically outperform their classical counterparts only on a handful of (often obscure) problems. Perhaps the best applications of quantum computers that have enticed investors until recently were quantum simulation and quantum cryptology (i.e. using QC to break encryption), which may have been simply insufficient to stimulate broad-scale public investments. In contrast ML and AI-type tasks may be regarded as the ``killer applications'' QC has been waiting for.  However, not only are ML and AI applications well motivated -- in recent times, arguments have been put forward that ML-type applications may be uniquely suited to be tackled by quantum technologies. For instance, ML-type applications deal with massive parallel processing of high dimensional data -- quantum computers seem to be good for this. Further, while most simulation and numerics tasks require data stability, which is incompatible with the noise modern days quantum devices undergo, ML applications always work with noisy data. This means that such an analysis makes sense only if it is robust to noise to start with, which is the often unspoken fact of ML: the important features are the robust features. Under such laxer set of constraints on the desired information processing, various current day technologies, such as quantum annealing methods may become a possible solution.
The two main flavours, or  directions of influence, in quantum ML thus have a natural synergistic effect further motivating that despite their quite fundamental differences, they should be investigated in close collaboration.
Naturally, at the moment, each individual sub-field of quantum ML comes with its own set of open problems, key issues which need to be resolved before any credible verdict on the future of quantum ML can be made. Most fit in one of the two quintessential categories of research into quantum-enhanced topic: a) what are the limits/how much of an edge over best classical solutions can be achieved, and b) could the proposals be implemented in practice in any reasonable term.
For most of the topics discussed, both questions above remain widely open. For instance, regarding quantum-enhancements using universal computation, only a few models have been beneficially quantized, and the exact problem they solve, even in theory, is not matching the best established methods used in practice. Regarding the second facet, the most impressive improvements (barring isolated exceptions) can be achieved only under a significant number of assumptions, such as quantum databases, and certain suitable properties the structure of the data-sets\footnote{In many proposals, the condition number of a matrix depending on the dataset explicitly appears in run-time, see section \ref{qcircuit}}.
Beyond particular issues which were occasionally pointed out in various parts of this review, we will forego providing an extensive list of specific open questions for each of the research lines, and refer the interested reader to the more specialized reviews for more detail \cite{2014_Wittek, 2014a_Schuld,2016_Biamonte,2017_deWolf, 2017_Ciliberto}.\vspace{0.2cm}\\
 
This leads us to the final topic of speculation of this outlook section: whether QC will truly be instrumental in the construction of genuine artificial (general) intelligence. On one hand, there is no doubt that quantum computers could help in heavily computational problems one typically encounters in, e.g., ML. In so far as AI reduces to sets of ML tasks, quantum computing may help. But AI is more than a sum of such specific-task-solving parts. Moreover, human brains are (usually) taken as a reference for systems capable of generating intelligent behaviour. Yet there is little, and no non-controversial, reason to believe genuine quantum effects play any critical part in their performance (rather, there is ample reasons to dismiss the relevance of quantum effects). In other words, quantum computers may not be necessary for general AI. The extent to which quantum mechanics has something to say about general AI will be subject of research in years to come. Nonetheless, already now, we can set aside any doubt that quantum computers and AI can help each other, to an extent which will not be disregarded.

\section*{Acknowledgements}
The authors are grateful to Walter Boyajian, Jens Clausen, Joseph Fitzsimons, Nicolai Friis, Alexey A. Melnikov, Davide Orsucci, Hendrik Poulsen Nautrup, Patrick Rebentrost, Katja Ried, Maria Schuld, Gael Sent\'{i}s, Omar Shehab, Sebastian Stabinger, Jordi Tura i Brugu\'{e}s, Petter Wittek and Sabine W\"{o}lk for helpful comments to various parts of the manuscript.

\bibliographystyle{ustphy}

\end{document}